\documentclass[5p,times]{elsarticle}

\usepackage{amsmath}
\usepackage[]{graphicx}
\DeclareGraphicsExtensions{.pdf,.jpeg,.png}
\usepackage{subfig}
\usepackage{xspace}
\usepackage{xcolor}
\usepackage{paralist}
\usepackage{color}
\usepackage{listings}
\usepackage{listings-rust}
\usepackage{url}
\usepackage[subtle]{savetrees}

\newcommand{\fig}[1]{Fig.~\ref{fig:#1}}
\newcommand{\tab}[1]{Table~\ref{tab:#1}}
\newcommand{\s}[1]{Section~\ref{sec:#1}}

\newcommand{\fakeparagraph}[1]{\smallskip\noindent\textbf{\textit{#1.}}}

\newif\ifhidenotes
\hidenotestrue

\ifhidenotes
\newcommand{\note}[1]{}
\newcommand{\noteam}[1]{}
\newcommand{\notegc}[1]{}
\newcommand{\rev}[1]{#1}
\else
\newcommand{\note}[1]{\footnote{\color{purple}{Note: #1}}}
\newcommand{\noteam}[1]{\footnote{\color{blue}{Ale: #1}}}
\newcommand{\notegc}[1]{\footnote{\color{red}{GPaolo: #1}}}
\newcommand{\rev}[1]{{\color{blue}{#1}}}
\fi

\newcommand{\system}{Renoir\xspace}
\newcommand{\systemtt}{\texttt{Renoir}\xspace}

\begin{document}

\title{The \system Dataflow Platform: Efficient Data Processing without
  Complexity}

\author[1]{Luca De Martini\corref{cor1}}
\ead{luca.demartini@polimi.it}
\author[1]{Alessandro Margara}
\ead{alessandro.margara@polimi.it}
\author[1]{Gianpaolo Cugola}
\ead{gianpaolo.cugola@polimi.it}
\author[1]{Marco Donadoni}
\ead{marco.donadoni@mail.polimi.it}
\author[1]{Edoardo Morassutto}
\ead{edoardo.morassutto@mail.polimi.it}

\affiliation[1]{organization={Politecnico di Milano},
            addressline={Piazza Leonardo da Vinci 32},
            postcode={20133},
            city={Milano},
            country={Italy}}
\cortext[cor1]{Corresponding author}

\begin{abstract}
  Today, data analysis drives the decision-making process in virtually every
  human activity.  This demands for software platforms that offer simple
  programming abstractions to express data analysis tasks and that can execute
  them in an efficient and scalable way.
  State-of-the-art solutions range from low-level programming primitives,
  which give control to the developer about communication and resource usage,
  but require significant effort to develop and optimize new algorithms, to
  high-level platforms that hide most of the complexities of parallel and
  distributed processing, but often at the cost of reduced efficiency.

  To reconcile these requirements, we developed \system, a novel distributed
  data processing platform written in Rust.
  \system provides a high-level dataflow programming model as mainstream data
  processing systems. It supports static and streaming data, it enables data
  transformations, grouping, aggregation, iterative computations, and
  time-based analytics, and it provides all these features incurring in a low overhead.

  In this paper, we present the programming model and the implementation
  details of \system.
  We evaluate it under heterogeneous workloads.  We compare it with
  state-of-the-art solutions for data analysis and high-performance computing,
  as well as alternative research products, which offer different programming
  abstractions and implementation strategies.
  \system programs are compact and easy to write: developers need not care
  about low-level concerns such as resource usage, data serialization,
  concurrency control, and communication. At the same time, \system consistently presents
  comparable or better performance than competing solutions, by a large margin
  in several scenarios.

  We conclude that \system offers a good tradeoff between simplicity and
  performance, allowing developers to easily express complex data analysis
  tasks and achieve high performance and scalability.
\end{abstract}



\maketitle


\emergencystretch 0.85em

\section{Introduction}
\label{sec:intro}

Today, companies heavily rely on data analytics to extract actionable
knowledge from static and dynamic (streaming) datasets.
Over the last decade, this need has driven the surge of several distributed
platforms designed to process data at scale~\cite{2023_margara_csur_model,
  zaharia:CACM:2016:spark, zaharia:SOSP:2013:Discretized_Streams,
  akidau:VLDB:2015:dataflow, carbone:IEEEB:2015:flink}.
Despite their differences, they all rely on the dataflow programming and
execution model.  Pioneered by the MapReduce
framework~\cite{dean:CACM:2008:mapreduce}, this model defines data analytics
jobs as directed graphs of operators, each applying a functional
transformation to its input data and feeding downstream operators with its
output.
This approach brings twofold benefits:
\begin{inparaenum}[(i)]
\item It enables a high degree of parallelism: different operators may execute
  simultaneously on the same or different hosts (task parallelism), and each
  operator may itself be decomposed into parallel instances, each one working
  independently on one partition of the input data (data parallelism).
\item It exposes a simple programming interface that abstracts away most of
  the complexity associated to the distribution of data and processing:
  developers focus on the behavior of operators and how the input data is
  partitioned across parallel instances, while the runtime automates
  deployment, scheduling, synchronization, and communication.  Some platforms
  further increase the level of abstraction by offering custom API or domain
  specific languages for some areas, such as relational data
  processing~\cite{abouzeid:VLDB:2009:HadoopDB,
    camacho-rodriguez:SIGMOD:2019:ApacheHive, armbrust:SIGMOD:2015:SparkSQL}
  or graph processing~\cite{gonzalez:SOSP:2014:GraphX}.
\end{inparaenum}

However, despite offering a simple and effective way to scale-out data
analytics, state-of-the-art platforms cannot provide a level of performance
that is comparable to custom programs optimized for the specific problem at
hand.  As recognized in recent
literature~\cite{reyes-ortiz:BigData:2015:Spark_vs_MPI,
  gonzalez:CCGrid:2017:Spark_vs_MPI,zeuch_analyzing_2019}, custom
implementations using hand-crafted code or low-level programming primitives,
such as MPI, can yield more than one order of magnitude performance
improvements.  But this comes at a price: a much greater difficulty in
software validation, debugging, and maintenance, as programmers are exposed to
concerns related to memory management, data serialization communication, and
synchronization.

In this paper, we investigate the possibility to reduce this gap. To do so, we
introduce \system, a new distributed data processing platform that provides a
lightweight and highly efficient implementation of the dataflow model.
In terms of programming model, \system retains the same ease of use as
mainstream dataflow platforms, but has the flexibility to write optimized
solutions when necessary.
It supports the analysis of static and dynamic data (batch and stream
processing jobs), it offers a rich library of operators for data
transformation, partitioning, aggregation, and iterative computations.
Moreover, developers may customize the behavior of operators via user-defined functions,
that are deeply integrated within the programming model.

Applications written in a few lines of code using \system have been shown to
deliver similar or even better performance than much more, complex custom MPI implementations.
This is made possible by some key design and implementation choices that
enable programs written in \system to be compiled to highly efficient code,
comparable to what can be achieved with hand-written optimized code.
In particular, \system abandons JVM-based languages, typically adopted by mainstream
competitors, in favor of Rust~\cite{klabnik:2018:rust}, a compiled programming
language that offers high-level abstractions at virtually no cost. This enables  \system to work alongside the Rust compiler to generate efficient code: it
avoids dynamic dispatching, it maximises the possibilities of inlining and
monomorphization, it adopts data structures that minimize the use of pointers
and indirections to increase memory locality.

Moreover, \system adopts a lightweight approach to resource management, which
leverages the services offered by the operating system when possible.  For
example, it co-locates operators that perform different steps of a processing
job on the same host, letting them compete for CPU time based on their dynamic
requirements, while it leverages the mechanisms embedded into TCP to implement
backpressure.
While originally developed for distributed deployment with multiple machines,
\system also proved effective in parallelizing computations within a
single-process, performing on par with dedicated software libraries such as
OpenMP.
\system has been used in practice to compete to the Grand Challenge in the
2022 ACM International Conference of Distributed and Event-Based Systems,
winning the performance award for the solution with the highest throughput and
lowest latency~\cite{de_martini:DEBS:2022:analysis}.

In summary, the careful co-design of the programming model and execution
platform makes \system unique in terms of its balance between simplicity,
expressivity, and performance.
As discussed above, mainstream dataflow platforms sacrifice performance for
simplicity.
Solutions that aim to further simplify use by offering higher-level API or
domain specific languages restrict expressivity and often rely on standard
platforms for the execution of tasks, which puts a bound to their level of
performance.
Other systems such as Timely Dataflow~\cite{timely_dataflow} aim to improve
performance, but offer more complex programming model that hamper usability.
Engines optimized for specific tasks such as Dask~\cite{rocklin_dask_2015}
offer simple programming models but cannot provide the same generality as
\system.


Our work brings several contributions to the research on distributed data
processing:
\begin{inparaenum}[(1)]
\item It introduces \system, a new dataflow platform that combines the
  simplicity of mainstream data analytic systems with a level of efficiency
  that is close or even better than custom low-level code.  \system is
  available as an open-source
  project\footnote{\url{https://github.com/deib-polimi/renoir}}.
\item It presents the key design and implementation choices that affect the
  efficiency and scalability of \system.
\item It presents an extensive experimental evaluation that analyzes the
  effectiveness of \system with highly heterogeneous workloads and compares it
  with mainstream dataflow platforms, low-level custom solutions, alternative
  research proposals, and libraries for parallel computations.
\end{inparaenum}

The paper is organized as follows.  \s{background} provides background on
distributed data processing platforms and Rust.  \s{api} and \s{impl} present
the programming model and the design of \system, and \s{eval} evaluates its
performance and scalability, comparing them with alternative data processing
platforms and custom MPI programs.  \s{related} discusses related work and
\s{conclusions} draws conclusive remarks.


\section{Background}
\label{sec:background}

This section presents the programming model of distributed data processing
platforms and the key features of Rust that \system exploits to attain
simplicity and efficiency.

\subsection{Distributed data processing}
\label{sec:background:dataflow}

Modern platforms for distributed data processing rely on a dataflow
programming model~\cite{akidau:VLDB:2015:dataflow, 2023_margara_csur_model}
first introduced by Google's MapReduce~\cite{dean:CACM:2008:mapreduce}.
Computation is organized into a directed graph, whose vertices represent
operators and edges represent the flow of data from operator to operator.
Since operators do not share any state, the model promotes distribution and
parallelism by deploying operators in multiple instances, each processing an
independent partition of the input data in parallel with the others, on the
same or on different hosts.


The famous example used to illustrate the model is ``word count'', a program
to count the number of occurrences of each word in a large set of documents.
It can be expressed using two operators, the first operates in parallel on
various partitions of the input documents splitting them in words and emitting
partial counts for each word.  These partial results are then regrouped by
word and passed to the second operator that sums the occurrences of each word.
Developers need only to express how to operate on an individual document
(first operator) and how to integrate partial results for each word (second
operator).  The runtime takes care of operator deployment, synchronization,
scheduling, and data communication: the most complex and critical aspects in
distributed applications.

The dataflow model accommodates stream processing computations with only minor
adjustments.  Due to the unbounded nature of streams, developers need to
specify when certain computations are triggered and what is their scope, which
is typically expressed using \emph{windows}.  For instance, developers could
implement a streaming word count computation over a window of one hour that
advances every ten minutes, meaning that the count occurs every ten minutes
and considers only new documents produced in the last hour.

Data processing systems implemented the dataflow model using two orthogonal
execution strategies.
Systems such as Hadoop~\cite{white:2010:hadoop} and Apache
Spark~\cite{zaharia:CACM:2016:spark} dynamically schedule operator instances
over the nodes of the compute infrastructure.  Communication between operators
occurs by saving intermediate results on some shared storage, with operators
deployed as close as possible to the input data they consume.
Systems such as Apache Flink~\cite{carbone:IEEEB:2015:flink} and Google
Dataflow~\cite{akidau:VLDB:2015:dataflow} deploy all operators instances
before starting the computation. Communication takes place as message passing
among instances.
\system adopts the second strategy, which enables lower latency for streaming
computations, as it does not incur the overhead of operator scheduling at
runtime.

\subsection{Rust}
\label{sec:background:rust}

\system heavily relies on some key features of the Rust programming language
to offer a high-level API with limited performance overhead.

\fakeparagraph{Generics and static dispatch}
In Rust, developers can express data structures and functions that are
\emph{generic} over one or more types.  For instance, all \system operators
consume and produce a generic \texttt{Stream<T>}, which represents a bounded
or unbounded dataset of a generic type \texttt{T}.
This high-level construct is implemented at virtually no cost by Rust, which
adopts static dispatching.  The compiler generates a separate version of each
generic structure or function for each different way in which it is
instantiated in the program, while invocations to generic functions are
translated into direct calls to the correct version~\cite{rust_traits}.

\fakeparagraph{Memory management}
Rust provides automatic and safe deallocation of memory without the overhead
of garbage collection.
It achieves this goal through an \emph{ownership and borrowing}
model~\cite{rust_concurrency}, which represents Rust's most distinctive
feature.  In Rust, every value has an \emph{owning scope} (for instance, a
function), and passing or returning a value transfers its ownership to a new
scope.  When a scope ends, all its owned values are automatically destroyed.
A scope can lend out a value to the functions it calls: the Rust compiler
checks that a lease does not outlive the borrowed object.
All together, this model allows Rust to fully check safety of memory accesses
at compile time, also avoiding the need for (costly) runtime garbage
collection.

\fakeparagraph{Iterators and closures}
The iterator pattern is heavily used in idiomatic Rust code and enables
chaining operations over a collection of items without manually implementing
the logic to traverse the collection. In practice, operations on collections
are implemented as \emph{iterator adapters} that take in input an iterator and
produce a new iterator.
Moreover, iterator adapters are often defined as higher-order functions that
accept \emph{closures} defining their behavior as parameters.
This iterator pattern strongly resembles the dataflow model discussed above.
For this reason, we used iterators as the blueprint for \system's model and
implementation, making its API intuitive both for Rust developers and for
users of data processing platforms.

\fakeparagraph{Traits and serialization}
Traits represent a collection of functionalities (methods) that any data type
implementing that trait should offer.  Traits are widely used in Rust to bound
generics, for instance to restrict the use of a generic function only to
parameters that implement certain traits.
\system leverages traits to transparently implement parameter passing among
distributed instances of operators.  More specifically, \system requires all
data types to implement the \texttt{Serialize} and \texttt{Deserialize}
traits, and automatically generates the code that efficiently implements these
traits.


\section{Programming Interface}
\label{sec:api}

\system offers a high-level programming interface that hides most of the
complexities related to data distribution, communication, serialization, and
synchronization.

\subsection{Streams}
\label{sec:api:streams}

Streams are the core programming abstraction of \system.  A generic
\texttt{Stream<T>} represents a dataset of elements of type \texttt{T}, which
can be any type that implements the \texttt{Serialize} and
\texttt{Deserialize} traits.  Since these traits can be automatically derived
at compile time by the Serde library~\cite{serde}, developers can use their
custom data types without manually implementing the serialization logic.
Streams model both static (bounded) datasets (e.g., the content of a file) and
dynamic (unbounded) datasets, where new elements get continuously appended
(e.g., data received from a TCP link).
Streams are created by \emph{sources}, processed by \emph{operators} that
produce output streams from input streams by applying functional
transformations, and collected by \emph{sinks}.  Streams can be partitioned,
enabling those partitions to be processed in parallel. The partitions can be
distributed on multiple hosts, each one being able to process multiple
partitions in parallel within the same process using threads.

\subsection{Creating and consuming streams}
\label{sec:api:sources_sinks}

In \system, a \texttt{StreamEnvironment} holds the system configuration and
generates streams from sources.
\system comes with a library of sources.  For instance, the following snippet
uses the \texttt{IteratorSource}, which takes an iterator in input and builds
a source that emits all the elements returned by the iterator. In the example,
the iterator is \texttt{0..100}, consequently the source will emit all
integers in the range from 0 to 99.

\smallskip
\begin{footnotesize}
\begin{lstlisting}[language=Rust,style=colouredRust] 
let env = StreamEnvironment::new(config); 
let source = IteratorSource::new(0..100);
let stream = env.stream(source);
\end{lstlisting}
\end{footnotesize}
\noindent
Similarly, the \texttt{ParallelIteratorSource} builds a source consisting of
multiple instances that emit elements in parallel.
It takes in input a closure with two parameters: the total number of parallel
instances (\texttt{instances} in the code snippet below) and a unique
identifier for each instance (\texttt{id}).  The closure is executed in
parallel on every instance, getting a different \texttt{id} from \texttt{0} to
\texttt{instances-1}.  The closure must return an iterator for each instance,
with the elements that the instance emits.
In the example below, each source instance runs in parallel with the others
and produces 10 integers, the first one starting from 0, the second one
starting from 10, and so on.


\smallskip
\begin{footnotesize}
\begin{lstlisting}[language=Rust,style=colouredRust]  
let source = ParallelIteratorSource::new(
   move |id, instances| {
      let start = 10*id;
      let end = 10*(id + 1);
      start..end
   });
\end{lstlisting}
\end{footnotesize}
\noindent
Since iterators are widespread in Rust, the sources above have wide
applicability. For example, iterators are used in the Apache Kafka API for
Rust, making it straightforward for developers to build sources that read
elements from Kafka topics.

\noindent
Sinks consume output data from \system and may be used to print the results or
store them into files or external systems, such as a database.
\system provides three main sinks: \texttt{for\_each} applies a function to
each and every element in the stream, \texttt{collect} gathers all elements in
a collection and returns it as output, \texttt{collect\_channel} returns a
multi-producer multi-consumer channel that can be used by external code to
receive the outputs.
For example, the following code snippet prints all elements in the stream:

\smallskip
\begin{footnotesize}
\begin{lstlisting}[language=Rust,style=colouredRust]  
stream.for_each(|i| println!("{i}"));
\end{lstlisting}
\end{footnotesize}

\subsection{Transforming streams with operators}
\label{sec:api:operators}

Operators define functional transformations of streams.  We first present
\emph{single stream} operators, which operate on one stream and produce one
stream. We distinguish stateless and stateful single stream operators and we
discuss how they are executed in parallel over partitions of the input stream.
Next, we generalize to \emph{multiple stream} operators that process data
coming from multiple streams or produce multiple streams.

\subsubsection{Single-stream operators}

Single stream operators ingest a single stream to produce a new stream.
Examples of single stream operators are \texttt{map}, \texttt{flat\_map},
\texttt{filter} and \texttt{fold}.
A \texttt{map} operator transforms each element of the input stream into one
element of the output stream, as specified in a user-defined closure.
For instance, the following code snippet transforms a stream of integers
doubling each element to produce the output stream.

\smallskip
\begin{footnotesize}
\begin{lstlisting}[language=Rust,style=colouredRust]
stream.map(|i| i * 2);
\end{lstlisting}
\end{footnotesize}
\noindent
A \texttt{flat\_map} operator may produce zero, one, or more elements in the
output stream for each element in the input stream.  For instance, for each
integer \texttt{i} in the input stream, the following code outputs three
integers: \texttt{i}, \texttt{i} multiplied by 2, and \texttt{i} multiplied by
3.  The developer packs the output elements produced when processing an input
element into a vector and \system automatically ``flattens'' these results into
the output stream.

\smallskip
\begin{footnotesize}
\begin{lstlisting}[language=Rust,style=colouredRust]
stream.flat_map(|i: u32| vec![i, i * 2, i * 3]);
\end{lstlisting}
\end{footnotesize}
\noindent
A \texttt{filter} operator takes a predicate and retains only the input
elements that satisfy it.  For instance, the code snippet below retains only
the even numbers from the input stream.

\begin{footnotesize}
\begin{lstlisting}[language=Rust,style=colouredRust]
stream.filter(|i: u32| v % 2 == 0);
\end{lstlisting}
\end{footnotesize}
\noindent
A \texttt{fold} operator combines all elements into an \emph{accumulator} by
applying a provided closure to each element in the stream.  The closure takes
as parameters a mutable reference to the accumulator and an input element, and
updates the value of the accumulator using the input.
For instance, the example below sums all input elements: the initial value of
the accumulator is \texttt{0} and each element \texttt{i} is added to the
accumulator \texttt{sum}.

\begin{footnotesize}
\begin{lstlisting}[language=Rust,style=colouredRust]
stream.fold(0, |sum: &mut u32, i: u32| *sum += i);
\end{lstlisting}
\end{footnotesize}
\noindent
The \texttt{reduce} operator provides a more compact way to
express the same kind of computations as \texttt{fold} when the elements in
the input and in the output streams are of the same types, as exemplified by
the code below, which uses a \texttt{reduce} operator to sum the elements of a
stream.

\smallskip
\begin{footnotesize}
\begin{lstlisting}[language=Rust,style=colouredRust]
stream.reduce(|sum, i| sum + i);
\end{lstlisting}
\end{footnotesize}
\noindent
\system also supports stateful operators, which can access and modify some
state during processing.  This way, developers may implement computations
where the evaluation of an element depends on the state of the system after
processing all previous elements in the input stream.
As an example, let us consider the \texttt{rich\_map} operator, which is the
stateful version of a \texttt{map} operator.  The evaluation of an element
depends on the input element and on the state of the operator.
The listing below adopts a \texttt{rich\_map} to output the difference between
the current element and the previous one.  The value of the previous element
is stored inside the \texttt{prev} variable, which is initialized to 0 and
moved inside the closure.  When processing a new element \texttt{x}, the
closure computes the difference (\texttt{diff}), updates the state
(\texttt{prev}) with the new element, and finally outputs the difference.

\begin{footnotesize}
\begin{lstlisting}[language=Rust,style=colouredRust]
stream.rich_map({
   let mut prev = 0;
   move |x: u32| {
      let diff = x - prev;
      prev = x;
      diff
   } });
\end{lstlisting}
\end{footnotesize}

\subsubsection{User defined functions}

As the above operators well exemplify, \system allows developers to
fully customize the logic of operators by passing user defined
functions in the form of closures.
In JVM-based dataflow systems, this degree of flexibility comes with the
performance overhead of dynamic dispatching to execute the specific code
supplied by the user.  On the contrary, \system leverages Rust monomorphization, which
translates operators to type-specific code, avoiding the runtime cost of
dynamic dispatching, while also offering opportunities for further
compile-time optimization such as inlining and loop fusion.
This degree of optimization is typically available only in specialized
engines, which work on specific data structured and offer predefined
operations that cannot be customized with user-defined code.  This is the case
of engines that expose a SQL-like API, (see \s{related}).
By working in a restricted domain, they can pre-compile each operation to
high-performance code.  The software architecture of \system extends these
performance benefits to user defined functions.

As closures can express arbitrarily complex operations, including calls to
external libraries or even external systems in the middle of the dataflow
computation, they bring vast potentials to extend the system without requiring
a deep knowledge of its internals.  This means that developers can customize
the behavior of \system by using its high-level API and their customization
will be compiled and optimized together with the framework code.



\subsubsection{Parallelism and partitioning}

In \system streams usually consist of multiple partitions. For instance, the
\texttt{ParallelIteratorSource} builds a partitioned stream, where each
partition holds the data produced by a different instance of the iterator.
%
%
Partitioning is key to improve performance, as it enables multiple instances
of the same operator to work in parallel, each on a different partition. For
instance, the \texttt{map} example described above processes data in parallel,
using as many instances of the \texttt{map} operator as the number of
partitions in the input stream.
This form of parallel execution is not possible in presence of operators, like
\texttt{fold} and \texttt{reduce}, which intrinsically need to operate on the
entire set of elements in the stream. To overcome this potential bottleneck,
in presence of associative operations, \system provides an optimized,
associative version of the \texttt{fold} and \texttt{reduce} operators, which
splits the computation in two stages. First, the operation is performed on
each partition, producing a set of intermediate results, then these partial
results are combined to produce the final results. The example below shows the
associative version of the summing job introduced in the previous section:

\smallskip
\begin{footnotesize}
\begin{lstlisting}[language=Rust,style=colouredRust]
stream.reduce_assoc(|sum, i| sum + i);
\end{lstlisting}
\end{footnotesize}
\noindent
In some cases, it may be necessary to control the way in which stream elements
are associated to partitions.  For instance, given a stream of sensor
readings, if we want to count the number of readings \emph{for each sensor},
we need to ensure that all the readings of a given sensor are always processed
by the same operator instance, which computes and stores the count.
To support these scenarios, \system allows developers to explicitly control
stream partitioning with the \texttt{group\_by} operator. It takes in input a
closure that computes a \emph{key} for each element in the stream and
repartitions the stream to guarantee that all elements having the same key
will be in the same partition.  This allows performing stateful operations
with the guarantee that the instance responsible for a given key will receive
all elements with that key.

Keys can be of any type that implements the \texttt{Hash} and \texttt{Eq}
traits.
As an example, the code below organizes an input stream of integers in two
partitions, even and odd, by associating each element with a key that is 0 for
even numbers and 1 for odd numbers.  Then, it sums all elements in each
partition. The result will be a stream with two partitions, each one made of
a single element, representing the sum of all even (respectively, odd) numbers
in the original stream.

\smallskip
\begin{footnotesize}
\begin{lstlisting}[language=Rust,style=colouredRust]
stream.group_by(|i| i % 2)
   .reduce(|sum, i| sum + i);  
\end{lstlisting}
\end{footnotesize}
\noindent
Since the summing operation is associative, \system allows obtaining the same
result in a more efficient way, using an optimized operator that combines
\texttt{group\_by} and \texttt{reduce} in a more parallel, associative way. In
particular, the following code:

\smallskip
\begin{footnotesize}
\begin{lstlisting}[language=Rust,style=colouredRust]
stream.group_by_reduce(
   |i| i % 2,
   |sum, i| *sum += i
);  
\end{lstlisting}
\end{footnotesize}
\noindent
creates a separate, keyed (sub)partitioning for each original partition of the
input stream, applies the summing closure to each one of those
(sub)partitions, producing a set of intermediate results (one for each key and
each original partition of the input stream), then combines these partial
results by key, producing the final stream composed of just two partitions,
each one made of a single element: the sum of all even (respectively, odd)
numbers in the original stream.
When the key partitioning is not needed anymore (or when we want to
re-partition a non-partitioned stream) the \texttt{shuffle} operator evenly
redistributes input elements across a number of partitions decided by a
configuration parameter or by a previous invocation of the
\texttt{max\_parallelism} operator.

\subsubsection{Multi-stream operators}

\system manages the definition of multiple streams within the same environment
through the \texttt{split}, \texttt{zip}, \texttt{merge}, and \texttt{join}
operators.

\noindent The \texttt{split} operator creates multiple copies of the same stream: each
copy is independent of the others and may undergo a different sequence of
transformations.
In the code below, \texttt{s1} and \texttt{s2} are two copies of
\texttt{stream}.

\noindent The \texttt{zip} operator combines two streams, associating each element of
the first stream with one of the second and producing a stream of pairs
(tuples with two elements).  Elements are paired in order of arrival.
In the code below, after traversing independent transformations (not shown)
\texttt{s1} and \texttt{s2} are combined together using \texttt{zip}.

\smallskip
\begin{footnotesize}
\begin{lstlisting}[language=Rust,style=colouredRust]
let mut splits = stream.split(2);
let s1 = splits.pop().unwrap();
let s2 = splits.pop().unwrap();
...
let s3 = s1.zip(s2);
\end{lstlisting}
\end{footnotesize}
\noindent
Likewise, the \texttt{merge} operator applies to streams that transport the
same type of elements to produce a new stream that outputs elements as they
arrive in the input streams.

\noindent The \texttt{join} operator matches elements of a stream to those of another
stream based on their value. It does so by using a closure to extract a key
from the elements of the first and the second stream and matching elements
with the same key. For instance, the listing below joins a stream of users and
a stream of purchases (both made of pairs with user id as first entry and user
or purchase information as second) using the user id as the joining key for
both streams.
\system supports inner, outer and left joins and different joining algorithms.

\smallskip
\begin{footnotesize}
\begin{lstlisting}[language=Rust,style=colouredRust]
s1.join(s2,
   |(u_id, user_info)| *u_id,     // u_id as key
   |(u_id, purchase_info)| *u_id  // u_id as key
).map(|((u_id, user_info), (_, purchase_info))|
   (u_id, user_info, purchase_info));
\end{lstlisting}
\end{footnotesize}

\subsection{Windows and time}
\label{sec:api:windows}

Windows identify finite portions of unbounded
datasets~\cite{arasu:VLDB:2006:CQL}.  As common in stream processing
systems~\cite{botan:VLDB:2010:Secret}, \system defines windows with two
parameters: \emph{size} determines how many elements they include and
\emph{slide} determines how frequently they are evaluated.  \system offers
both \emph{count} windows, where size and slide are expressed in terms of
number of elements, and \emph{time} windows, where size and slide are
expressed in terms of time.  After defining the windowing on data, the next
operator will apply its logic to each window to produce elements that will be
sent along the stream.
For instance, the code below uses a count window to compute, every 2 elements,
the sum of the last 5 elements received.

\smallskip
\begin{footnotesize}
\begin{lstlisting}[language=Rust,style=colouredRust]
stream.window_all(CountWindow::sliding(5, 2)).sum();
\end{lstlisting}
\end{footnotesize}
\noindent
Windowing can also be performed on a stream that has been partitioned by key,
using a \texttt{group\_by} operator, in this case, the windowing logic is
applied independently for each partition.  For instance, the code below
applies one window to even numbers and one to odd numbers.  The example uses
time windows that are evaluated every 20\;ms over the elements received
in the last 100\;ms.

\smallskip
\begin{footnotesize}
\begin{lstlisting}[language=Rust,style=colouredRust]
let window_def = ProcessingTimeWindow::sliding(
   Duration::from_millis(100),
   Duration::from_millis(20));
stream.group_by(|v| v % 2)
   .window(window_def)
   .max(); // Compute the max of the window
\end{lstlisting}
\end{footnotesize}
\noindent
When dealing with time windows, \system supports two definitions of time:
\emph{processing} and \emph{event} time.  Processing time is the wall clock
time of the machine computing the window.  For instance, when executing the
code snippet above, the process responsible for the partition of even numbers
computes the sum of elements received in the last 100\;ms according to the
clock of the machine hosting that process.
However, many scenarios need to decouple application time from execution
time~\cite{akidau:VLDB:2015:dataflow} to guarantee consistent results even in
the case of delays or when processing historical data.
To handle these cases, \system supports event time semantics. First, a
timestamp is associated to the elements using the \texttt{add\_timestamps}
operator. This is typically done at the source. Then, the window can be
defined using the \texttt{EventTimeWindow}.
 
\system also supports \emph{transaction windows}, whose opening and closing
logic is based on the content of elements actually received through the
stream. With this kind of windows, the user specifies a closure that
determines when the current window should be opened and closed.  For instance,
the following code snippet defines a windowing logic that closes a window (and
opens a new one) upon receiving an element greater than 100.  The windowing
logic seamlessly integrates with other operators like \texttt{group\_by} and
\texttt{sum}.

\smallskip
\begin{footnotesize}
\begin{lstlisting}[language=Rust,style=colouredRust]
let window_def = TransactionWindow::new(|v|
   if v > 100 {
      TxCommand::Commit
   } else {
      TxCommand::None
   });
stream.group_by(|v| v % 2)
   .window(window_def)
   .sum(); // Compute the sum of the window
\end{lstlisting}
\end{footnotesize}
\noindent
The examples presented above exploited pre-defined operators on windows, such
as \texttt{sum} and \texttt{max}.
To implement custom operators, \system provides two approaches.
The first approach exposes an accumulator interface, such that the result of a
computation over a window can be calculated incrementally as new elements
enter the window one by one.
The second approach exposes the entire content of the window when it closes,
for those operators that cannot be implemented incrementally.

\subsection{Iterations}
\label{sec:api:iterations}

Several algorithms for data analytics are iterative in nature.  For instance,
many machine learning algorithms iteratively refine a solution until certain
quality criteria are met.  \system supports iterative computations with two
operators.

The \texttt{iterate} operator repeats a chain of operators until a terminating
condition is met or a maximum number of iterations is reached.  In the first
iteration, the chain consumes elements from the input stream, while at each
subsequent iteration, the chain operates on the results of the previous
iteration.  It holds a state variable that is updated at each iteration using
a local (per partition) and a global folding logic, specified via closures. In
the end, the operator returns two streams: one with the final value of the
state variable, the other with the elements exiting the last iteration.
For instance, the following code snippet repeats the \texttt{map} operator,
that multiplies all elements by 2 at each iteration and computes their sum in
the state variable.  The iteration terminates when either 100 iterations have
been executed or the sum is greater than 1000.
%

\smallskip
\begin{minipage}{\columnwidth}
\begin{footnotesize}
\begin{lstlisting}[language=Rust,style=colouredRust]
let (state, items) = s.iterate(
   100, // max iterations
   0,   // initial state
   |s, state| s.map(|n| n * 2), // body
   |l_state: &mut i32, n|
      *l_state += n, // local fold (sum),
   |state, l_state|
      *state += l_state, // global fold (sum),
   |state| state < 1000, // terminating condition
);
\end{lstlisting}
\end{footnotesize}
\end{minipage}
\noindent
The \texttt{replay} operator takes the same parameters, but instead of feeding
the output of the current iteration as input of the next one, it replays the
input stream until the termination condition is reached and returns the final
value of the state variable.


\section{Design and Implementation}
\label{sec:impl}

\system is implemented as a Rust framework that offers the API discussed in
\s{api}.  It is designed to scale horizontally by exploiting the resources of
different machines, which we denote as \emph{hosts}.  Each host runs a
\emph{worker} process and each worker adopts multiple threads to run the
computation in parallel.

To run a data processing job on a set of \emph{hosts}, developers:
\begin{inparaenum}[(i)]
\item write and compile a Rust \emph{driver program}, which defines the job
  using \system API;
\item provide a configuration file that specifies the list of hosts and the
  computational resources (number of CPU cores) available;
\item run the driver program, which starts the computation on the hosts.
\end{inparaenum}
The driver program reads the configuration file and uses \texttt{ssh/scp} to:
\begin{inparaenum}[(i)]
\item connect to the hosts;
\item send them the program executable, if needed; 
\item spawn one worker process per host.
\end{inparaenum}
Workers connect to each other and coordinate to collectively execute the job.
This workflow is inspired by MPI, the standard for compute-intensive
tasks~\cite{gropp:1999:MPI}.

\subsection{Job translation and deployment}
\label{sec:impl:deployment}

To illustrate the process of translating a job into executable tasks and
deploying them onto threads, we use the classic word count example, which
counts the occurrences of each word in a large document.
The following code snippet shows its implementation in \system\footnote{The
  proposed implementation works best for illustrating the translation process.
  More compact and efficient implementations are possible and will be used as
  part of our evaluation of \system in \s{eval}.}.

\begin{footnotesize}
\begin{lstlisting}[language=Rust,style=colouredRust]
let result = env
  .stream_file(&file_path)
  .flat_map(|line| split_words(line))
  .group_by(|word: &String| word.clone())
  .map(|_| 1)
  .reduce(|(count, w)| *count += w)
  .collect_vec();
\end{lstlisting}
\end{footnotesize}
\noindent
\begin{figure}[tbp]
  \centering
  \includegraphics[width=0.8\columnwidth]{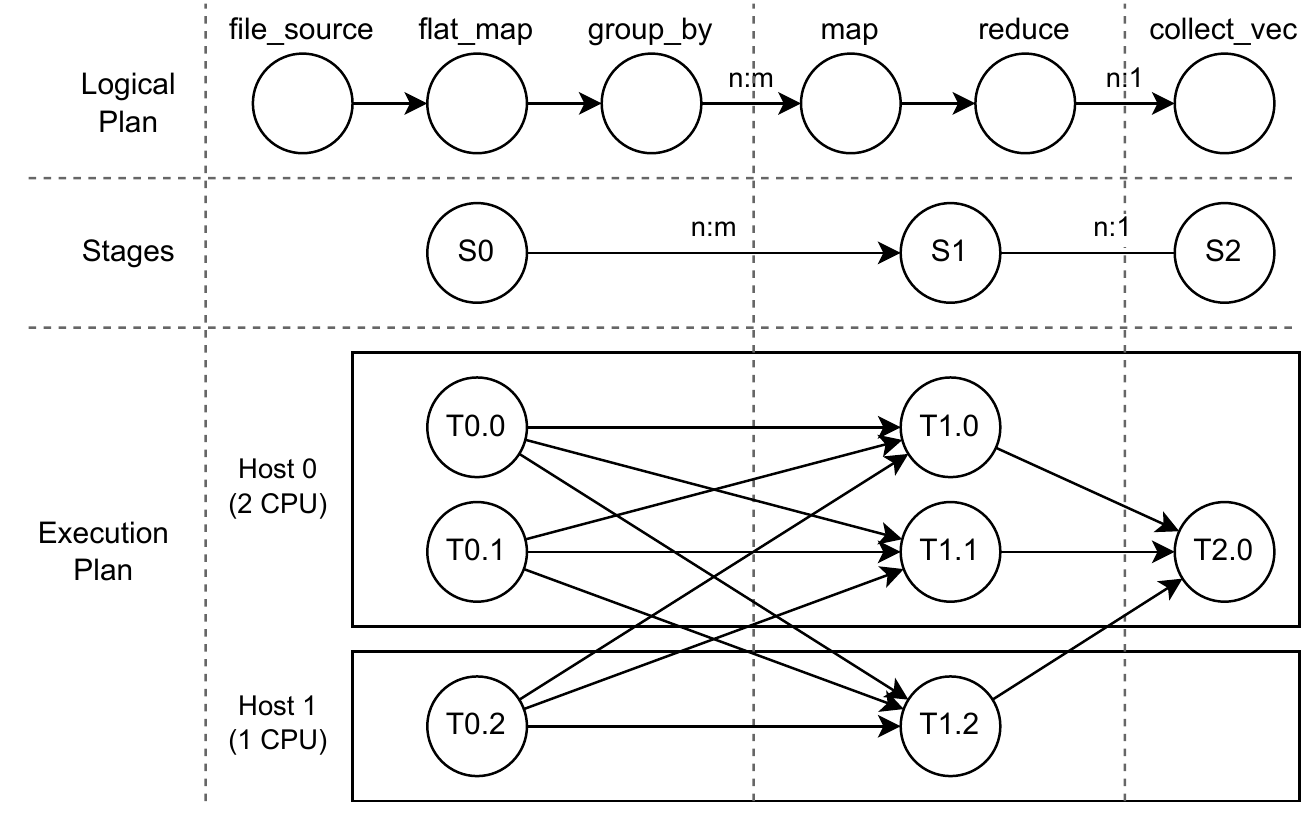}
  \caption{Deployment of the word count example.}
  \label{fig:deployment}
\end{figure}
The \texttt{stream\_file} helper method creates a parallel source that reads a
file and produces a stream of text lines; \texttt{flat\_map} extracts the
words from each line (this is done by the \texttt{split\_words} closure
passed to the \texttt{flat\_map} operator); \texttt{group\_by} groups
identical words together (the key is the entire word); \texttt{map} transforms
each word into number 1; \texttt{reduce} sums up all numbers per partition
(i.e., per word); finally, \texttt{collect\_vec} gathers the final results
(that is, the counts for each word) into a vector.

The translation and deployment process takes place when the driver program is
executed and works in four steps, as illustrated in \fig{deployment}:
\begin{inparaenum}[(i)]
\item the job is analyzed to extract its \emph{logical plan};
\item the logical plan is organized into \emph{stages} of computation;
\item each stage is instantiated as one or more \emph{tasks};
\item tasks are deployed onto hosts as independent threads.
\end{inparaenum}

\fakeparagraph{Logical plan}
The logical plan is a graph representation of the job, where vertices are
operators and edges are flows of data between operators.  It contains the
logical operators that transform the input streams generated by sources into
the output streams consumed by sinks.
In most cases, there is a one-to-one mapping between the operators defined in
the driver program using \system API and the vertices in the logical plan, as
in the word count example presented in \fig{deployment}.
However, some high-level operators part of \system API translate into multiple
operators in the logical plan.  For instance, a \texttt{group\_by\_reduce}
translates to three operators: a \texttt{key\_by} that organizes data by key,
a local \texttt{reduce} performed within each partition, and a global
\texttt{reduce} that combines all partial results for each key together.

\fakeparagraph{Stages}
Operators in the logical plan are combined into \emph{stages}: each stage
consists of contiguous operators that do not change the partitioning of data.
For instance, the word count example in \fig{deployment} contains three
stages: the first one (\texttt{S0}) starts at the source and terminates when
the \texttt{group\_by} operator repartitions data by word; the second one
(\texttt{S1}) performs the mapping and reduction in parallel for each word;
the third one (\texttt{S2}) brings the results of all partitions together.

Since stages represent the minimal unit of deployment and execution in
\system, by combining operators into stages we avoid inter-task communication
when it is not strictly necessary, that is, when the partitioning of data does
not change when moving from the upstream operator to the downstream operator.
In this case, passing of data from one operator to the next one takes place
within the same stage as a normal function invocation.
Additionally, this choice results in the code of stages being a monomorphized
version of the Job code provided by the programmer, allowing for inlining and
other compiler optimizations that involve the code of multiple operators.

\fakeparagraph{Execution plan and task deployment}
As anticipated above, each stage is instantiated multiple times into units of
execution that we denote as \emph{tasks}, which are deployed as independent
threads within the workers running on hosts.
%
%
For instance, \fig{deployment} shows an execution plan deployed onto two
hosts, where \texttt{Host 0} runs two tasks for each stage (for instance,
tasks \texttt{T0.0} and \texttt{T0.1} for stage \texttt{S0}) and \texttt{Host
  1} runs one task for each stage (for instance, task \texttt{T0.2} for stage
\texttt{S0}).
%

Knowing the number of tasks per stage to allocate into each host, each worker
can autonomously determine, in a deterministic way, the set of tasks it is
responsible for and how they should be connected to build the execution plan.
This removes the need for coordination and synchronization at initialization
time: each worker can start working independently from the others and the
connections between different hosts can happen asynchronously.

\subsection{Use of resources}
\label{sec:impl:resources}


%
By default, \system instantiates one task for each stage for each CPU
core. Consequently, a host with $n$ cores will execute $n$ tasks for each
stage.  For instance, in \fig{deployment}, \texttt{Host 0} has 2 CPU cores and
is assigned 2 tasks for each stage, and \texttt{Host 1} has 1 CPU core and is
assigned a single task for each stage.
While this default behavior can be changed by setting the number of tasks to
be instantiated for each stage at each host in the configuration file, our
experiments show that this default strategy most often yields the best
performance.
Indeed, by instantiating tasks as kernel threads, \system delegates task
scheduling to the operating system, and the default strategy leaves a high
degree of flexibility to the scheduler, which may adapt task interleaving to
the heterogeneous demands of the different processing stages and of the
different tasks within the same stage.
If a stage is particularly resource demanding, its tasks may obtain all CPU
cores for a large fraction of execution time; likewise, if data is not evenly
distributed, tasks associated to larger partitions can get a larger fraction
of execution time.

The downside of this approach is that it overcommits resources by spawning
multiple threads for each CPU core, which increases the frequency of context
switching. Our empirical evaluation shows that this is not a problem,
especially considering that the overall design of \system contributes to
alleviate this cost: for instance, as we explain in \s{impl:communication},
tasks exchange batches of data instead of individual elements, thus ensuring
that a task acquires CPU resources only when it has enough work to justify the
context switch.

Other stream processing platforms such as Flink and Kafka Streams
allocate a similar number of threads, but they use JVM threads, paying the
overhead of the JVM architectural layer.


\subsection{Communication}
\label{sec:impl:communication}

In \system, tasks communicate in one of two ways: in-memory channels or TCP
sockets.
Tasks running on the same host exploit shared memory through multiple-producer
single-consumer (MPSC) channels\footnote{\url{https://github.com/zesterer/flume}.}
to achieve fast communication avoiding serialization.
Tasks that run on different hosts use TCP channels.



To reduce the overhead of inter-task communication (both within the same host
and across hosts), \system supports batching of data elements.
With batching, subsequent elements that need to be delivered to the same
recipient task are grouped in a batch.
\system supports two batching policies: \emph{fixed} and \emph{adaptive}.
With fixed batching, a batch is sent when it reaches the exact size that was
specified.  This policy guarantees that a fixed number of elements are
delivered together over a channel, but may increase latency, as it needs to
wait to complete a batch before sending it to the recipient.
Adaptive batching also limits latency by sending a batch if a maximum timeout
expires after the last batch was sent, regardless of the number of elements in
the batch.
Developers can configure the system with different batch sizes and different
timeouts for adaptive batching, depending on their needs, potentially choosing
different batching policies for different parts of the job graph.

The combined effect of using batching and in-memory channels is the fact that
sending a batch of messages to a local worker consists of a single channel
operation that puts the reference to the batch on the queue, transferring its
ownership without any serialization or memory transfer.
As anticipated, the use of batching is also beneficial for task scheduling.
Indeed, when a task is scheduled for execution, it is guaranteed to have a
minimum number of elements ready, which reduces the frequency of context
switching.
In addition, as data is always moved in batches, operators are compiled to
code that operates on input vectors and produces output vectors of elements,
and the compiler is free to merge multiple instructions into faster vectorized
alternatives when possible.

\subsubsection{Flow control}
\label{sec:impl:communication:flowcontrol}

In designing inter-task communication, we adopted an approach that is similar
to what we presented for task scheduling and resource allocation: we delegate
as much as possible to the operating system without replicating its
functionalities within our framework.
In particular, we delegate flow control to the underlying TCP implementation:
if the receiving task cannot sustain the rate of data coming from the sending
task, the sender will be automatically suspended until the receiver has
processed previous data.
This design differs from the typical approach of alternative dataflow engines,
which implement flow control within the framework.  For instance, Apache Flink
adopts a mechanism denoted as back
pressure\footnote{\url{https://nightlies.apache.org/flink/flink-docs-stable/docs/ops/monitoring/back_pressure/}}.

\begin{figure}[hptb]
  \centering
  \includegraphics[width=0.6\columnwidth]{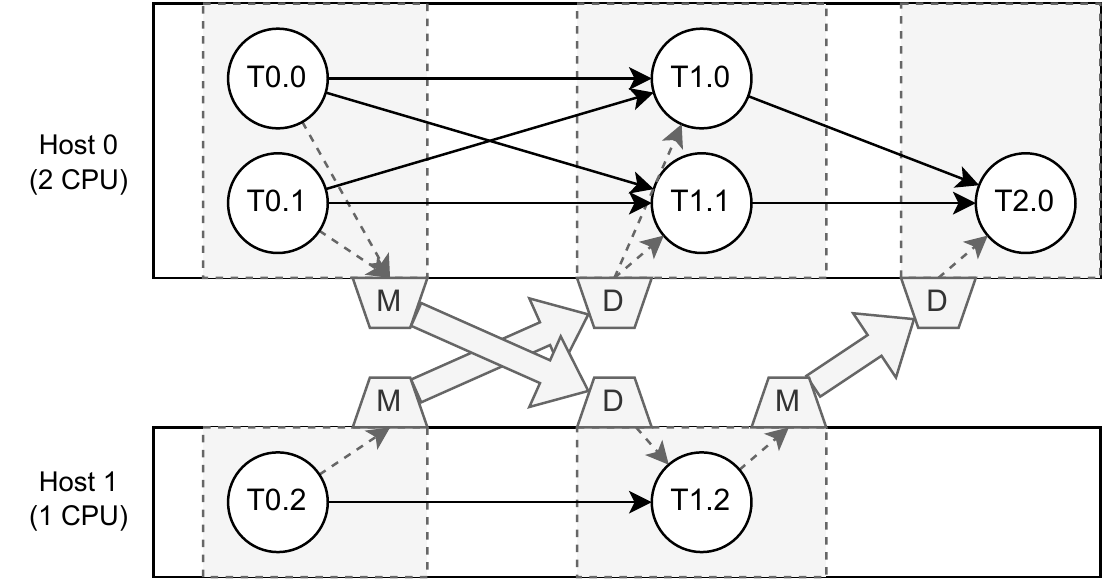}
  \caption{Communication between tasks in the word count example.}
  \label{fig:communication}
\end{figure}

To limit the number of TCP connections, \system forces all tasks running on a
host and belonging to the same stage to share the same TCP channel to
communicate with tasks running on a different host and part of the downstream
stage.
For instance, \fig{communication} shows the communication channels for
the word count example in \fig{deployment}.
\texttt{Host 0} has two tasks for stage \texttt{0}: \texttt{T0.0} and
\texttt{T0.1}.  They communicate with the tasks for stage \texttt{1} deployed
on the same host (\texttt{T1.0} and \texttt{T1.1}) using in memory channels
(black arrows), while they share the same TCP connection (large gray arrows) to
communicate with task \texttt{T1.2}, deployed on \texttt{Host 1}.
Likewise, \texttt{T0.2} communicates with \texttt{T1.2} using an in-memory
channel and with \texttt{T1.0} and \texttt{T1.1} over a single TCP connection.
Elements \texttt{M} and \texttt{D} in \fig{communication} represent
\texttt{multiplexer} and \texttt{demultiplexer} components that allow multiple
tasks to share the same TCP channel for communication.

\begin{figure}[hptb]
  \centering
  \includegraphics[width=\columnwidth]{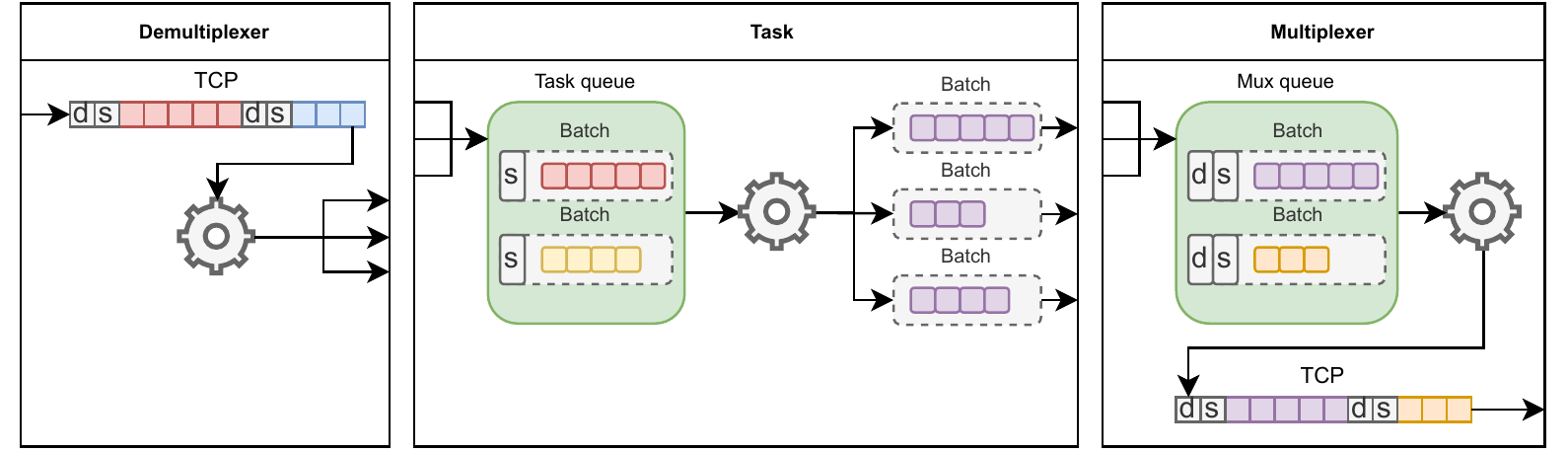}
  \caption{Communication between tasks on different processes.}
  \label{fig:communication-detail}
\end{figure}

\fig{communication-detail} provide further details on this mechanism by
focusing on the architectural components involved in the communication.
As anticipated above, each \system worker includes a \texttt{demultiplexer}
for each stage: this component receives batches of input elements for that
stage from a TCP channel.  Each batch is annotated with the sending task and
the destination task (\texttt{s} and \texttt{d} in
\fig{communication-detail}).  The \texttt{demultiplexer} exploits this
information to dispatch batches to recipient tasks.  Each
\texttt{demultiplexer} runs on a separate thread.
Next, each task (as exemplified in the middle block in
\fig{communication-detail}) receives incoming batches in a \texttt{task
  queue}, processes them according the specific logic of the task, and
delivers them to a \texttt{multiplexer} component (one per host and stage).
Each \texttt{multiplexer} runs on a separate thread.  It stores incoming
batches in a \texttt{mux queue}, serializes them using the binary
serialization format \texttt{bincode}~\cite{bincode}, and forwards them to
remote tasks of the next stage using a TCP channel.

\subsubsection{Timestamped streams and watermarks}
\label{sec:impl:communication:watermark}

When using event time, sources associate a \emph{timestamp} metadata to each
element in the stream they generate, and the runtime needs to guarantee
timestamp order during processing.  However, as the tasks of a stage evolve in
parallel, they may not guarantee order.
\system solves this problem with
\emph{watermarks}~\cite{carbone:IEEEB:2015:flink}, an established mechanism in
dataflow platforms.
Watermarks are special elements periodically emitted by sources that contain a
single timestamp $t$ indicating that no elements with timestamp lower than $t$
will be produced in the future.  Under event time semantics, tasks are
required to process data in order, so they buffer and reorder incoming
elements before processing them.  In particular, when a task \texttt{T} in a
stage \texttt{S} receives a watermark greater than $t$ from all incoming
channels, it can be sure that it will not receive any other element with
timestamp lower than or equal to $t$.  At that point, it can process all
elements up to timestamp $t$, and propagate the watermark $t$ downstream.

\subsubsection{Iterations}
\label{sec:impl:communication:iterations}

Iterations (see \s{api:iterations}) enable repeating a set of operations (the
\emph{body} of the iteration) until a given condition is met.
This requires distributed coordination across the tasks that implement the
iteration body to decide whether to start a new iteration.

\system implements this coordination logic using two implicit operators: an
\texttt{Iteration} operator is put before the body and an
\texttt{IterationLeader} operator is added after the body.
The \texttt{Iteration} operator implements a barrier logic to synchronize all
body tasks at each iteration.
The \texttt{IterationLeader} collects updates from all hosts and computes a
global state to decide whether the iteration should continue.  In that case,
the new state is broadcast to all \texttt{Iteration} operators through a
\emph{feedback link} and made available to all tasks.  Inputs for the next
iteration are also sent with feedback links, and the tasks in the body wait
for the barrier from the \texttt{Iteration} operator before processing inputs
for the next iteration.

\rev{
\subsection{Discussion}
\label{sec:impl:abstractions}
  To reach our current level of performance, we carefully designed and
  step-wise refined our system to exploit the benefits that come from the Rust
  programming language.

  In JVM-based dataflow engines, operators accept the code defining their
  behavior in the form of functional interfaces.  Developers define the logic
  of operators by creating custom implementations of these interfaces, and the
  engine selects the correct implementation at runtime through dynamic
  dispatch.
  The first versions of \system relied on similar design strategies, but we
  soon moved from dynamic dispatch to statically compiled generics to capture
  user-defined logic for operators, resulting in significant performance
  improvements.
  Rust generics are monomorphized and compiled to type-specific and optimized
  code for each version of the generic call.
  Starting from this insight, we decided to make operators generics also with
  repect to their upstream operators.  This way, operators that are chained to
  form stages of execution are compiled together, allowing optimizations such
  as inlining that cross the boundaries of individual operators.
  Moreover, being \system a library that is compiled together with
  user-defined code, compile-time optimizations can also take place between
  library code and user code.
  We used the higher-kind polymorphism capabilities of Rust to enable generics
  even in presence of complex operators.  As a concrete example, the windowing
  logic is generic over both the windowing strategy and the type of the state
  being accumulated during window evaluation.  We implemented it as a Rust
  Trait with generic associated types, which enable developers to easily
  extend the API with new strategies and new accumulators, while still
  avoiding dynamic dispatching.

  In summary, our implementation entirely avoids dynamic dispatching for task
  execution, communication, and serialization, and enables the compiler to
  produce type-specific code with holistic optimizations that span library
  code and user-defined code across multiple operators.
  Thanks to the safety guarantees and strict aliasing rule typical of the Rust
  language, the Rust compiler may automatically inline and perform loop fusion
  and fission more freely than the C++ compilers, and this becomes even more
  relevant on modern hardware where the compiler can exploit vectorized
  instructions.
}


\section{Evaluation}
\label{sec:eval}

\newcommand{\flink}{\texttt{Flink}\xspace}
\newcommand{\mpi}{\texttt{MPI}\xspace}
\newcommand{\timely}{\texttt{Timely}\xspace}
\newcommand{\omp}{\texttt{OMP}\xspace}
\newcommand{\rayon}{\texttt{Rayon}\xspace}

\newcommand{\wc}{\texttt{wc}\xspace}
\newcommand{\winwc}{\texttt{win-wc}\xspace}
\newcommand{\kmeans}{\texttt{k-means}\xspace}
\newcommand{\collisions}{\texttt{coll}\xspace}
\newcommand{\collatz}{\texttt{collatz}\xspace}
\newcommand{\enumtriangles}{\texttt{tri}\xspace}
\newcommand{\pagerank}{\texttt{pagerank}\xspace}
\newcommand{\transitive}{\texttt{tr-clos}\xspace}
\newcommand{\connected}{\texttt{conn}\xspace}
\newcommand{\nexmark}{\texttt{nexmark}\xspace}

The goal of this evaluation is to assess the performance and ease of use of
\system, in absolute terms and in comparison with:
\begin{inparaenum}[(i)]
  \item state-of-the-art data processing platforms, and
  \item lower-level programming primitives used to implement high-performance
  parallel and distributed computations.
\end{inparaenum}

To do so, we implemented various data processing tasks, which cover the key
functionalities offered by \system and other state-of-the-art data processing
platforms.  We measure performance in terms of throughput, latency, and
(horizontal and vertical) scalability.  We evaluate the complexity of each
implementation both quantitatively, by counting the lines of code, and
qualitatively, by observing the aspects that the different programming
interfaces can abstract away.
The section is organized as follows: \s{eval:setup} presents the setup we use
in our experiments.  \s{eval:programming} compares the programming model of
\system with that of alternative solutions.  \s{eval:perf_batch} and
\s{eval:perf_stream} measure the performance and horizontal scalability for
batch and stream processing workloads, respectively.  \s{eval:perf_single}
measures vertical scalability within a single machine.  Finally,
\s{eval:discussion} summarizes and discusses our findings.

\subsection{Experiment setup}
\label{sec:eval:setup}

We present the experiment setup in terms of systems under test, benchmarks
adopted, hardware and software configurations, and evaluation methodology.

\subsubsection{Systems under test}

We compare \systemtt with the following alternative solutions for distributed
data processing and high-performance parallel and distributed computations.

\fakeparagraph{Apache Flink} Apache Flink (\flink from now on) is a
state-of-the-art dataflow processing system, widely used in industrial
settings.  It is written in Java and offers a high-level API to define batch
and stream processing computations and deploy them on a cluster of nodes.
We consider Flink as representative of high-level data processing platforms,
because it is frequently adopted as a reference in the recent literature, and
offers a level of performance that is comparable or better than competing
commercial systems~\cite{marcu:CLUSTER:16:SparkVsFlink}.

\fakeparagraph{OpenMPI} OpenMPI (\mpi from now on) is an implementation of the
Message Passing Interface specification for C/C++.  It is used in
high-performance computing and scientific computations to distribute a
computational workload among multiple machines in a cluster.
We consider \mpi as representative of the level of performance that can be
achieved with custom C++ solutions and low-level communication primitives for
data distribution.

\fakeparagraph{Timely Dataflow} First introduced in the Naiad
system~\cite{murray:SOSP:2013:naiad}, Timely Dataflow (\timely from now on) is
a generalization of the dataflow model to better express computations that
iteratively and incrementally update some mutable state.  Like the dataflow
model, it is designed to run parallel computations on a cluster of nodes.  The
implementation we consider for our comparison is written in
Rust\footnote{\url{https://github.com/TimelyDataflow/timely-dataflow}}.  To
implement some benchmarks, we use Differential
Dataflow~\cite{murray:CACM:2016:incremental}, a higher-level programming
interface for incremental computations written on top of \timely.
We consider \timely for two reasons.  First, it shares with \systemtt the goal
of finding a balance between the expressiveness and ease of use of the
programming model and the performance of the implementing platform.  As such,
it represents another point in the design space of data processing systems.
Second, it relies on the same Rust programming language as \systemtt, allowing
for a comparison of programming interfaces and system design that build on a
common ground.
%
%
We found that fully utilizing the potential of the system requires a deep
understanding of how its timestamping logic and that improper use of
timestamps often lead to degraded performance.  For this reason, we compare
\systemtt and \timely on a subset of benchmarks for which we found an
implementation from the authors, thus ensuring a proper use of the system.

\fakeparagraph{OpenMP} OpenMP\footnote{\url{https://www.openmp.org}} (\omp
from now on) is a specification of an API considered the de-facto standard for
high-performance parallel computations. We adopt the implementation of \omp
provided by the gcc compiler and we use it as a reference comparison for
the performance of \systemtt within a single machine.

\fakeparagraph{Rayon}
Rayon\footnote{\url{https://docs.rs/rayon/latest/rayon/}} (\rayon from now on)
is a Rust library for parallel processing.  It is one of the most popular
tools used to perform data-parallel computations on a single host in Rust.
We consider \rayon for two reasons.  First, it enables comparing the
programming interface of \systemtt with a simple and idiomatic way to express
data-parallel computations in Rust.  Second, it allows us to measure the
overhead of \systemtt when used within a single host and to compare it with a
widely adopted library designed specifically for this purpose.

\subsubsection{Benchmarks}

The set of benchmarks has been chosen to highlight different patterns of
computations that are common in batch and stream processing applications.

\fakeparagraph{Word count} Word count (\wc) is the classic example used to
present the dataflow model, as it well emphasizes key features of the model
such as data parallelism and repartitioning of data by key.
The task consists in reading words from a file and counting the occurrences of
each word. The input used for this task contains 4GB of books in plain text
format from the project Gutenberg repository of
books~\cite{project_gutenberg}.

\fakeparagraph{Vehicle collisions} The vehicle collisions (\collisions)
benchmark requires computing multiple queries of increasing complexity over a
large set of data. The input used for this benchmark is a public dataset of
vehicle collisions~\cite{nyc_opendata} in the form of a CSV file containing
4.2GB of data.
The queries are the following:
\begin{inparaenum}
  \item compute the number of lethal accidents per week;
  \item compute the number of accidents and percentage of lethal accidents per
  contributing factor;
  \item compute the number of accidents and average number of lethal accidents
  per week per borough.
\end{inparaenum}

\fakeparagraph{K-Means} K-Means (\kmeans) is a clustering algorithm that
partitions a set of $d$-dimensional points in $k$ non-overlapping clusters. It
is an iterative\rev{, machine learning} algorithm that closes in to a local optimum
at each iteration. We use a dataset of up to 100 million 2D points (2GB of
data).

\fakeparagraph{Connected components} Connected components (\connected) takes
in input a graph and computes the maximal sub-graphs whose nodes are all
reachable from each other.  It is an iterative graph algorithm that performs a
join operation inside the loop body.  We use a dataset of 20k nodes and 5M
edges, where nodes are represented as integer numbers.

\fakeparagraph{Transitive closure} Transitive closure (\transitive) is another
iterative graph algorithm, which computes the transitive closure of a relation
graph.  Differently from the \connected example, the number of edges in the
set grows at each iteration, reaching \(O(n^2)\) nodes at the end of the
execution.  We use a dataset of 2k nodes and 3k edges, where nodes are
represented as integer numbers.

\fakeparagraph{Enumeration of triangles} Enumeration of triangles
(\enumtriangles) is an example of a non-iterative graph algorithm: it computes
the number of node triplets that are directly connected in an undirected
graph.  We use a dataset of 1.5k nodes and 900k edges, where nodes are
represented as integer numbers.

\fakeparagraph{Pagerank} Pagerank (\pagerank) is a well known graph algorithm
used to estimate the importance of a node in a graph. Each node starts with
the same \emph{rank} and at each iteration, the rank is redistributed along
the edges to other nodes.  We use a dataset of 80k nodes and 2.5M edges,
where nodes are represented as integer numbers.

\fakeparagraph{Collatz conjecture} The collatz conjecture (\collatz) benchmark
computes the collatz conjecture steps for all the numbers from 1 to 1~billion
to find which number takes the highest number of steps before converging to
1. This is an iterative algorithm that does not require intermediate
synchronization. Accordingly, it is embarrassingly parallel but the workload
for each number is very different, which makes this algorithm well suited to
evaluate the ability of a solution to cope with unbalanced workloads. We use
it to compare solutions for parallel computations.

\fakeparagraph{Nexmark} The Nexmark (\nexmark) benchmark
suite~\cite{tucker:2008:nexmark} is increasingly being used for benchmarking
stream processing platforms.  It includes streaming queries with various
complexities, ranging from simple stateless transformations of the input
elements to stateful joins of multiple streams.

\subsubsection{Hardware and software configuration}

Unless otherwise specified, we run the experiments on an AWS cluster composed
of c5.2xlarge instances, equipped with 4-cores/8-threads processors and 16GB
of RAM each, running Ubuntu server 22.04, residing in the us-east-2 zone, and
communicating through the internal AWS network with an average ping time of
0.1ms.
\systemtt, \timely and \rayon programs are compiled with rustc~1.66.1 in
release mode with thin link-time optimization active and
\texttt{cpu-target=native}.
We use \flink~1.16.0 executed on the OpenJDK~11.0.17, with 12GB of RAM
allocated to TaskManagers.  To offer a fair comparison, we disable
checkpointing to durable storage in \flink, as \systemtt currently does not
support persistence or fault-tolerance.
We compile \mpi and \omp programs with gcc~11.3.0 using OpenMPI~4.1.2 and
OpenMP~4.5 and maximum optimization level.

\subsubsection{Metrics and methodology}
\label{sec:eval:method}

We use finite input datasets and measure the total execution time for both
batch and stream processing tasks, which measures the maximum throughput of
data a system can handle. For the streaming benchmarks we also measure latency
of processing, defined as the difference between the arrival time, at the
source, of the input element that triggered a certain output and the time at
which that output element was delivered to the sink. To measure this form of
latency, we use a single source and a single sink deployed on the same
physical machine, and exploit the real-time clock of this machine to compute
the timing.
Each experiment is executed at least 4 times, discarding the result of the
first run, which is used as a warm-up, allowing the operating system to cache
the input in memory.
We measure horizontal scalability from 1 to 4 hosts (that is, from 8 to 32
cores).
For the batch benchmarks, our measurements include the cost of job deployment,
calculating the time between submission of the job and its completion, rather
than only measuring the processing time. Indeed, in real-world settings,
deployment time contributes to the cost of running a dataflow application. For
the streaming benchmarks, which represent long running, continuous
computations, we only consider the time required to process data, excluding
job deployment.

\subsection{Programming model}
\label{sec:eval:programming}

\begin{table*}[htp]
  \scriptsize
  \centering
  \begin{tabular}[t]{|l c c c c|}
    \hline
    \multicolumn{5}{|c|}{\textbf{Batch (\s{eval:perf_batch})}} \\
    \hline
                   & \systemtt & \flink & \mpi & \timely       \\
    \hline
    \wc            & 28        & 26     & 138  & 93            \\
    \collisions    & 192       & 139    & 503  & n.a.          \\
    \kmeans        & 125       & 158    & 222  & n.a.          \\
    \pagerank      & 59        & 125    & 74   & 73            \\
    \connected     & 70        & 97     & 85   & n.a.          \\
    \enumtriangles & 44        & 159    & 204  & n.a.          \\
    \transitive    & 39        & 82     & 162  & n.a.          \\
    \hline
  \end{tabular}
  \hfill
  \begin{tabular}[t]{|l c c c|}
    \hline
    \multicolumn{4}{|c|}{\textbf{Streaming (\s{eval:perf_stream})}} \\
    \hline
                             & \systemtt & \flink & \timely         \\
    \hline
    \texttt{nexmark\_common} & 67        & 80     & 217             \\
    \texttt{nexmark\_Q0}     & 3         & 11     & 7               \\
    \texttt{nexmark\_Q1}     & 9         & 17     & 9               \\
    \texttt{nexmark\_Q2}     & 6         & 8      & 11              \\
    \texttt{nexmark\_Q3}     & 23        & 15     & 59              \\
    \texttt{nexmark\_Q4}     & 58        & 21     & 125             \\
    \texttt{nexmark\_Q5}     & 20        & 39     & 119             \\
    \texttt{nexmark\_Q6}     & 64        & 17     & 128             \\
    \texttt{nexmark\_Q7}     & 17        & 19     & 70              \\
    \texttt{nexmark\_Q8}     & 29        & 24     & 65              \\
    \hline
  \end{tabular}
  \hfill
  \begin{tabular}[t]{|l c c c|}
    \hline
    \multicolumn{4}{|c|}{\textbf{Single host (\s{eval:perf_single})}} \\
    \hline
             & \systemtt & \omp & \rayon                              \\
    \hline
    \wc      & 29        & 84   & 39                                  \\
    \kmeans  & 125       & 142  & 131                                 \\
    \collatz & 30        & 48   & 23                                  \\
    \hline
  \end{tabular}
  \caption{Lines of code used to implement each benchmark.}
  \label{tab:eval:loc}
\end{table*}

In this section, we analyze the programming models of the systems under test.
We recognize that assessing code complexity is difficult and highly
subjective, and we approach the problem by
\begin{inparaenum}[(i)]
  \item measuring the number of lines of code for each benchmark as a
  coarse-grained indication of complexity;
  \item reporting the key features that some programming models expose to the
  developers and may contribute to code complexity.
\end{inparaenum}

\begin{figure}[hptb]
  \centering
  \includegraphics[width=\columnwidth]{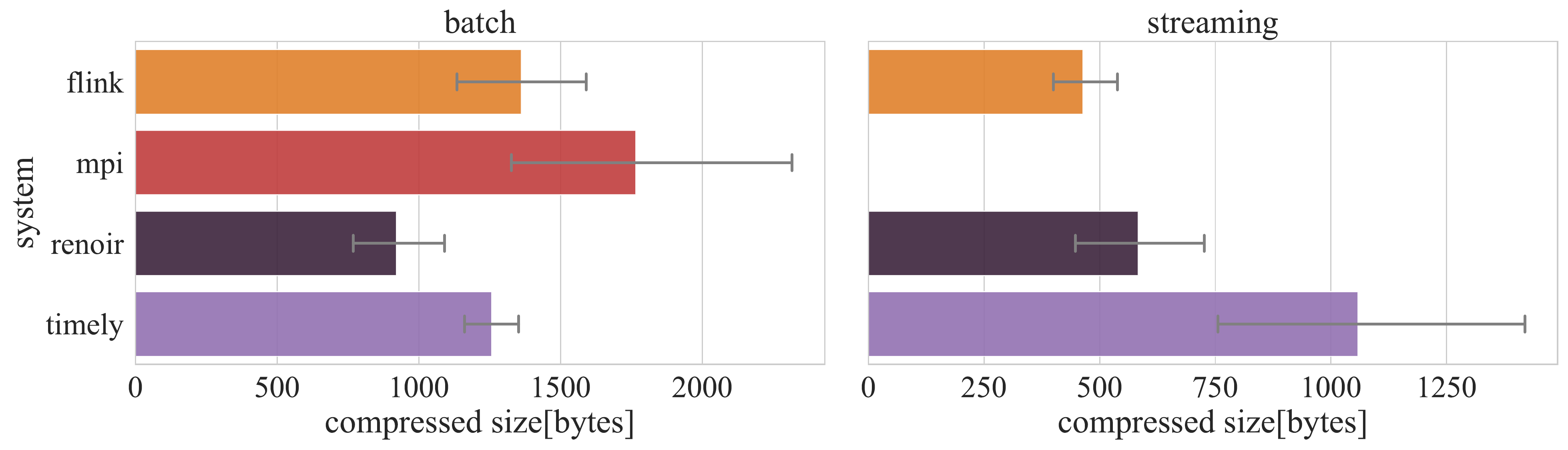}
  \caption{Bytes of the Gzip compressed source files: average and 90\%
    confidence interval.}
  \label{fig:eval:gzip_size}
\end{figure}

\tab{eval:loc} reports the lines of code for each benchmark.  It follows the
same structure as the remainder of the paper.  First, it presents batch
processing workloads, which we use to compare \systemtt, \flink, \mpi, and
\timely in \s{eval:perf_batch}.  Then, it presents the \nexmark stream
processing benchmark, which we use to compare \systemtt, \flink
\footnote{For the \nexmark benchmarks we report the lines of code of the
  queries using the Flink SQL API}, and \timely in \s{eval:perf_stream}.
Finally, it presents benchmarks for parallel computations on a single host,
which we use to compare \systemtt, \omp, and \rayon in \s{eval:perf_single}.
To ensure a fair comparison, we adopted the following approach: we excluded
comments, imports, input parsing and output formatting, we formatted the Rust,
Java and C/C++ code using the default formatter provided with the respective
language extensions in Visual Studio Code.
In addition, \fig{eval:gzip_size} reports the average size in bytes of the
source files for each solution after being compressed with gzip: the
compression algorithm limits the contribution of common language keywords, and
partially masks the differences in verbosity between languages.

For both batch and streaming benchmarks, solutions written in \systemtt have
roughly the same numbers of lines of code as those written in \flink.
Overall, the \flink versions are slightly longer, with a total of 1037 lines of
code to implement all batch and streaming benchmarks, with respect to 853
lines of code for \systemtt.  Similar differences appear in the size of the
compressed source files (\fig{eval:gzip_size}), indicating that the gap may
not be completely attributed to the different verbosity of the programming
languages adopted.

Solutions written in \systemtt present a similar structure as those in \flink,
but also closely resemble the syntax of Rust standard iterators, making it
easy for developers to port sequential Rust code to \systemtt.
Some differences between \systemtt and \flink appear in iterative algorithms.
For instance, in \kmeans, \flink uses broadcast variables to propagate shared
state, while \systemtt operators can interact with the shared state through a
smart pointer passed to the closure that defines the inner loop.

Conversely, \mpi requires more coding effort and results in significantly more
lines of code than \systemtt: 4.9$\times$ more lines in \wc, 2.6$\times$ in
\collisions, 1.77$\times$ in \kmeans, 4.6$\times$ in \enumtriangles.  The
compressed source files (\fig{eval:gzip_size}) are about twice as large in
\mpi than in \systemtt.  Only \pagerank presents almost the same lines of
code, as in this case the programming model of \mpi, based on mutable state,
helps simplifying the implementation.
Most importantly, \mpi developers need to deal with low-level concerns that
are abstracted away in \systemtt and \flink: they need to select the data
structures that encode input data and store intermediate results, they need to
decide and implement serialization and deserialization strategies, they need
to decide how communication and serialization overlaps with processing.
As we will show in the following, exposing these concerns gives a high degree
of freedom to developers, but may also be an obstacle to achieve high
performance, as code optimization may become a difficult task.  In fact,
\systemtt can outperform \mpi in several benchmarks, mainly due to better
serialization and deserialization strategies that exploit procedural macros,
which would be hard to replicate in \mpi.

In terms of code safety, the low-level communication primitives of \mpi must
be used with care to prevent deadlocks and data races.
Additionally, C++ requires developers to manually allocate and manage the
memory used to store data moving through the system, a complex task vulnerable
to errors such as memory leaks or invalid references.  Rust (used to develop
\systemtt and \timely) avoids most of these issues with its automatic memory
model and without incurring the overhead of garbage collection as in Java
(used to develop \flink).

The programming model of \timely represents a different trade-off between ease
of use and performance.  It generalizes the dataflow model by exposing the
management of timestamps and watermarks to the application, thus allowing it
to control how the computation and its results evolve over time.
However, its generality comes at the price of additional complexity, which
reflects in a higher number of lines of code.
Our experience also shows that properly handling the programming abstractions
it exposes (e.g., timestamps to govern the evolution of a computation) may not
be easy, and improper use may lead to poor performance.
In principle, the architecture of \systemtt could support the same
abstraction, but we decided to hide the watermarking logic from the
application layer to simplify the programming model.
In practice, we only used \timely with benchmarks for which we could find an
available implementation, to be confident about their correctness and avoid
misuse of the programming model, which could be detrimental for performance.
In some cases, we had to adapt our implementation to make it comparable with
that of \timely.  For instance, the \timely implementation of \pagerank adopts
a higher-level library (differential dataflow) that hides some of the
complexities of \timely, but introduces additional constraints as the
impossibility of using floating point numbers to express ranks.

When compared with libraries for parallel computations, we observe that
\systemtt has nearly the same number of lines of code as \rayon.  Indeed, both
systems mimic the interface of standard Rust iterators, leading to compact and
paradigmatic Rust code.
The higher number of lines of code in \omp are mainly due to the language being
used (C++), which brings the same verbosity and safety concerns discussed for
\mpi.  However, \omp code is simpler than the equivalent \mpi code, as the use
of shared memory avoids serialization and inter-process communication
concerns.

\rev{Notice that the programming and execution model of \systemtt and \mpi
  requires compiling and running a separate executable for each data
  processing job, whereas platforms such as \flink can run multiple jobs
  simultaneously, potentially optimizing the provisioning of resources.}

%
%

\subsection{Performance: batch workloads}
\label{sec:eval:perf_batch}

\begin{figure*}[t]
  \centering
  \subfloat[\wc]{%
    \includegraphics[width=0.16\linewidth]{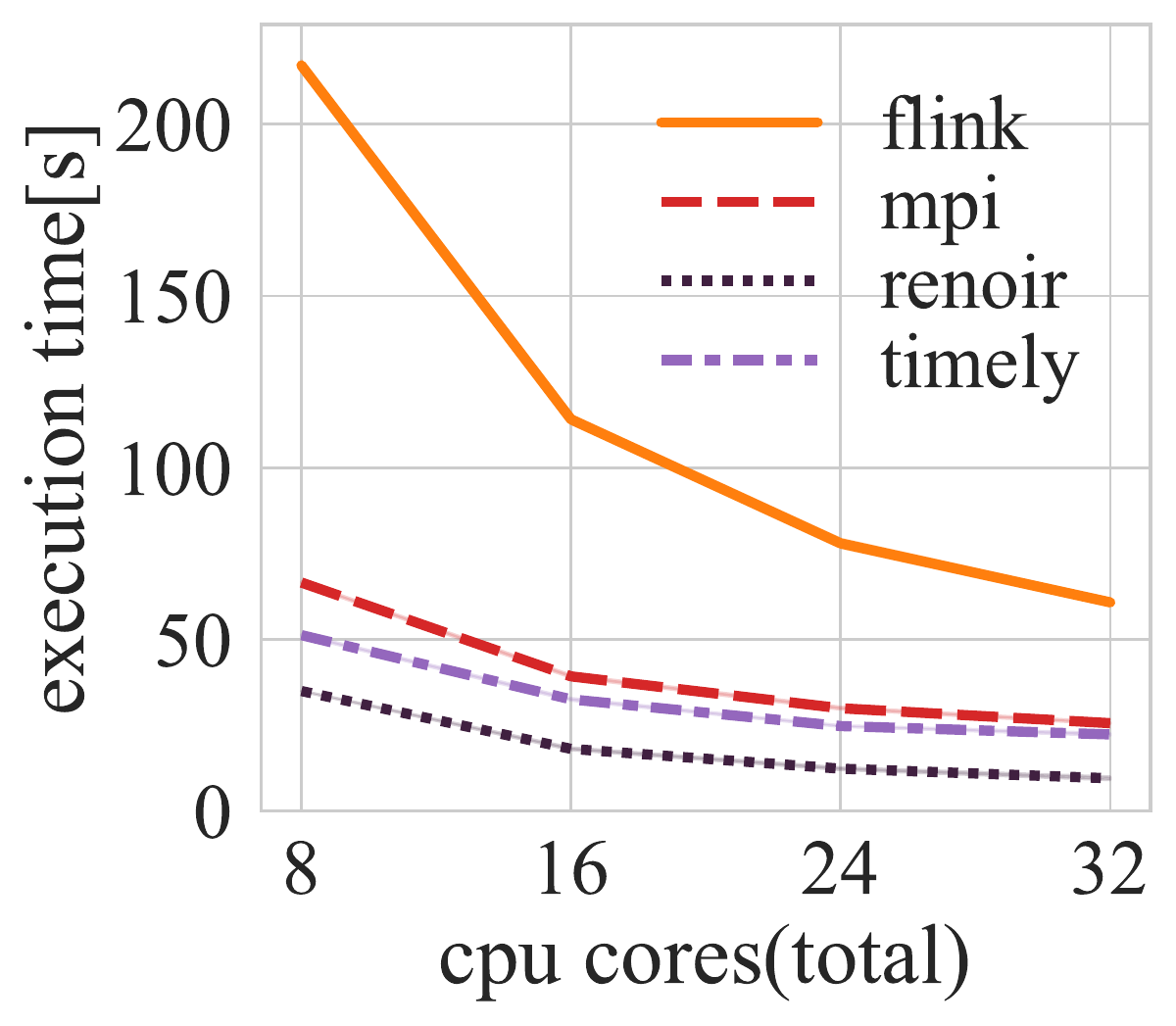}%
    \includegraphics[width=0.16\linewidth]{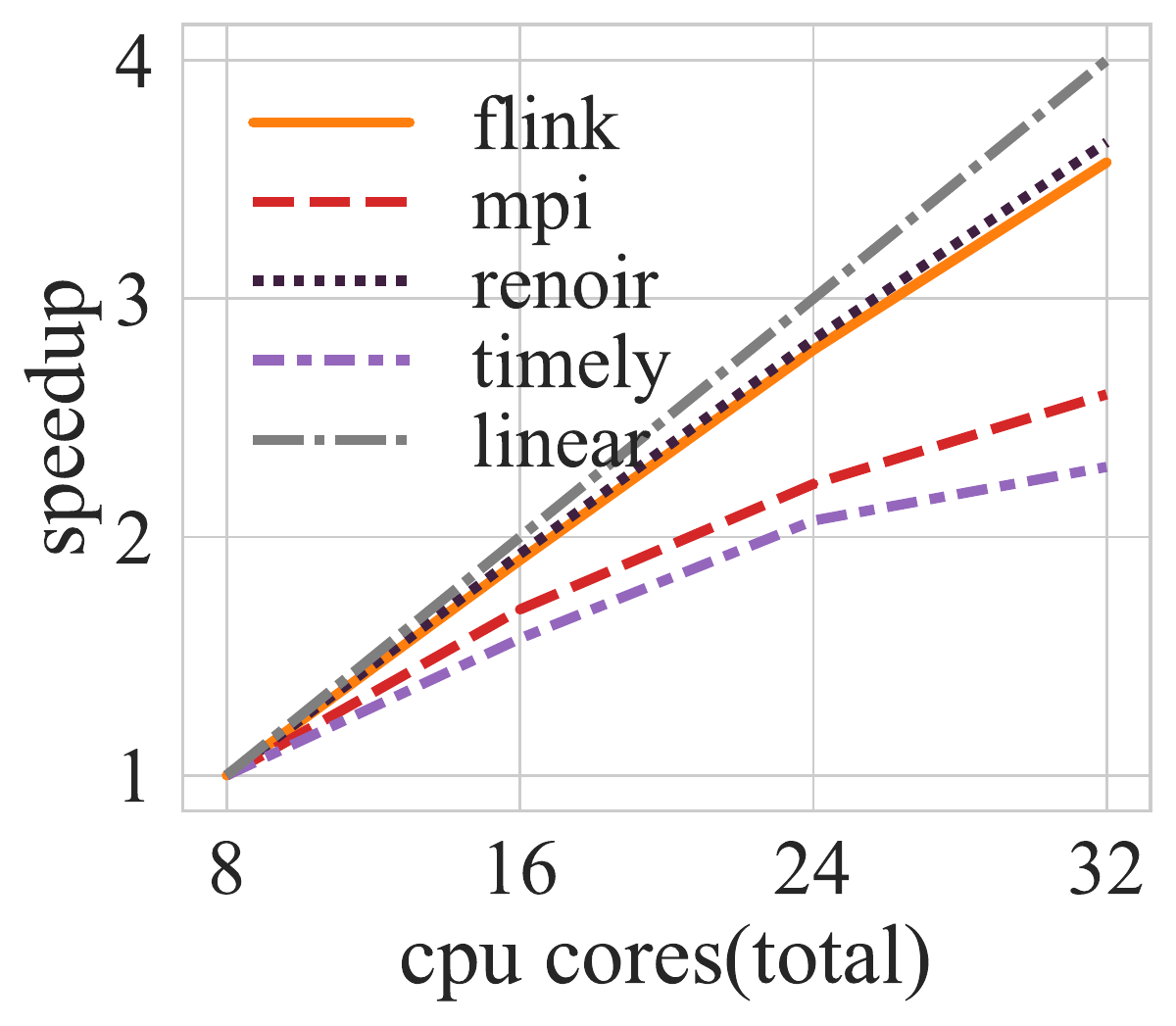}%
    \label{fig:eval:wc}}
  \hfill
  \subfloat[\wc (optimized)]{%
    \includegraphics[width=0.16\linewidth]{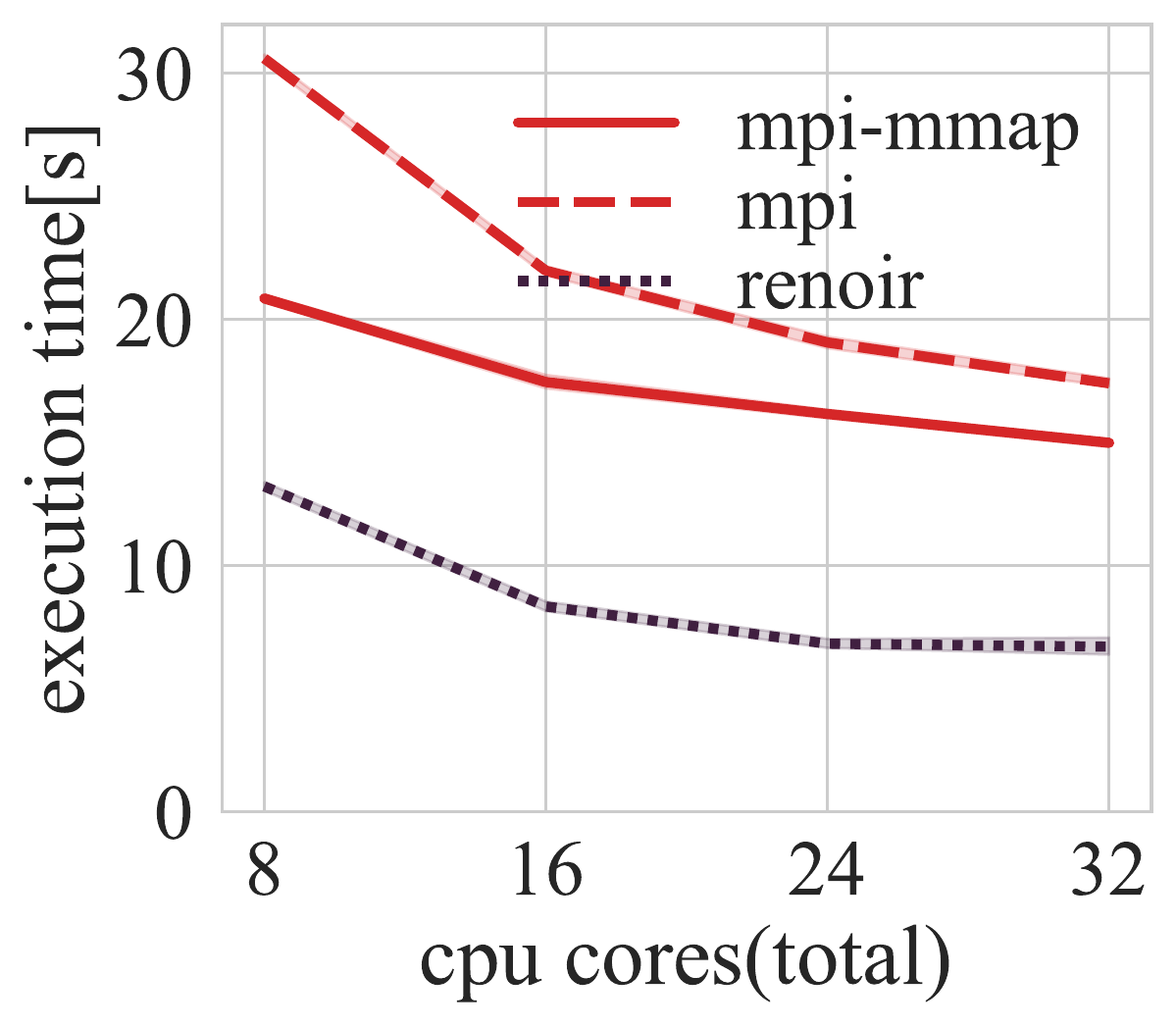}%
    \includegraphics[width=0.16\linewidth]{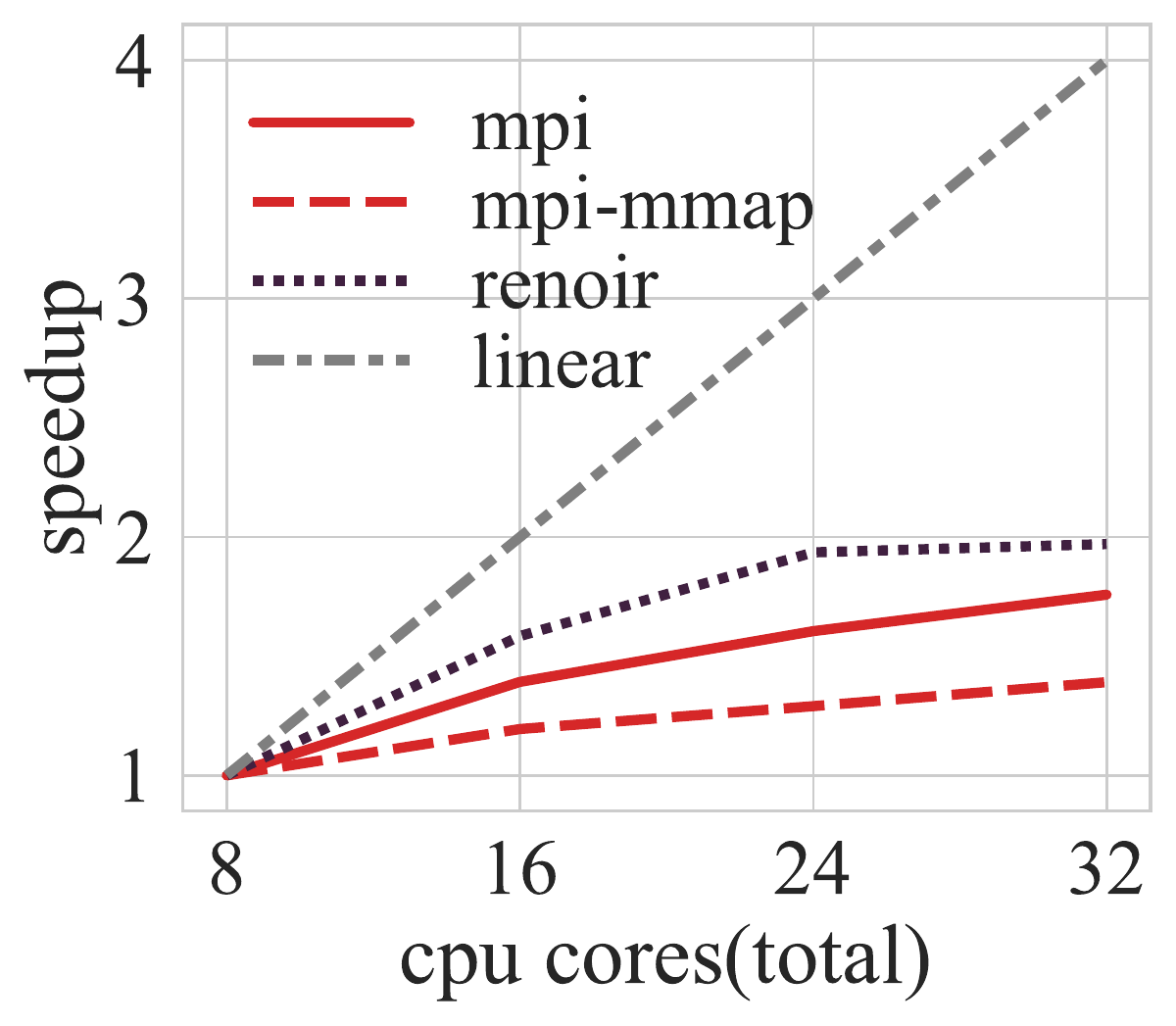}%
    \label{fig:eval:wc-opt}}
  \hfill
  \subfloat[\collisions]{%
    \includegraphics[width=0.16\linewidth]{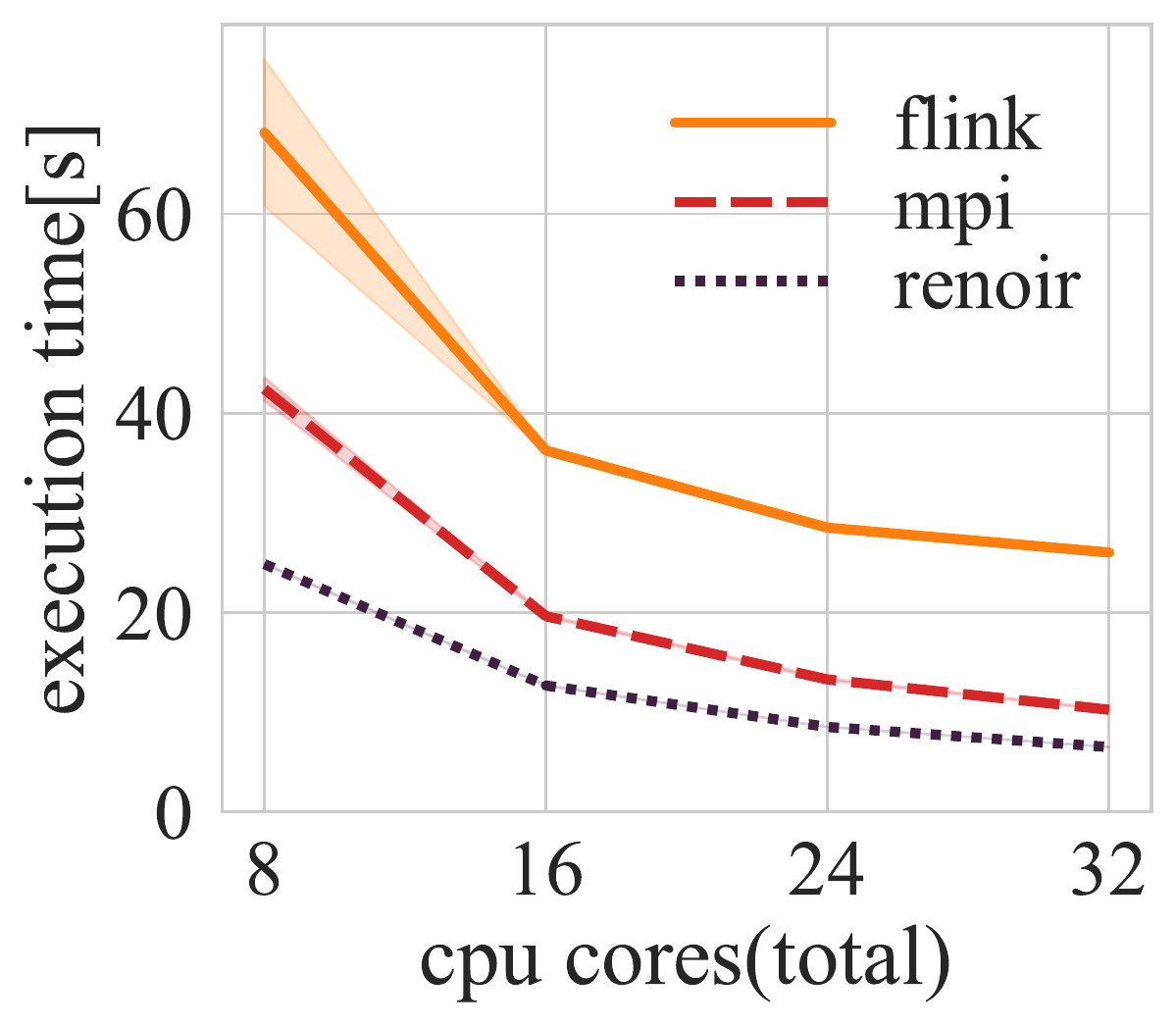}%
    \includegraphics[width=0.16\linewidth]{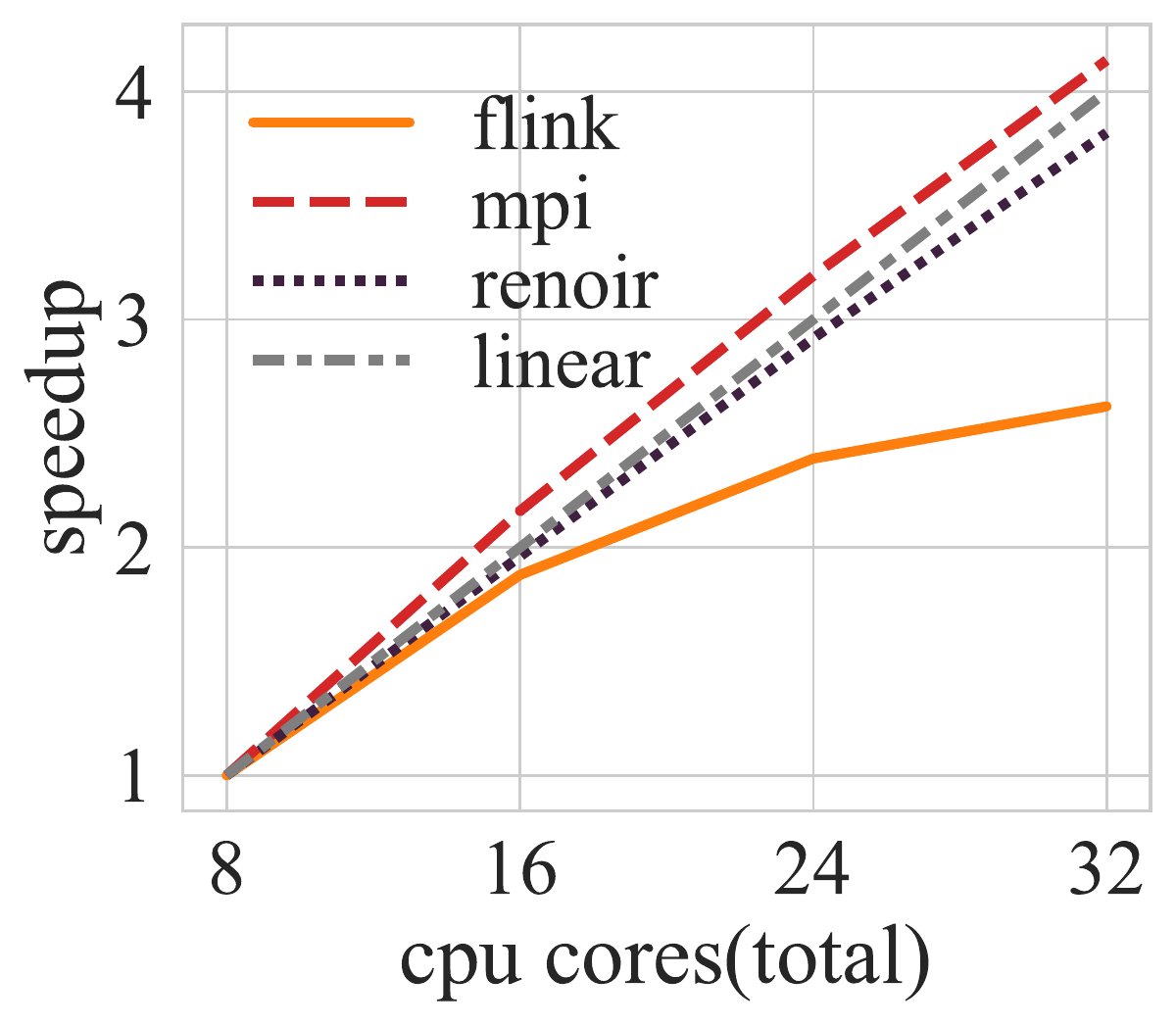}%
    \label{fig:eval:collisions}}\\
  \subfloat[\kmeans (30 centroids, 10M points)]{%
    \includegraphics[width=0.16\linewidth]{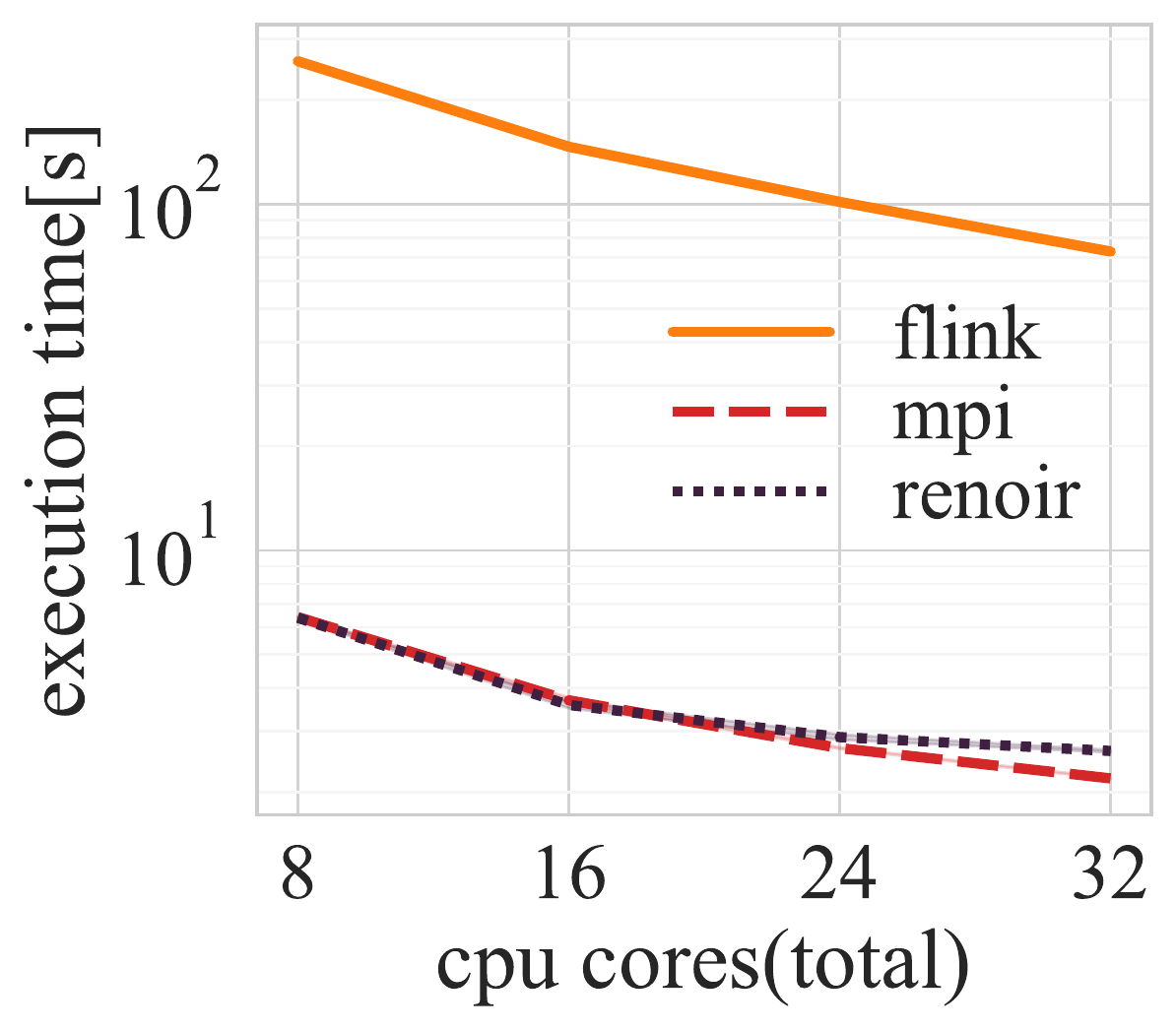}%
    \includegraphics[width=0.16\linewidth]{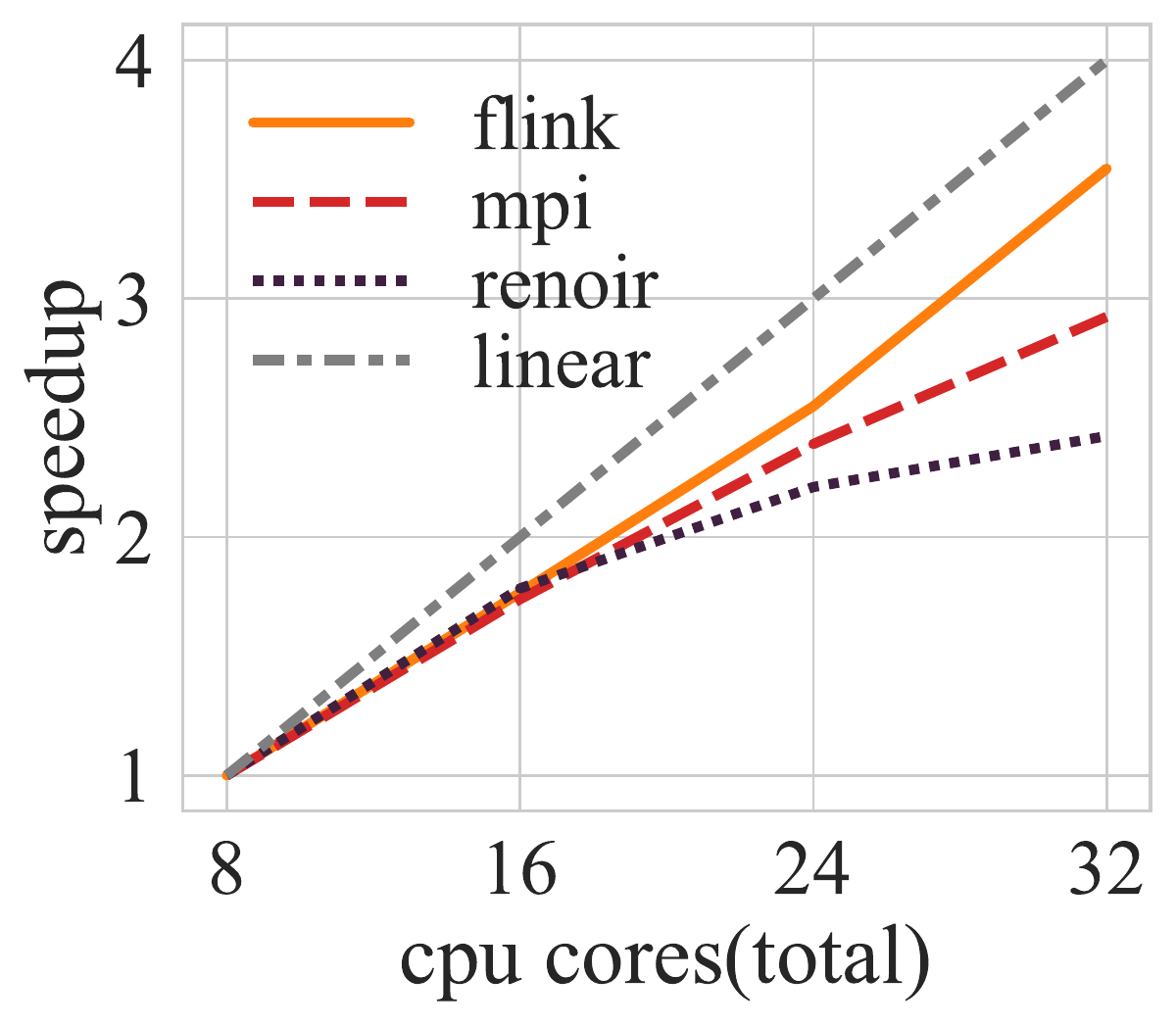}%
    \label{fig:eval:kmeans_30_10}}
  \hfill
  \subfloat[\kmeans (300 centroids, 10M points)]{%
    \includegraphics[width=0.16\linewidth]{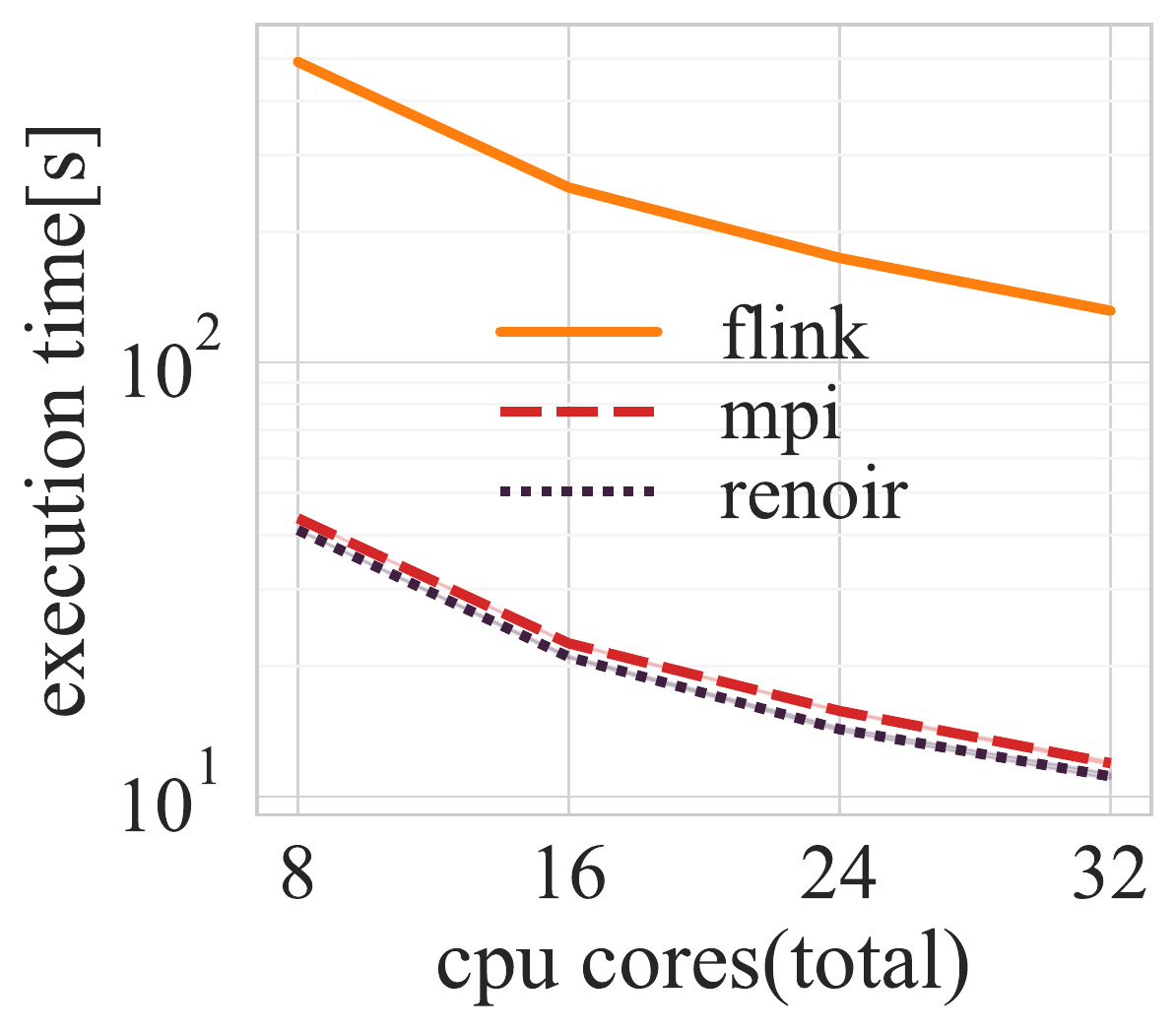}%
    \includegraphics[width=0.16\linewidth]{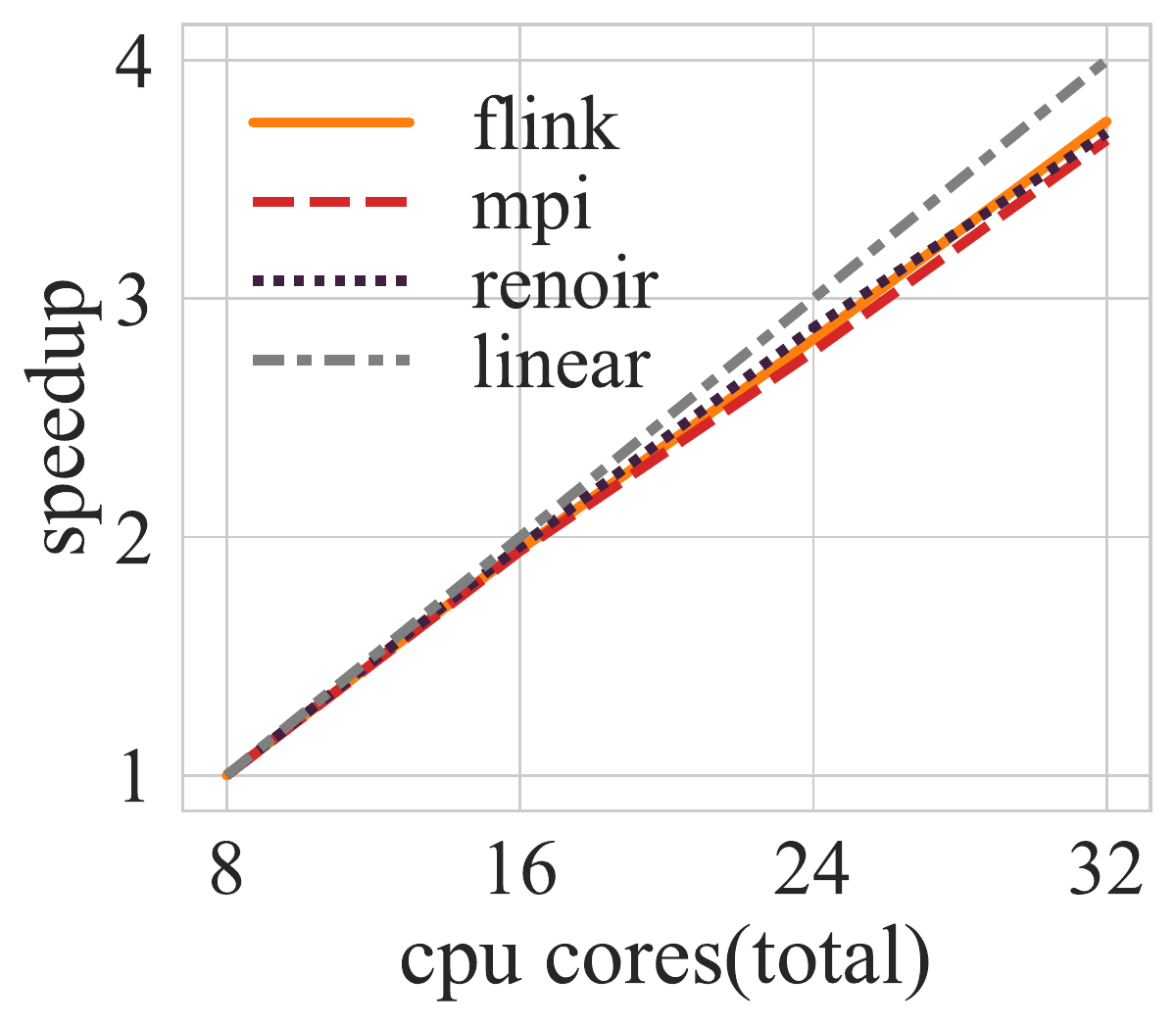}%
    \label{fig:eval:kmeans_300_10}}
  \hfill
  \subfloat[\kmeans (30 centroids, 100M points)]{%
    \includegraphics[width=0.16\linewidth]{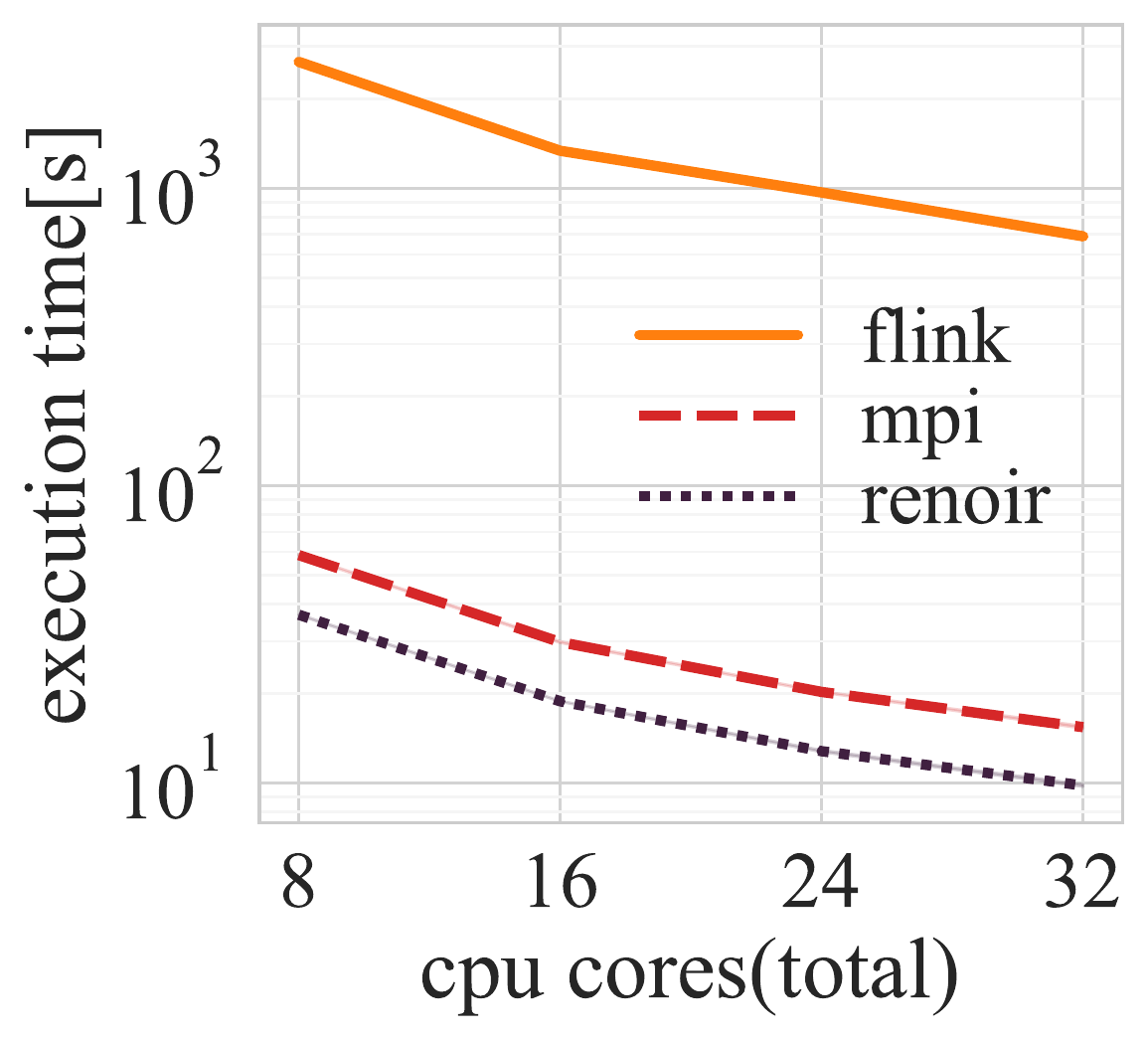}%
    \includegraphics[width=0.16\linewidth]{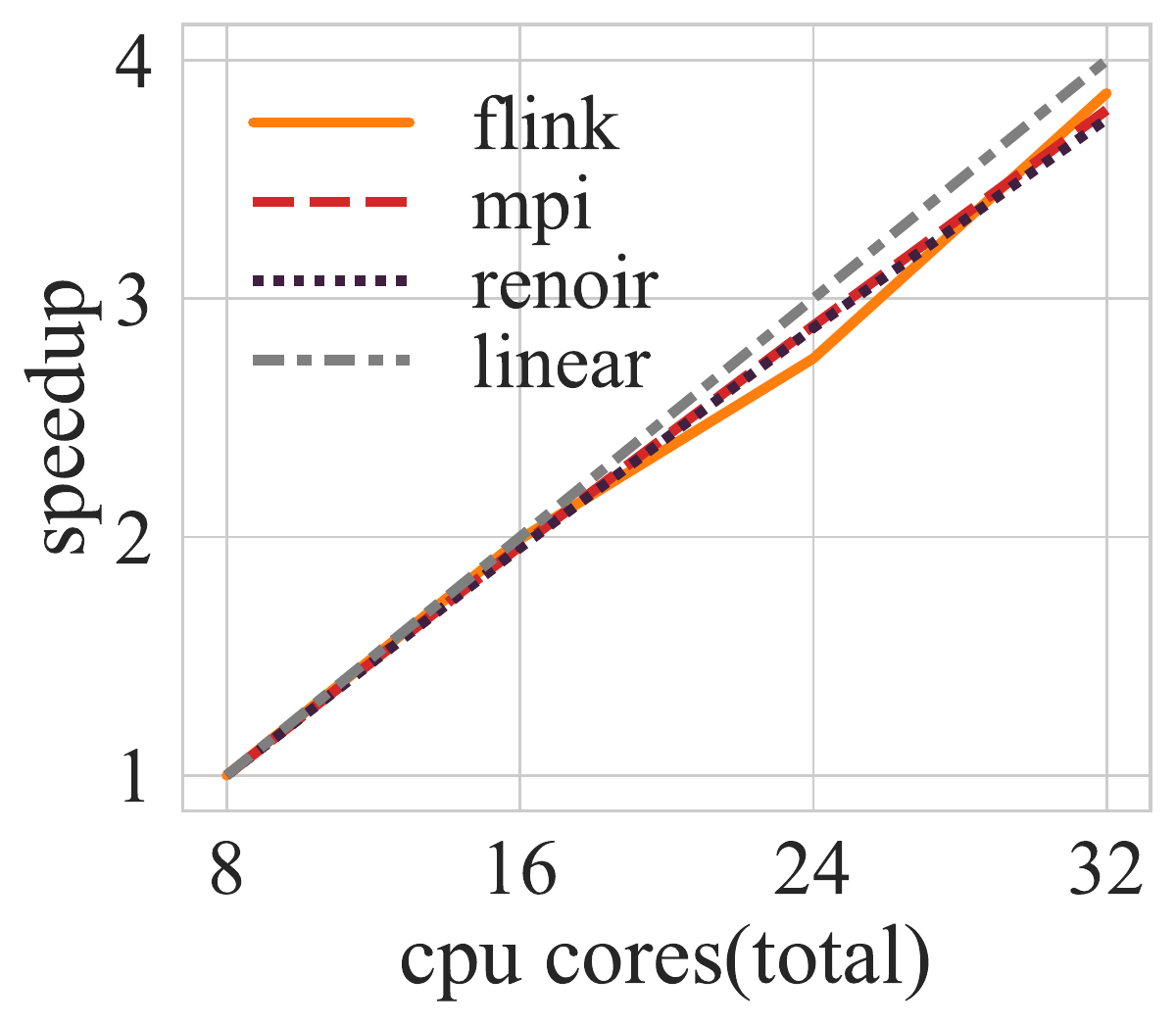}%
    \label{fig:eval:kmeans_30_100}}\\
  \subfloat[\pagerank: comparison with \mpi]{%
    \includegraphics[width=0.16\linewidth]{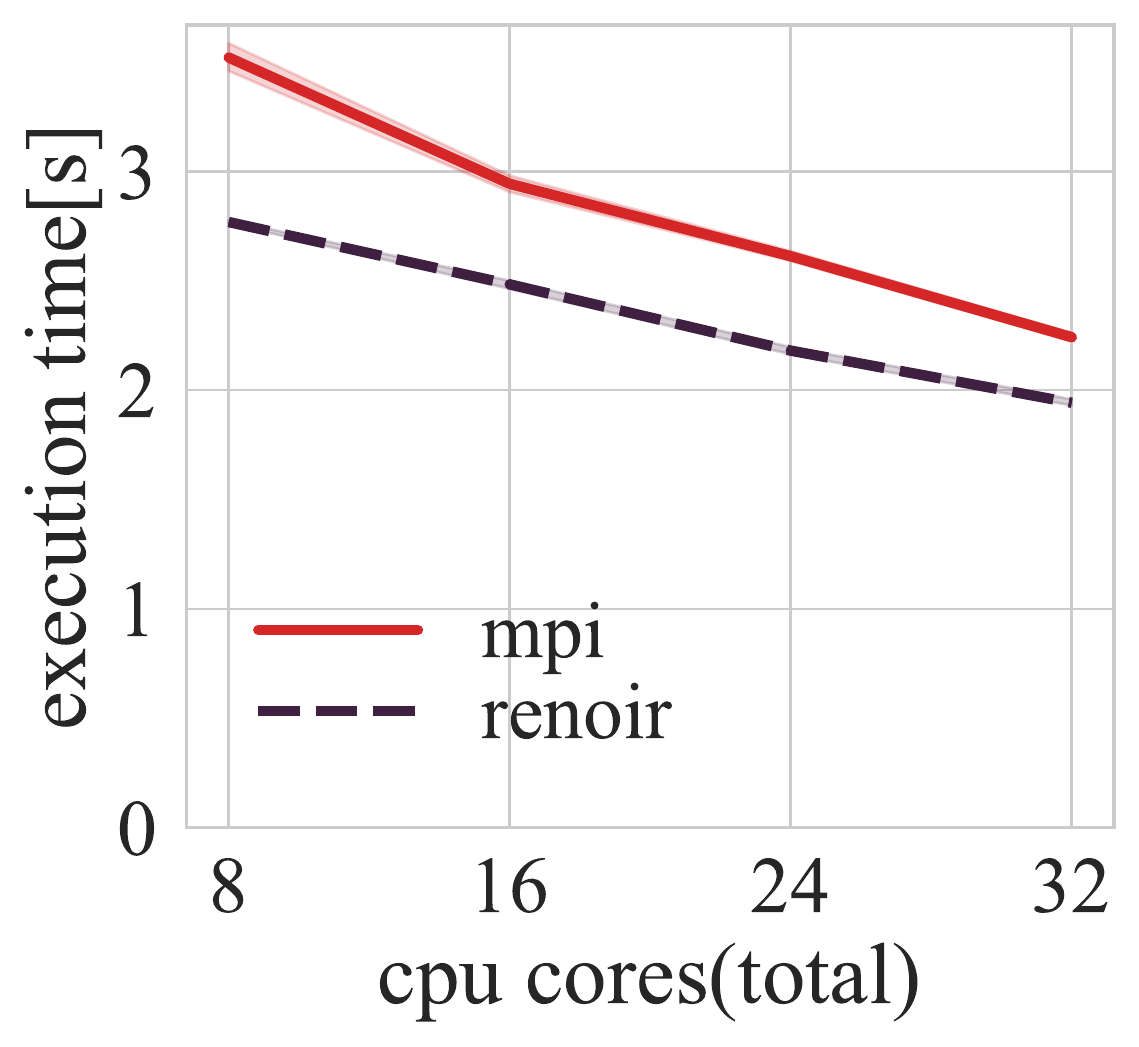}%
    \includegraphics[width=0.16\linewidth]{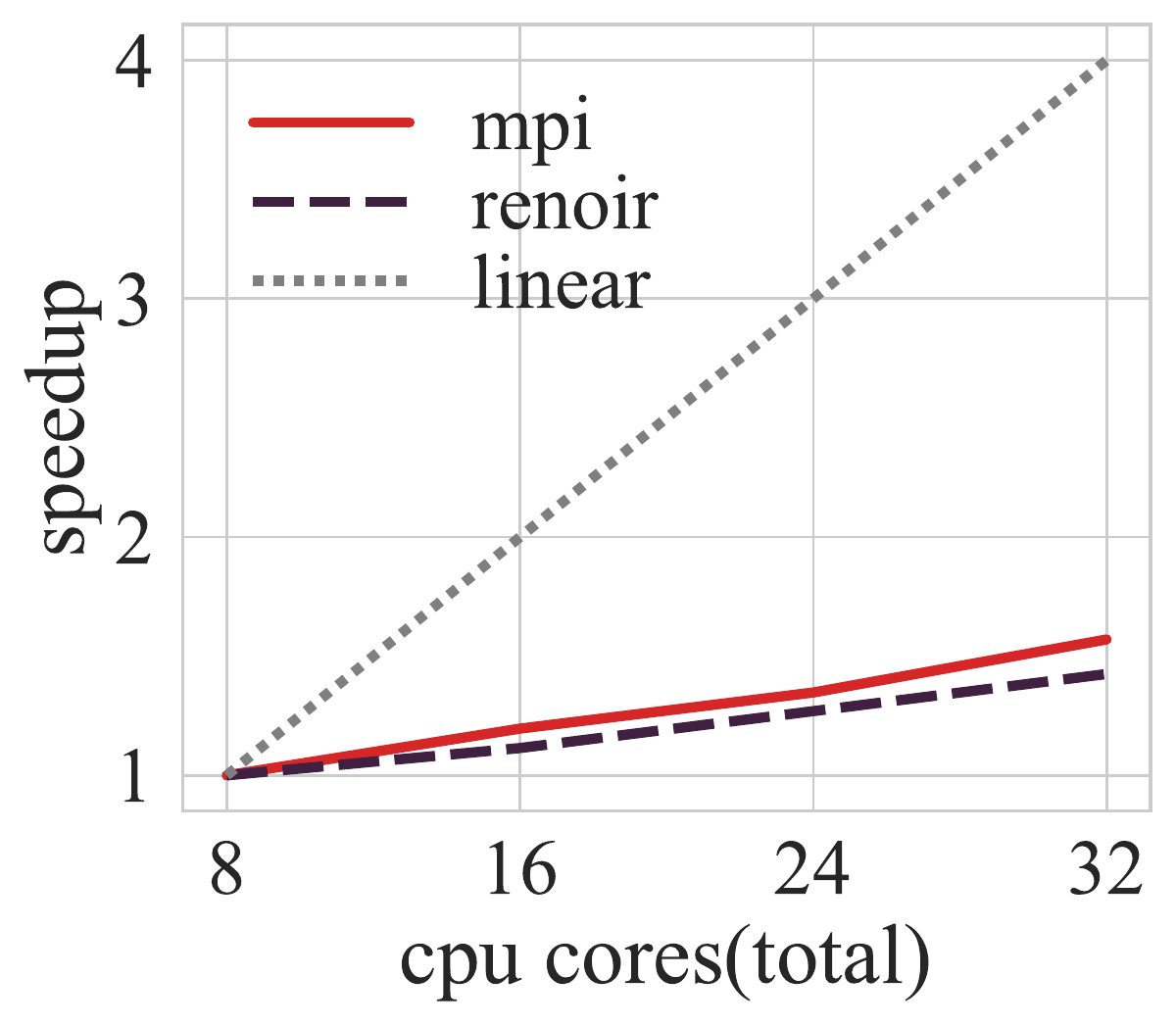}%
    \label{fig:eval:pagerank-mpi}}
  \hfill
  \subfloat[\pagerank: comparison with \flink]{%
    \includegraphics[width=0.16\linewidth]{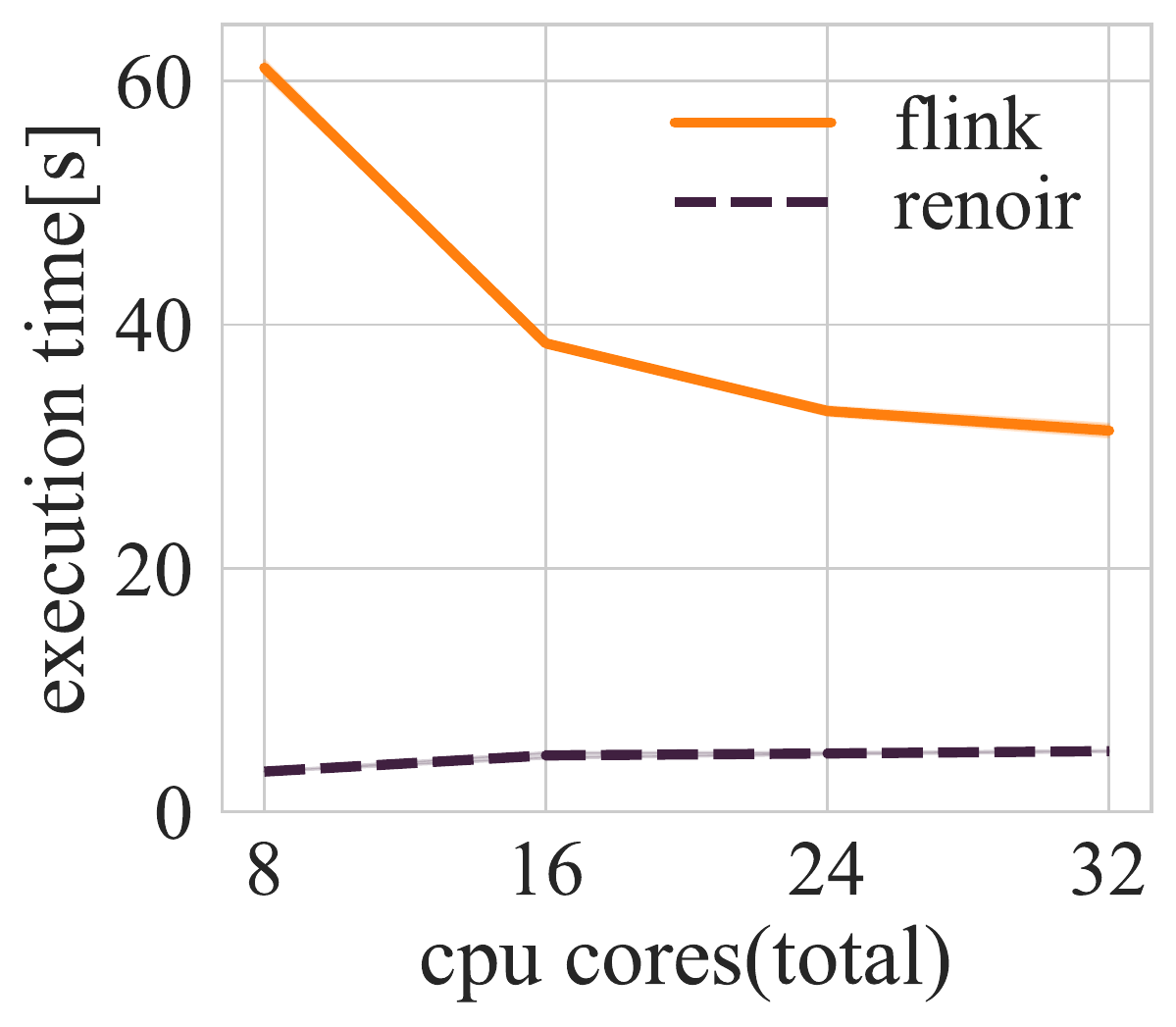}%
    \includegraphics[width=0.16\linewidth]{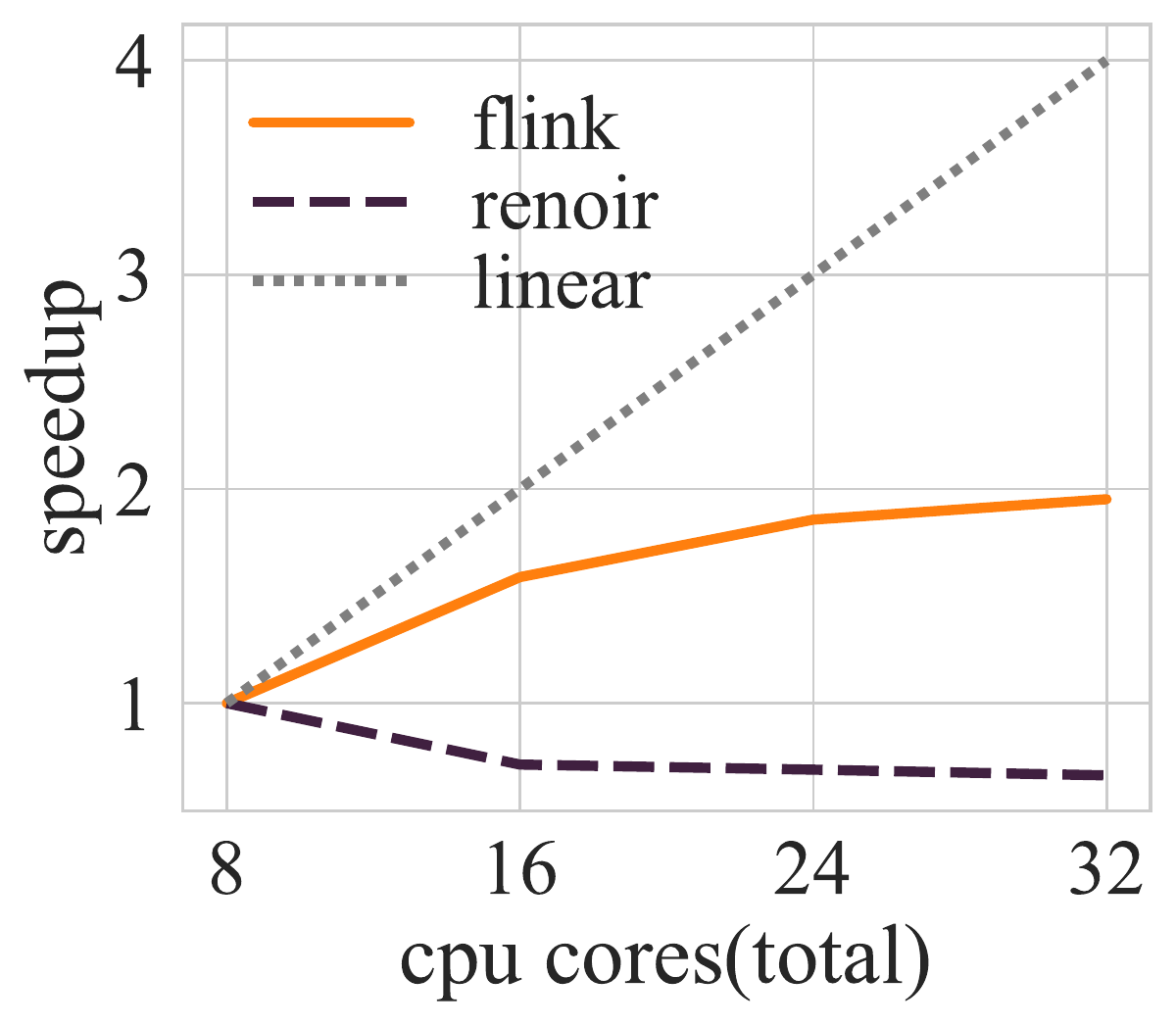}%
    \label{fig:eval:pagerank-flink}}
  \hfill
  \subfloat[\pagerank: comparison with \timely]{%
    \includegraphics[width=0.16\linewidth]{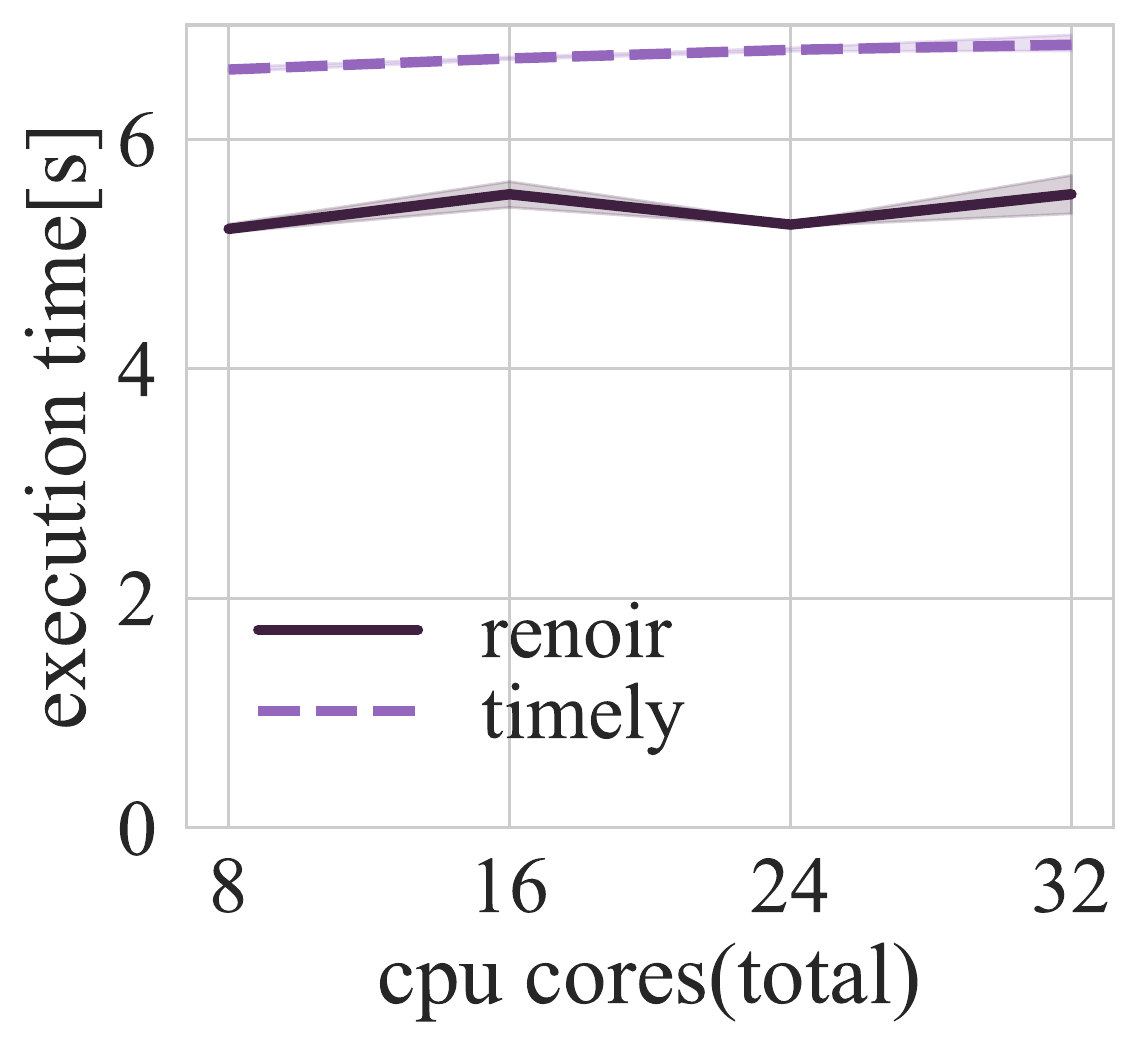}%
    \includegraphics[width=0.16\linewidth]{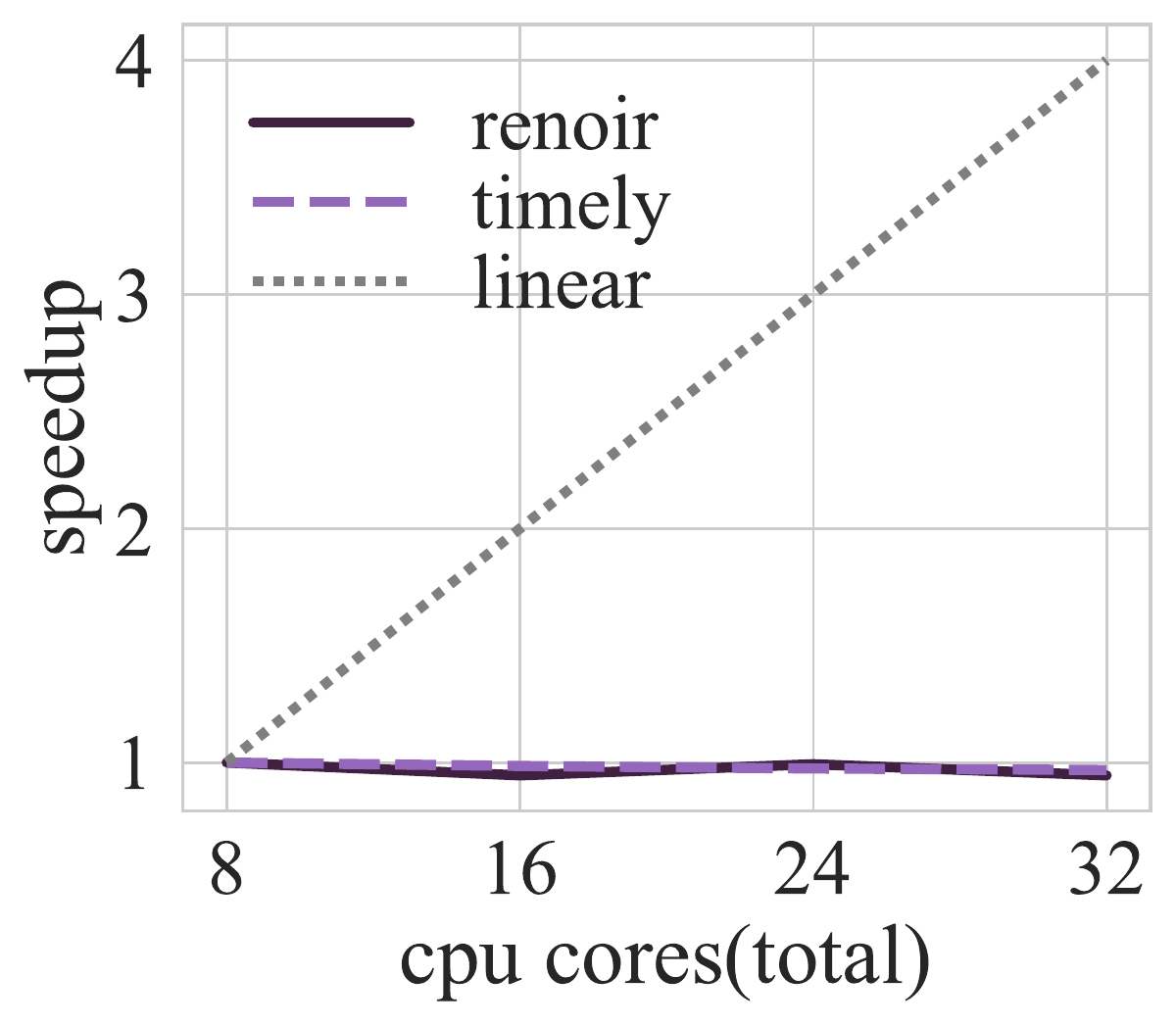}%
    \label{fig:eval:pagerank-timely}}\\
  \subfloat[\connected]{%
    \includegraphics[width=0.16\linewidth]{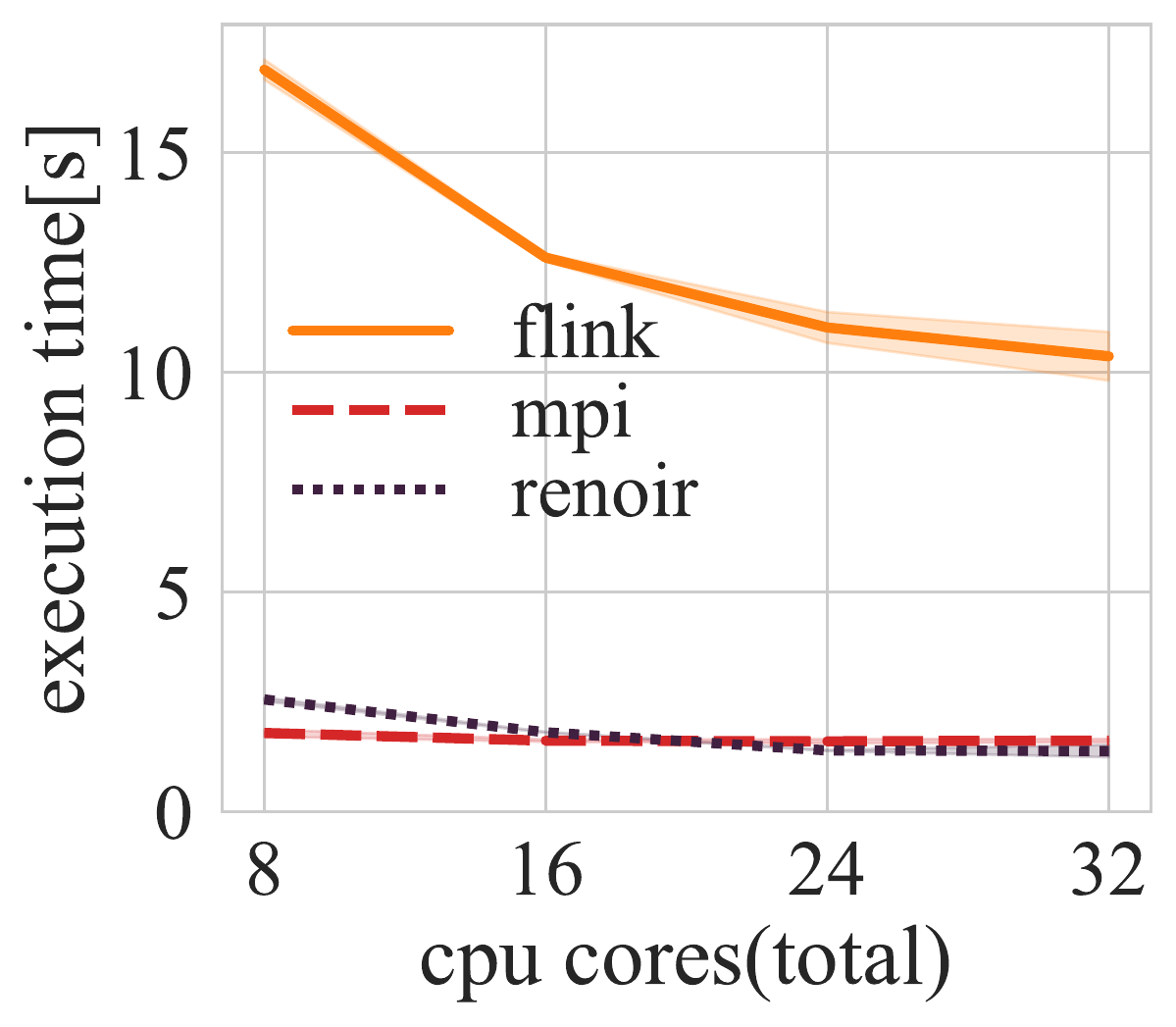}%
    \includegraphics[width=0.16\linewidth]{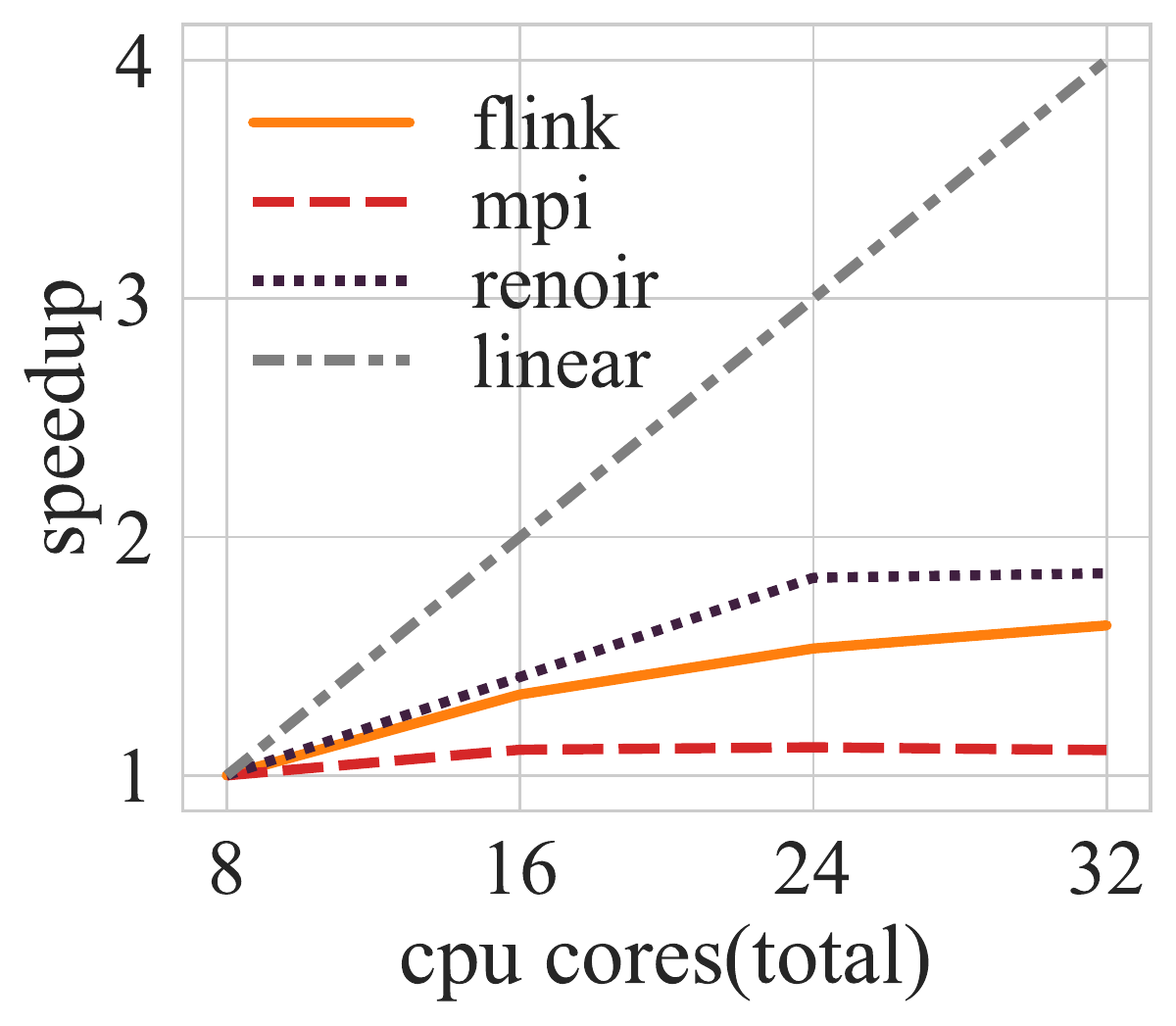}%
    \label{fig:eval:connected}}
  \hfill
  \subfloat[\enumtriangles]{%
    \includegraphics[width=0.16\linewidth]{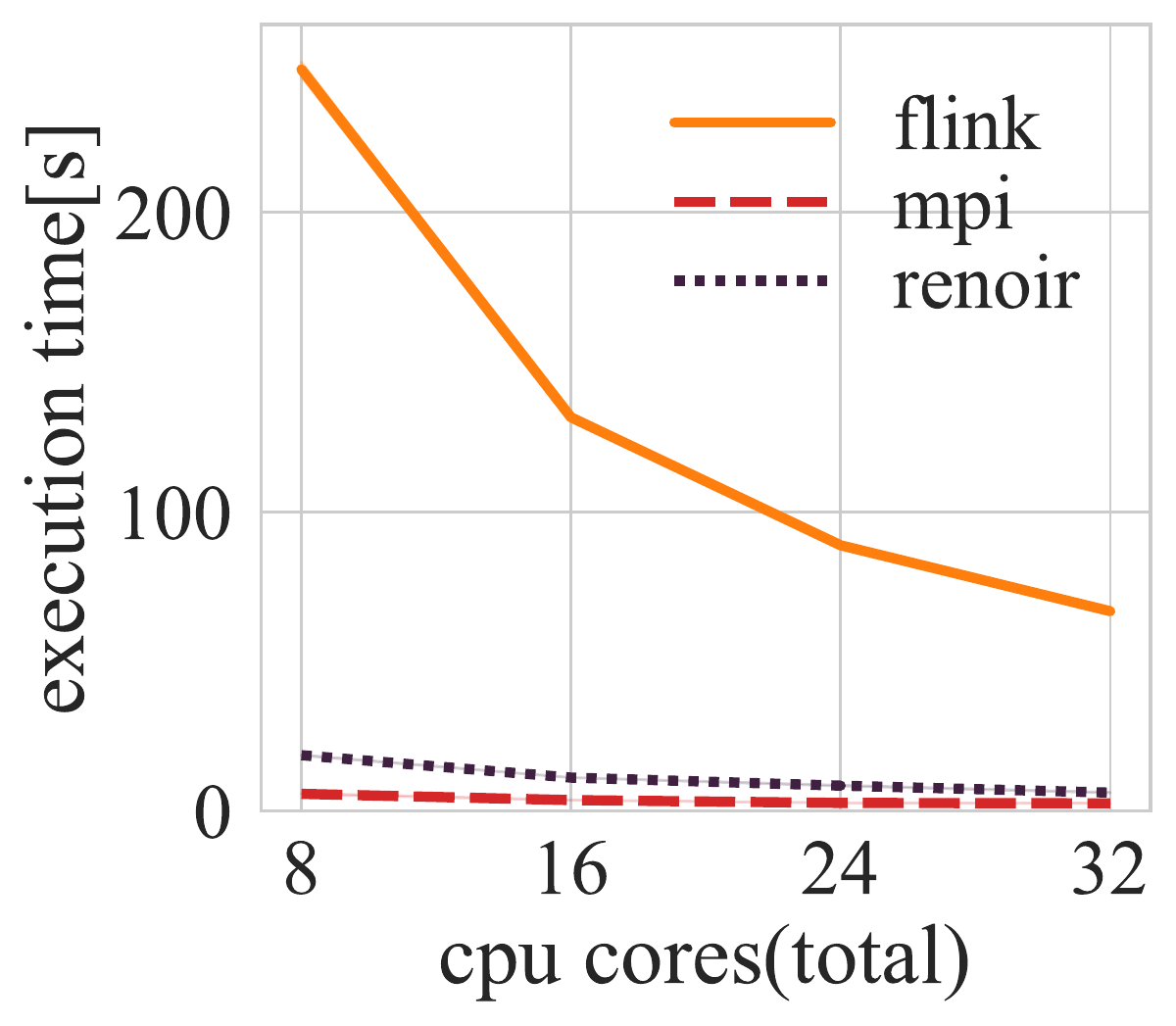}%
    \includegraphics[width=0.16\linewidth]{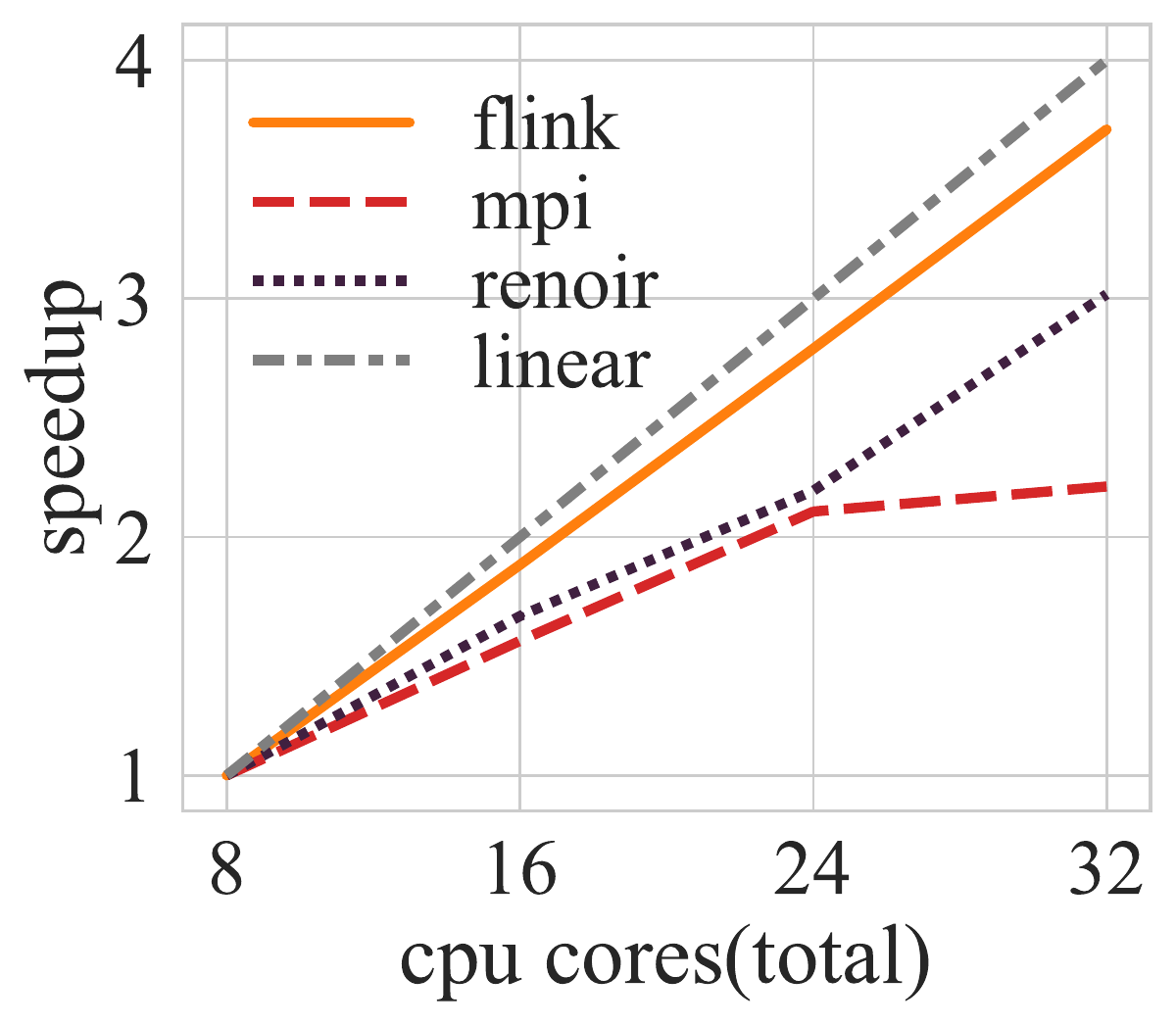}%
    \label{fig:eval:enumtriangles}}
  \hfill
  \subfloat[\transitive]{%
    \includegraphics[width=0.16\linewidth]{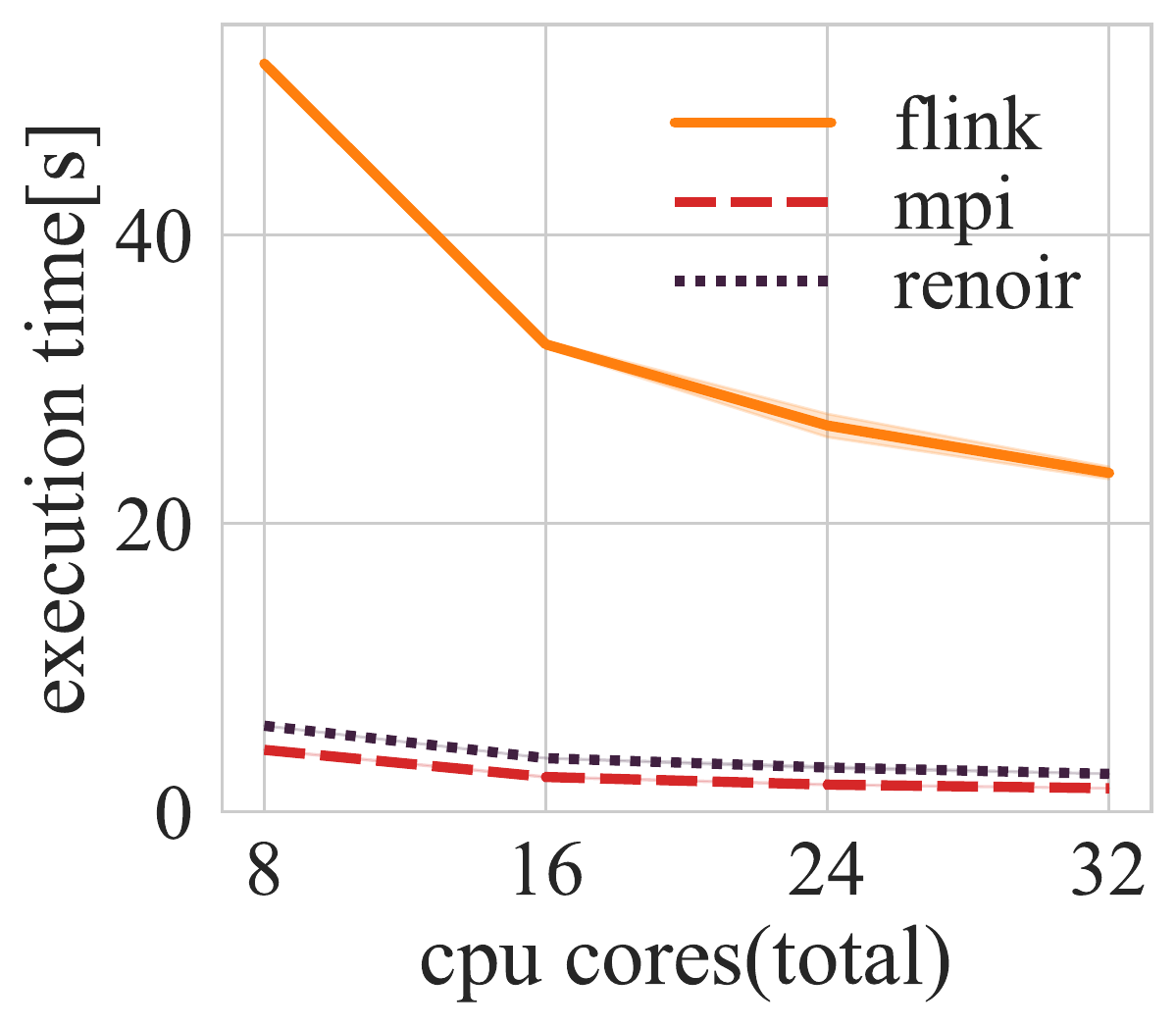}%
    \includegraphics[width=0.16\linewidth]{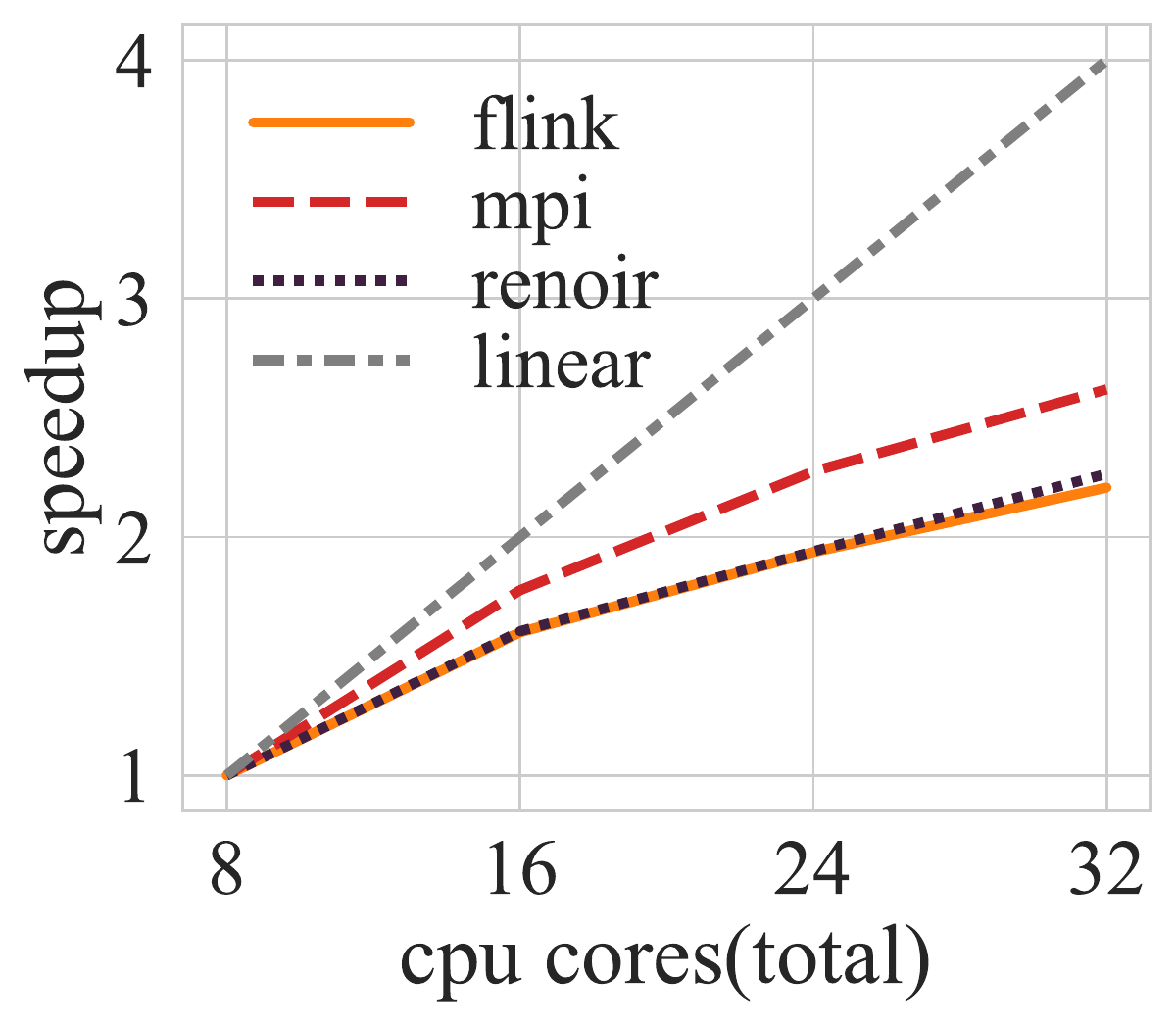}%
    \label{fig:eval:transitive}}
  \caption{Performance with batch workloads. For each workload we show
    execution time (left) and scalability (right).}
  \label{fig:eval:batch}
\end{figure*}

In this section, we evaluate the performance and scalability of \systemtt and
alternative solutions for batch processing workloads.  For each workload, we
measure the execution time while moving from one to 4 hosts.  \fig{eval:batch}
presents the results we measured.

\subsubsection{Word count (\wc)}

\fig{eval:wc} shows the execution time and scalability for \wc.  \systemtt
completes the task in 34.97s on a single host, and in 9.56s on 4 hosts.
In comparison, \flink is more than 6$\times$ slower: 217.14s on a single
host and 60.79s on 4 hosts.

We optimized the \mpi code in many ways.
In the reduction phase, \systemtt and \flink partition the dataset by word and
perform the reduction in parallel, before collecting all the results in a
single process and saving the results.  Given the limited size of the partial
results, in the \mpi implementation, we skip the intermediate phase and
collect all partial results directly in a single process, saving one
communication step.  We also made different experiments to exploit
thread-level parallelism using OpenMP, but we obtained better results with one
MPI process per core, with each process using a single thread.
Despite these custom-made optimizations, \mpi is still about 2.5$\times$
slower than \systemtt.  A detailed analysis showed a bottleneck when reading
data from file using functions from the standard C++ library and parsing using
regular expressions (the same we do in \flink and \systemtt).

Thus, we implemented an additional version with ad-hoc file reading (by
mapping the file in memory with \texttt{mmap}) and a simplified parser that
only considers 7-bit ASCII instead of UTF-8 encoded text.
This version is presented in \fig{eval:wc-opt} and is labeled
\texttt{MPI-mmap}.  It reduces the gap with \systemtt, but at the cost of
additional code complexity and reduced generality and reusability, and remains
about 1.5$\times$ slower with 4 hosts.  We attribute this difference to the
different serialization strategy: \mpi uses fixed-size arrays to represent
strings whereas \systemtt uses a more compact binary serialization format.
\fig{eval:wc-opt} also shows the performance of an optimized version of \wc in
\systemtt, which exploits the same strategy as \mpi by skipping the
intermediate reduction phase.  This version only requires 10 additional lines
of code: in particular, it does not exploit a \texttt{group\_by\_count}
operator, but it implements the count using an associative \texttt{fold}.
This small change reduces the execution time by nearly 30\%, and \systemtt can
complete the task in 6.7s with 4 hosts.

Going back to the comparison in \fig{eval:wc}, \timely completes the task in
51.24s on one host and 22.33s on 4 hosts, which means up to 2.3$\times$
higher execution time than \systemtt.
\timely adopts a different execution model with respect to \systemtt, where
each worker thread is responsible for part of the dataflow graph, it loops
through the operators within that part of the graph, and it executes one
operator at a time.  This architectural difference leads to different
performance and scalability characteristics in the experiments we performed.

\begin{figure}[t]
  \rev{
  \centering
  \subfloat[\wc]{%
    \includegraphics[width=0.16\textwidth]{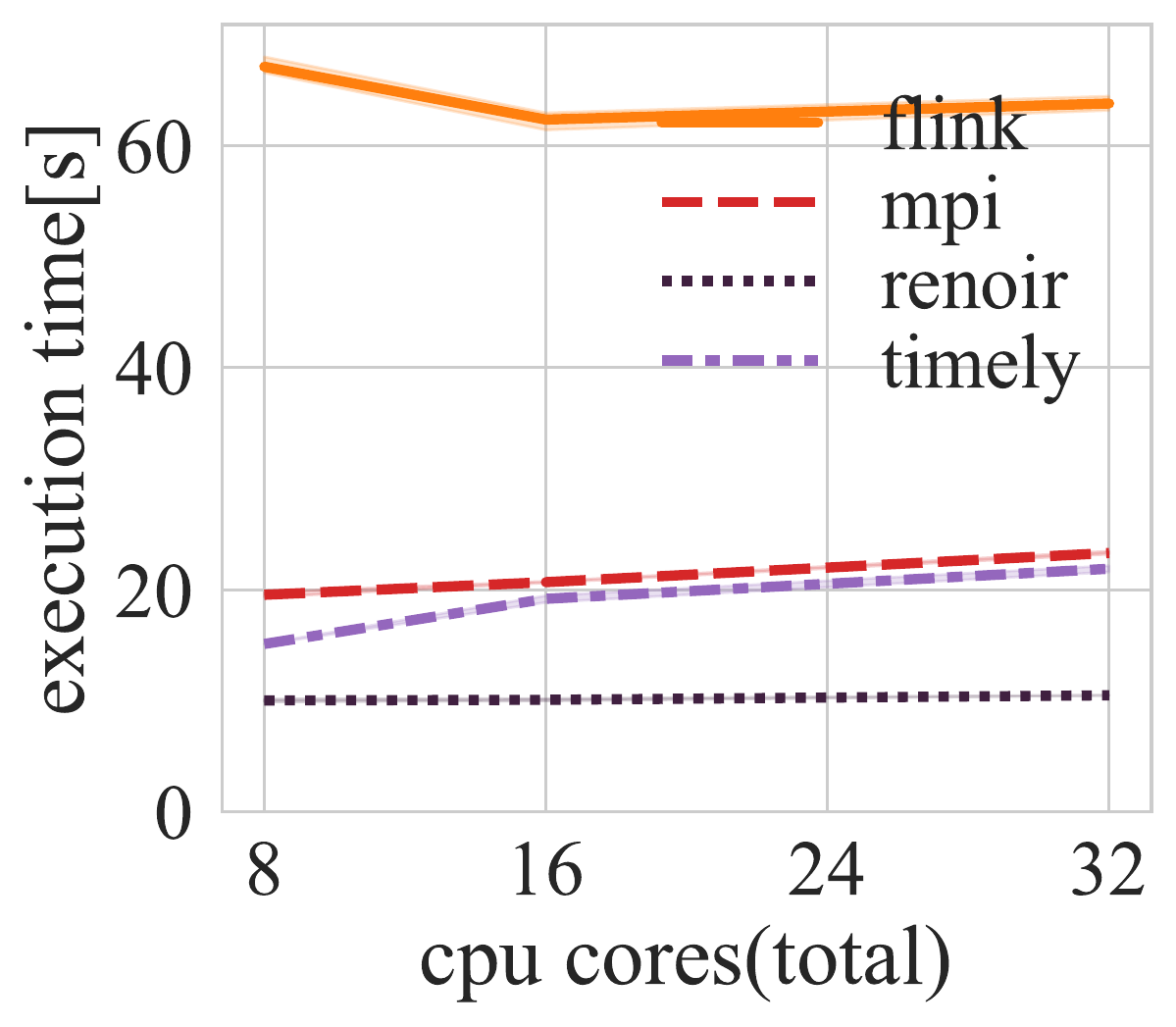}%
    \label{fig:eval:wc_weak}}
  \subfloat[\wc (optimized)]{%
    \includegraphics[width=0.16\textwidth]{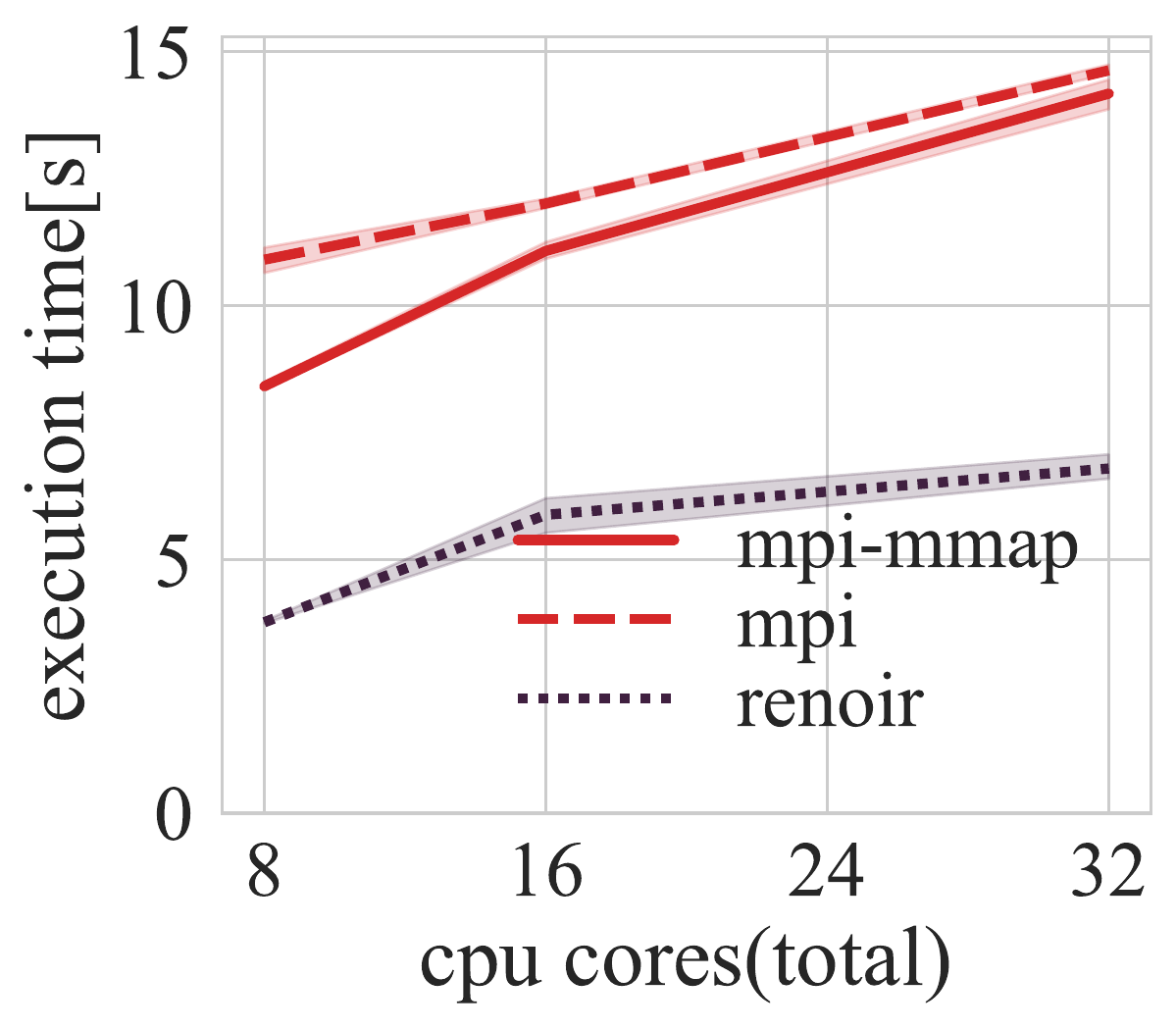}%
    \label{fig:eval:wc_opt_weak}}
  \caption{Weak scalability for \wc (1GB per host).}
  \label{fig:eval:weak}}
\end{figure}

In terms of horizontal scaling, \systemtt and \flink achieve near linear
scalability: when moving from 1 to 4 hosts, we measure a speedup of
3.65$\times$ for \systemtt and 3.57$\times$ for \flink.  Indeed, the most
expensive operations, namely reading and parsing the file, and performing a
partial count, are executed in parallel without synchronization across
processes.
\mpi has a speedup of 2.59$\times$ and the \texttt{MPI-mmap} version has a
scalability of only 1.39$\times$: again, we suspect this may be due to a less
efficient serialization strategy that introduces more network
traffic. However, further optimizing serialization for the specific problem at
hand would require significant effort, while \systemtt offers better
performance without any complexity about serialization being exposed to the
developer.
\timely presents a similar speedup of 2.29$\times$.  This is a general
characteristic we observed in our experiments: \timely exhibits a lower
scalability than \systemtt.
\rev{As an additional metric of the performance of the systems, we measured
  weak scalability for \wc, reporting the results in \fig{eval:weak}.  In this
  experiment, we used a dataset of 1GB when running on a single host and
  increased the size of the dataset proportionally to the number of hosts,
  thus keeping a ratio of 1GB per host.
  When considering the non-optimized version of the benchmark
  (\fig{eval:wc_weak}), all systems under test show near constant execution
  time when moving to more hosts.  For instance, the execution time time of
  \systemtt increases by only 5\% when moving from 1 to 4 hosts.
  When considering the optimized version (\fig{eval:wc_opt_weak}), both
  \systemtt and \mpi increase the execution time by more than 65\% when moving
  from 1 to 4 hosts.  This is due to the limited complexity of the task, which
  makes the overhead of communication and coordination across hosts more
  visible.}
  
\subsubsection{Vehicle collisions (\collisions)}

This workload presents two differences with respect to \wc:
\begin{inparaenum}[(i)]
  \item it computes three distinct output results starting from the same input
  data;
  \item each computation involves more operators.
\end{inparaenum}

\systemtt remains the fastest system, completing the task in 24.87s on a
single host and in 6.51s on 4 hosts.  In comparison, \flink requires
68.15s on one host and 26.02s on 4 hosts, while \mpi requires 42.41s on
a single host and 10.25s on 4 hosts.
With respect to \wc, the increased computational complexity of the operators
partially masks the overhead of \flink, reducing the gap with respect to \mpi
and \systemtt.
One possible explanation for the lower performance of \mpi is that the
computations required to obtain the three output results are executed one
after the other: indeed, implementing the strategies to run them in parallel
would require a radical change of the code and would expose the additional
complexity of managing parallel execution.
Conversely, in \flink and \systemtt, the computations of different output
results can be executed in parallel without any additional code complexity.
Furthermore, intermediate results can be reused: in the specific case of
\collisions, \flink only parses the input once, and reuses the parsed input
for all computations.  We avoided this optimization in \systemtt to enable a
more fair comparison with \mpi.

\subsubsection{KMeans (\kmeans)}

We used \kmeans to assess the performance of \systemtt and competing platforms
in executing a \rev{machine learning} algorithm.  Recall that the algorithm takes in
input 2D points and clusters them around a set of centroids.  At each
iteration, the algorithm analyzes all points and computes the distance with
the current set of centroids.  It updates the position of centroids until they
become stable.
New centroids are broadcast at each iteration.
We performed various experiments changing the size of the input and the number
of centroids.  We set a maximum number of 30 iterations to limit the execution
time of experiments.

\fig{eval:kmeans_30_10} shows the performance and scalability of the systems
under test when using an input of 10M points (about 200MB of data) and 30
centroids.
\systemtt and \mpi show comparable performance: their execution time is almost
identical on a single host, but \mpi scales slighly better and becomes about
30\% faster on 4 hosts.
\flink is significantly less efficient (notice the log-scale in
\fig{eval:kmeans_30_10}, \fig{eval:kmeans_300_10}, \fig{eval:kmeans_30_100})
in performing this iterative task, with an execution time that is from
25$\times$ to 40$\times$ higher than \systemtt.

\fig{eval:kmeans_300_10} shows the performance of the three systems when
increasing the number of centroids from 30 to 300.  Increasing the number of
centroids adds computational complexity, as input points need to be compared
with each centroid at each iteration.
This decreases the gap between \flink and the other two systems, as the cost
of communication and synchronization becomes smaller in comparison with the
cost of computing.  \systemtt remains the fastest system, with \mpi being
marginally (at most 10\%) slower, and \flink reducing its gap but still
remaining more than 10$\times$ slower than \systemtt.

\fig{eval:kmeans_30_100} shows the performance of the three systems when
increasing the number of points from 10M to 100M (about 2GB of data).
This workload stresses communication, as points are transferred at each
iteration.  Also in this case, \systemtt shows a level of performance that is
comparable to \mpi, with a slightly better scalability, while \flink is about
70 times slower, indicating the effectiveness of \systemtt in terms of
communication.

\subsubsection{Pagerank (\pagerank)}

The \pagerank benchmark may be implemented in different ways.  We present
three different implementations, each of them mimicking a reference
implementation in one of the other systems under test.  This enables us to
explore a larger area of design and implementation strategies, while offering
a comparison with other systems that is as fair as possible.

The first approach stores the current rank of nodes in a mutable state.  The
list of adjacent nodes is replicated within each process as an additional
immutable state.  At each iteration, each process distributes part of its
current rank to adjacent nodes, which will use it to update their rank in the
next iteration.
We implemented this approach in \mpi and we present a performance comparison
in \fig{eval:pagerank-mpi}.
\mpi uses messages between nodes to distribute the current rank.  \systemtt
mimics the same behavior by storing the current rank in a stateful operator (a
\texttt{rich\_map}) and keeping a single copy of the list of adjacent nodes
per process; the rank is distributed to adjacent nodes by propagating it back
in the feedback stream of an iteration.
This is the most efficient implementation of the algorithm: \systemtt and \mpi
show similar performance, with \systemtt completing the task about 10\% faster
than \mpi.

The second approach mimics the reference implementation found in the official
repository of \flink.  \flink considers both the list of adjacent nodes and
the current rank of nodes as two streams that are joined together at each
iteration to produce the new rank for each node (a new stream).  It then
compares the current and previous rank to check convergence and terminate the
loop.
\fig{eval:pagerank-flink} shows the performance of this implementation for
\flink and \systemtt.  Clearly, this implementation is not optimal for
\systemtt, which can store the list of adjacent nodes as part of the state of
each process, as exemplified in the first approach.  Yet, \systemtt remains at
least 6$\times$ faster than \flink when using the same implementation
strategy.  However, the need to repeatedly join the input streams and transfer
them over the network affects scalability, which in this case is even
negative.

\rev{The ability to replicate the strategies used in \mpi speaks of the
  flexibility of the \systemtt execution model.  \systemtt spawns a process
  for each host, which shares the same address space with all the tasks
  running on the same host. Combined with the safety guarantees of the Rust
  programming language, this allows developers to implement optimized
  patterns of memory access.  In the specific case of \pagerank, all tasks
  within a host can access the same copy of the list of adjacent nodes.  This
  is simply not possibile in engines like \flink, where users code is shipped
  to remote executors that are not under the control of developers.}

Finally, \fig{eval:pagerank-timely} uses the reference implementation of
\pagerank for Differential Dataflow, an abstraction built on top of \timely.
We mimic the same approach in \systemtt and we adopt the same workload to
ensure a fair comparison.  Specifically, this version of \pagerank uses
integer numbers instead of floating point numbers to represent the rank of the
nodes.  The choice to use integers is forced by \timely, which requires that
the types used in the feedback of a loop have mathematical characteristics
which floats do not have.
%
%
Interestingly, this implementation shows a speedup close to zero for \systemtt
but also for \timely.  Also in this case, \systemtt is consistently at least
20\% faster than \timely.

\subsubsection{Connected components (\connected)}

\connected is an iterative algorithm to compute the connected components of a
graph.  The algorithm iteratively updates the component to which each node
belongs: a components $c$ is represented by the numerical identifier of the
smallest node currently part of $c$.  If the algorithm discovers that a node
$n$ is connected to a component $c$ with an identifier smaller than $n$, it
assigns $n$ to $c$ and propagates the new association to the next iteration,
where nodes directly connected to $n$ are evaluated and possibly included into
$c$.
\flink natively supports this type of iterations that continuously update a
mutable state.  They are defined as \emph{delta-iterations} in the \flink
documentation.  In \systemtt, we implement a similar logic using the
\texttt{iterate} operator: we store the association of nodes to components in
the iteration state, and we propagate to the next iteration only the
associations that have changed during the current iteration.

\fig{eval:connected} shows the execution time and scalability of the systems
under test.  Scalability in \connected is limited by the need to propagate
state changes to all hosts at each iteration.
This is visible for all the systems under test, but in particular for \mpi,
which has a maximum speedup of only 1.1$\times$ when moving from 1 to 4
hosts.  In comparison, \flink and \systemtt have a speedup of 1.63$\times$ and
1.85$\times$, respectively.
In absolute terms, \systemtt is almost identical to \mpi on a single host
(1.81s vs 1.80s) but becomes faster on 4 hosts (1.38s vs 1.62s).
\flink remains about an order of magnitude slower with  total execution time
that moves from 16.89s on a single host to 10.37s on 4 hosts.

\subsubsection{Enumerate triangles (\enumtriangles)}

\enumtriangles is a graph algorithm that does not require multiple iterative
steps, as it needs to verify a local property: which triples of nodes are
directly connected with each other, forming a triangle.

\flink implements this algorithm with a join operation to complete the
triangle.  Instead, \mpi stores the complete adjacency list in memory, which
is more efficient as it enables random access.  In principle, \systemtt could
also exploit shared state within processes to optimize the job.  However, we
decided to implement the same approach used by \flink, which is easier to
express in a dataflow engine.

\fig{eval:enumtriangles} shows the execution time and scalability of the
systems under test when computing \enumtriangles.  \mpi is the fastest
systems, more than 3$\times$ faster than \systemtt on a single host.  Thanks
to a better scalability (3$\times$ vs 2.2$\times$ speedup), \systemtt reduces
the gap on 4 hosts, where \systemtt and \mpi complete the task in 6.19s and
2.61s, respectively.
\flink remains significantly slower, with an execution time of 66.8s on 4
hosts.

\subsubsection{Transitive closure (\transitive)}

\transitive is an iterative graph algorithm that presents different
characteristics with respect to \connected.  In particular, it iteratively
enriches a partial result that is proportional to the number of edges
(quadratic with respect to the number of nodes).

\fig{eval:transitive} shows the execution time and scalability of the systems
under test when performing this task.  \systemtt shows execution times from
5.98s on 1 host to 2.63s on 4 hosts (2.26$\times$ speedup), while \flink
is about 8 times slower, with a maximum speedup of 2.4$\times$ when moving
from 1 to 4 hosts.  Consistently with what we reported for \enumtriangles, the
\mpi implementation yields better results by saving the adjacency list as
mutable state, leading to an execution time of 1.64s on 4 hosts
(2.61$\times$ speedup).

\subsection{Performance: streaming workloads}
\label{sec:eval:perf_stream}

In this section, we evaluate the performance of \systemtt and alternative
solutions for stream processing workloads using \nexmark.
We consider all original \nexmark queries (Q1--Q8) and the passthrough query
Q0, which measures the monitoring overhead, that is, the time for analyzing
the entire input data without performing any concrete data transformation.

We compare \systemtt with other stream processing platforms: \flink and
\timely.  We exclude from this analysis \mpi as it is not designed for
streaming workloads.
We adopt the reference \flink implementation available in the \nexmark
repository\footnote{\url{https://github.com/nexmark/nexmark}}, and the \timely
implementation made available by the authors as part of the Megaphone
project\footnote{\url{https://github.com/strymon-system/megaphone}}.
For \systemtt, we generate input data using a parallel iterator source, making
sure that it is consistent with the specification of the benchmark.

\begin{figure}[t]
  \centering
  \includegraphics[width=0.9\columnwidth]{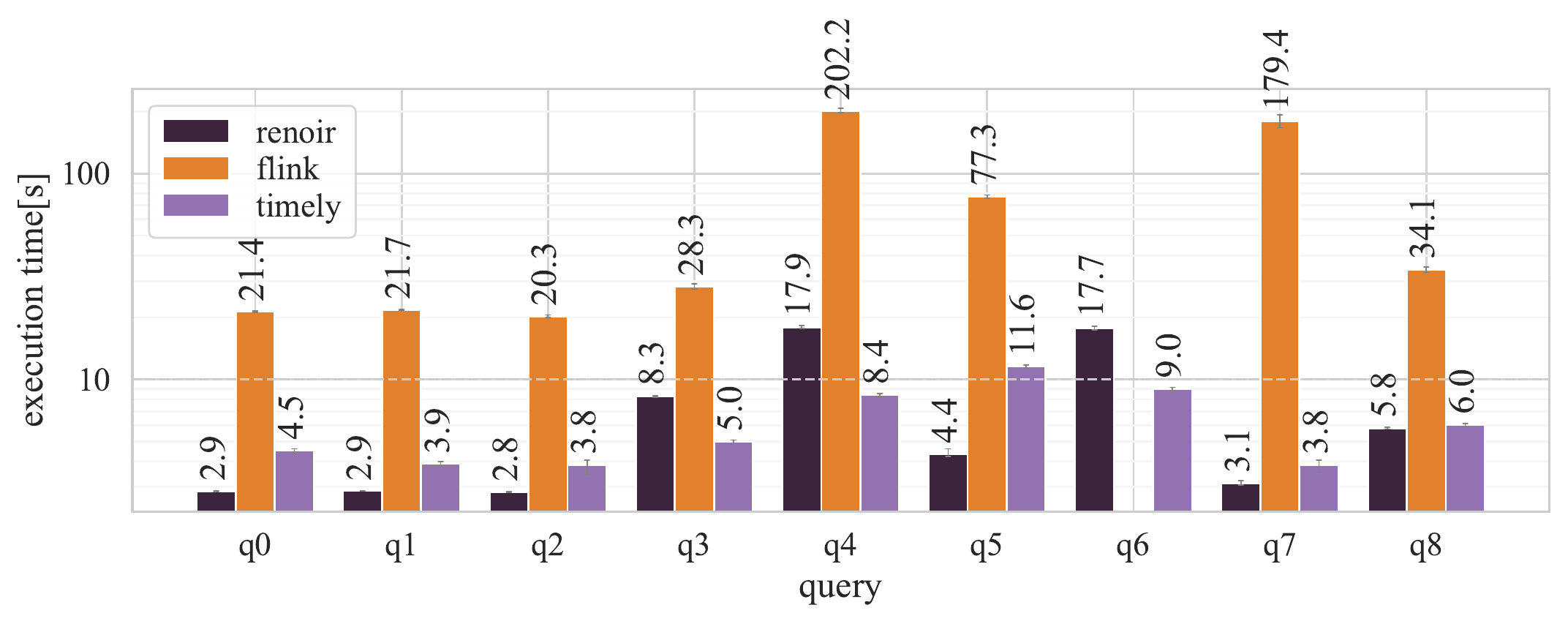}
  \caption{\nexmark queries: execution time with 32 cores.}
  \label{fig:eval:nexmark_time}
\end{figure}

\fig{eval:nexmark_time} shows the time required to process the entire workload
for all queries using 4 hosts (32 cores). \systemtt consistently outperforms
\flink, completing queries from 3.4$\times$ to 57.8$\times$ faster, depending
on the query. \flink cannot even terminate query \texttt{Q6}, which \systemtt
computes in about 17.7s.
The execution times of \systemtt and \timely are comparable, and \systemtt is
faster in 6 out of 9 queries.  In queries Q4 and Q6 \timely is about twice as
fast as \systemtt.  We attribute these results to the different ways in which
the two systems implement the windowing logic required in the queries: the
\timely implementation adopts a custom windowing implementation that
simplifies the way in which watermarks are handled.  While we could replicate
the same approach in \systemtt by writing ad-hoc windowing operators for the
specific queries, we decided to use the standard window operator offered by
the framework, which simplifies the code.

\begin{figure}[t]
  \centering
  \subfloat[Query Q2]{%
    \includegraphics[width=0.32\columnwidth]{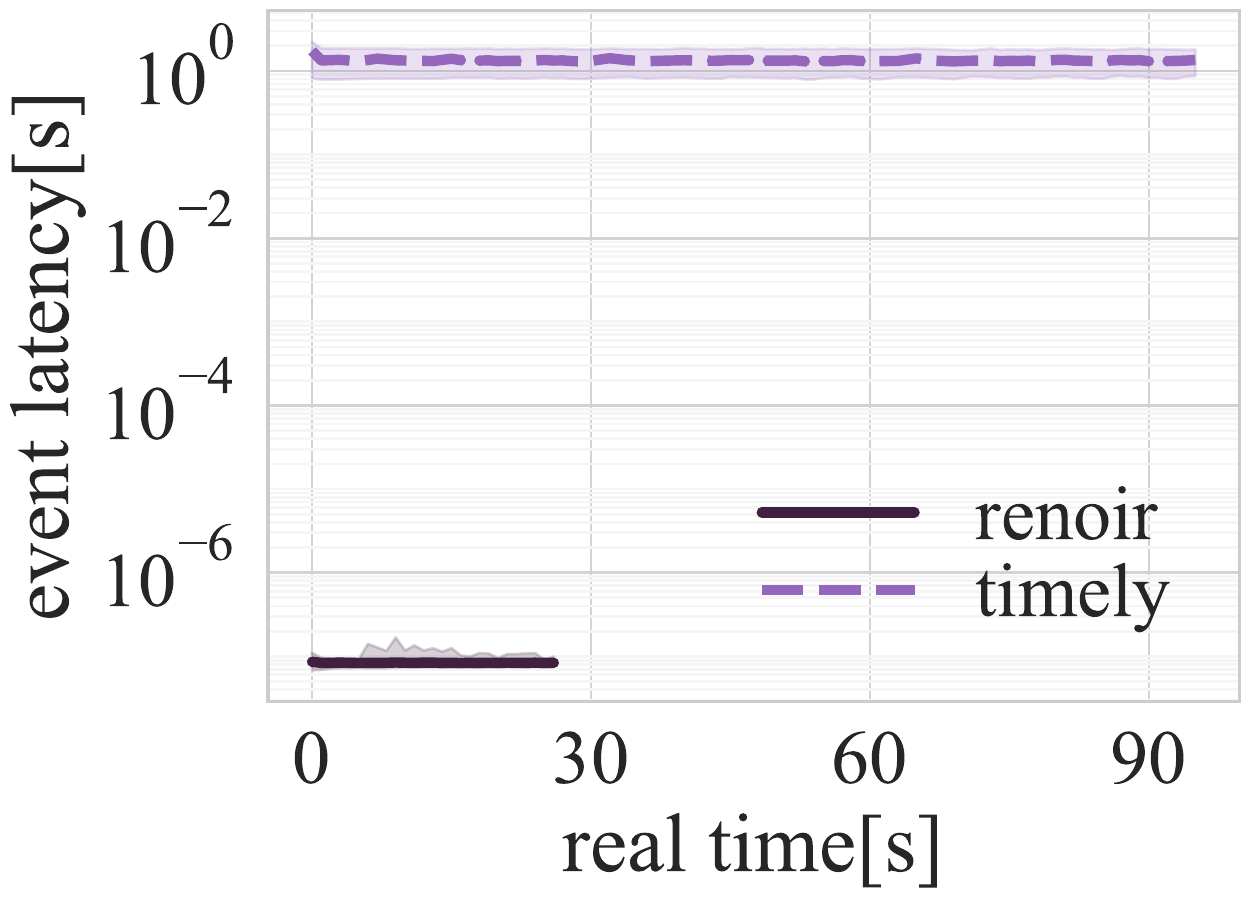}%
    \label{fig:eval:nexmark_latency_q2}}
  \subfloat[Query Q3]{%
    \includegraphics[width=0.32\columnwidth]{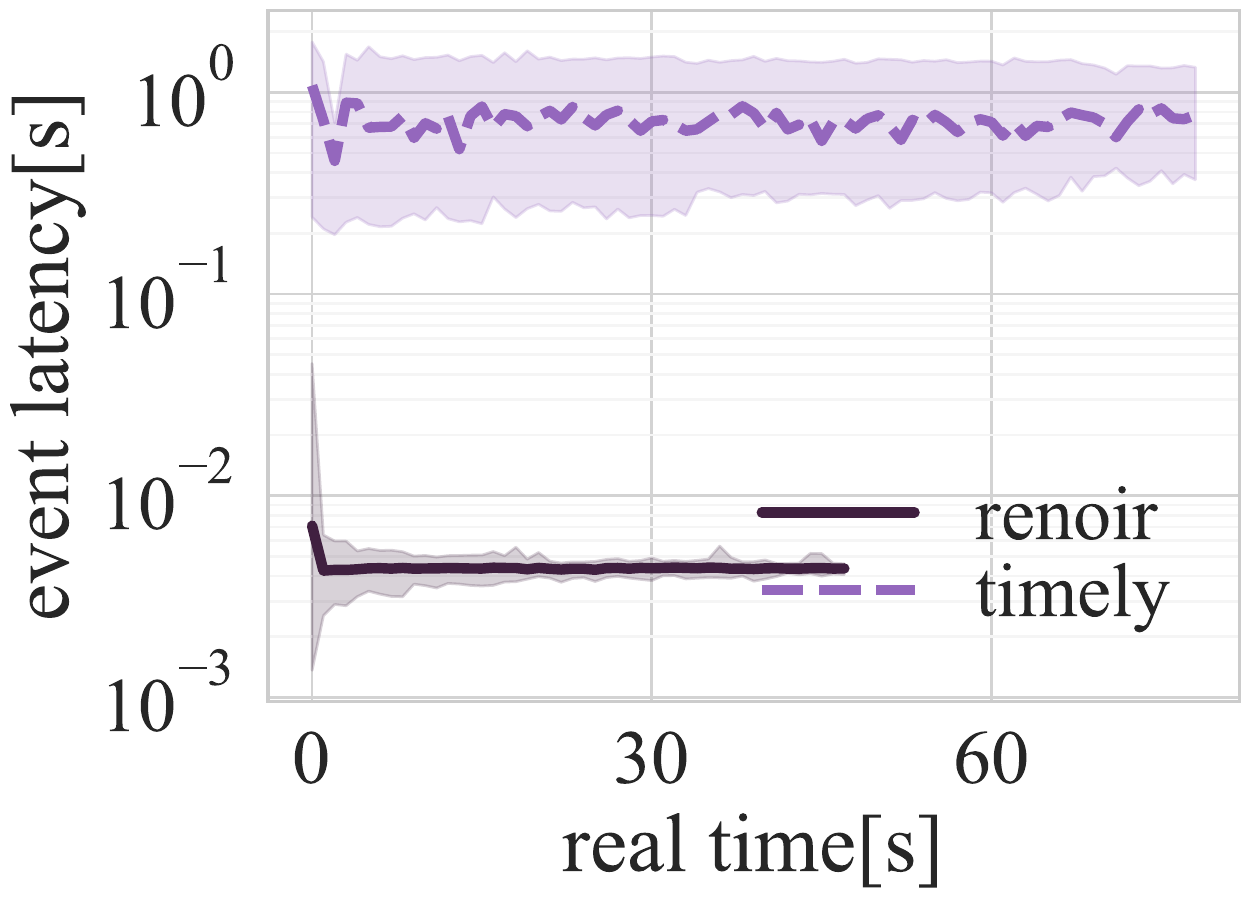}%
    \label{fig:eval:nexmark_latency_q3}}
  \subfloat[Query Q5]{%
    \includegraphics[width=0.32\columnwidth]{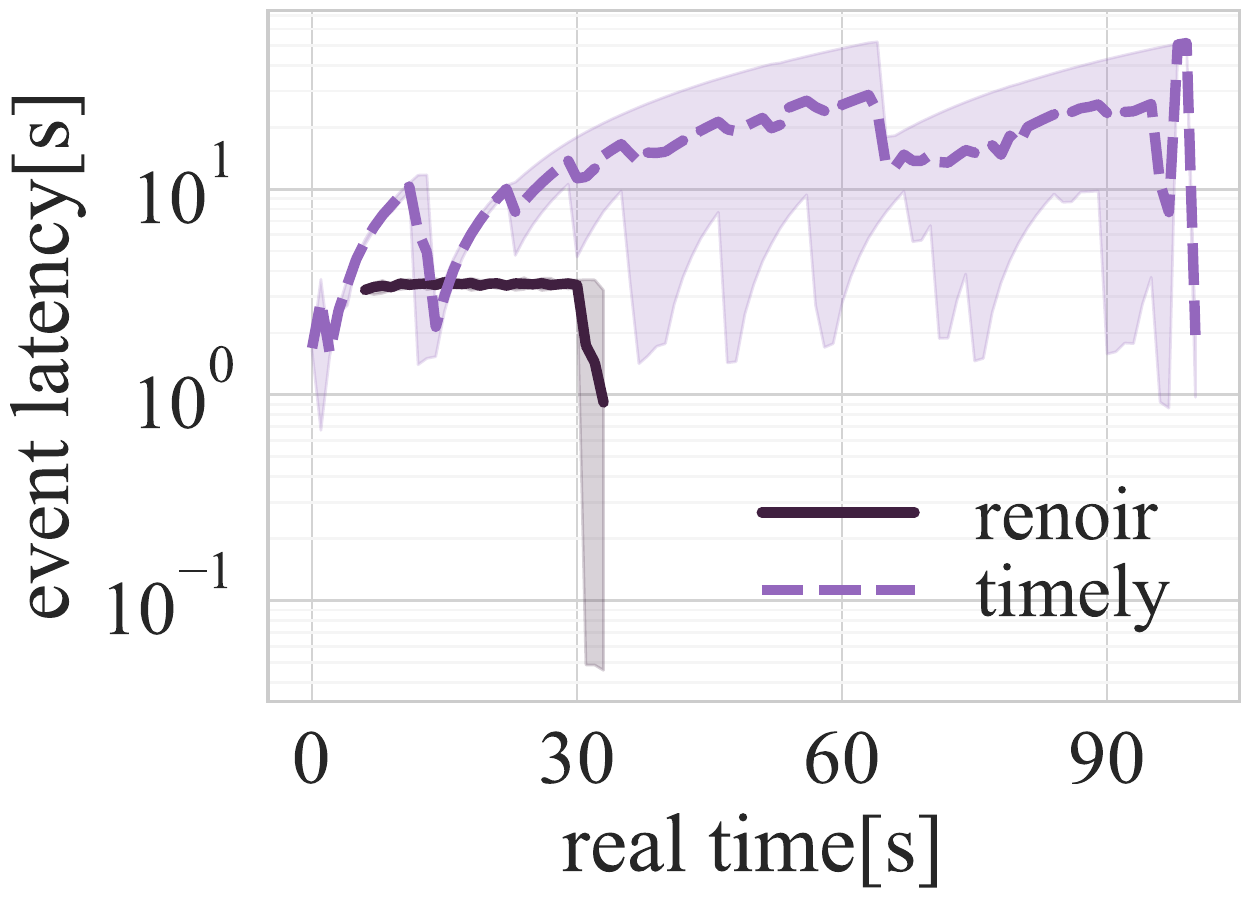}%
    \label{fig:eval:nexmark_latency_q5}}
  \caption{\nexmark: latency for queries Q2, Q3, Q5.}
  \label{fig:eval:nexmark_latency}
\end{figure}

\fig{eval:nexmark_latency} shows the latency of \systemtt for three
representative queries with heterogeneous characteristics (Q2, Q3, and Q5),
and compares it with \timely, the system that presents a similar execution
time.
\fig{eval:nexmark_latency} plots the mean latency over time in 1s windows,
with the 99th percentile band (shaded area around each line).
Recall that, to measure latency, we use a single source and a single sink
placed on the same machine, consequently, the results in
\fig{eval:nexmark_latency} are not directly comparable with those in
\fig{eval:nexmark_time}.

For \timely, due to the cooperative scheduling implementation, developers need
to explicitly use timestamps to indicate how frequently to alternate between
sending (batches of) events and processing them, which clearly affects
latency.
We experimented with different fixed and dynamic intervals as detailed below,
\fig{eval:nexmark_latency} reports the results when processing 100ms of
events at each round.

Query Q2 (\fig{eval:nexmark_latency_q2}) requires a single stage of stateless
computations.  In this setting, \systemtt never splits the input data into
partitions after reading from the sources, and executes the entire computation
sequentially, which leads to a latency of about 80ns.  Instead, the latency
in \timely is dominated by the evaluation interval, which is set to 100ms:
reducing this interval to 1ms indeed improves latency, but it still remains
around 10ms, while the overall execution time further increases to more than
140s.  In comparison, the execution time of \systemtt is below 30s.

Query Q3 (\fig{eval:nexmark_latency_q3}) includes multiple stages of
computation and a join.
In \systemtt, this introduces inter-stage communication, which increases the
latency to about 5.5ms.  The latency of \timely is about 900ms.  We
considered an evaluation interval of 100ms as in Query Q2, which leads to a
throughput that is comparable to that of \systemtt.

In query Q5 (\fig{eval:nexmark_latency_q5}), the latency is dominated by the
presence of windows.  The average latency is below 3s for \systemtt and over
12.5s for \timely.  Also, the latency is more stable in \systemtt, with a
standard deviation of 1.1s compared to 14.3s of \timely.

\subsection{Scalability on a single host}
\label{sec:eval:perf_single}

\begin{figure}[t]
  \centering
  \subfloat[\collatz]{%
    \includegraphics[width=0.33\columnwidth]{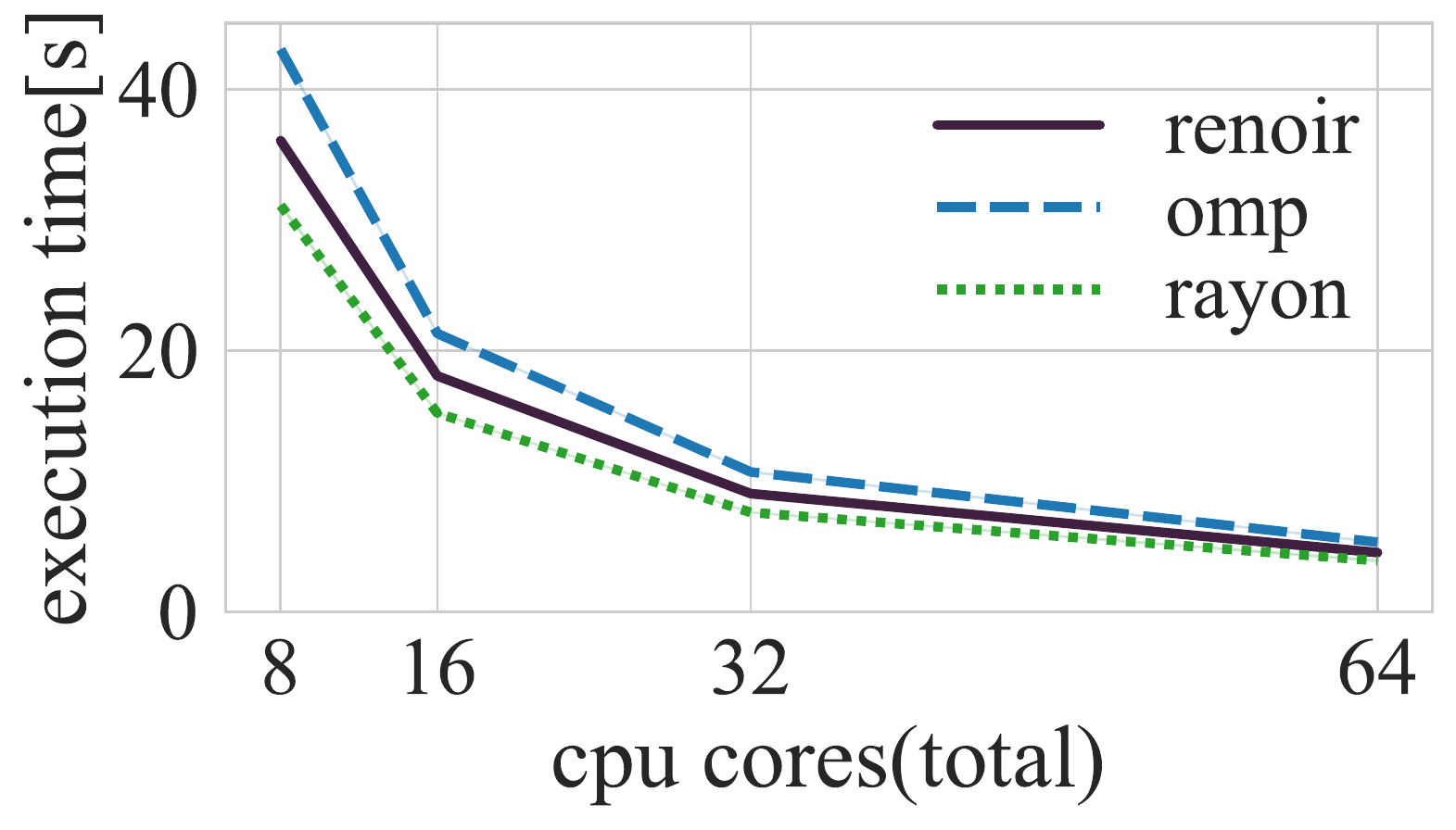}%
    \label{fig:eval:vertical:collatz}}
  \qquad
  \subfloat[\wc]{%
    \includegraphics[width=0.33\columnwidth]{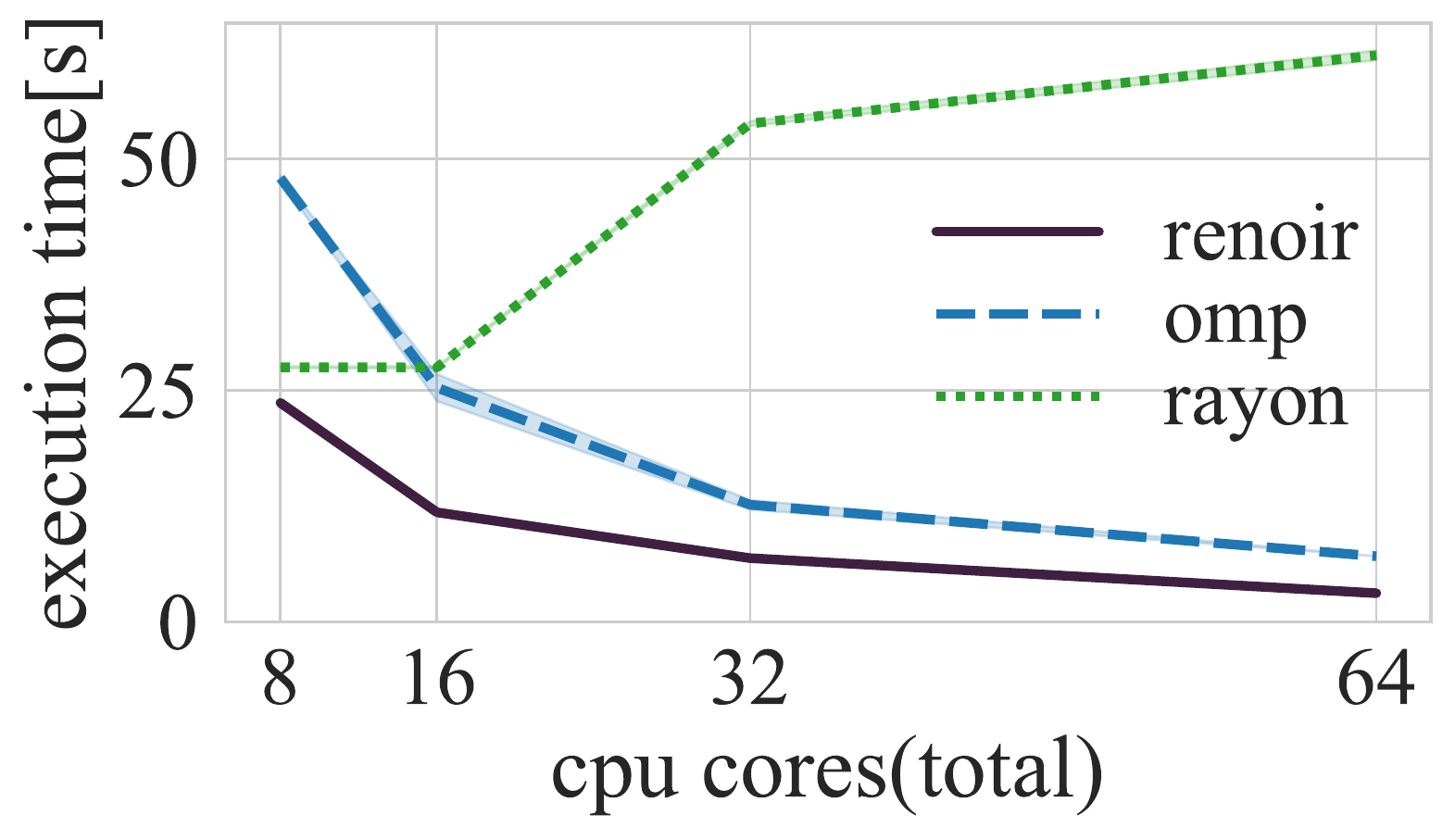}
    \label{fig:eval:vertical:wc}}\\
  \subfloat[\kmeans: 300 centroids 10M points (left), 30 centroids 10M
    points (center), 30 centroids 100M points(right)]{%
    \includegraphics[width=0.33\columnwidth]{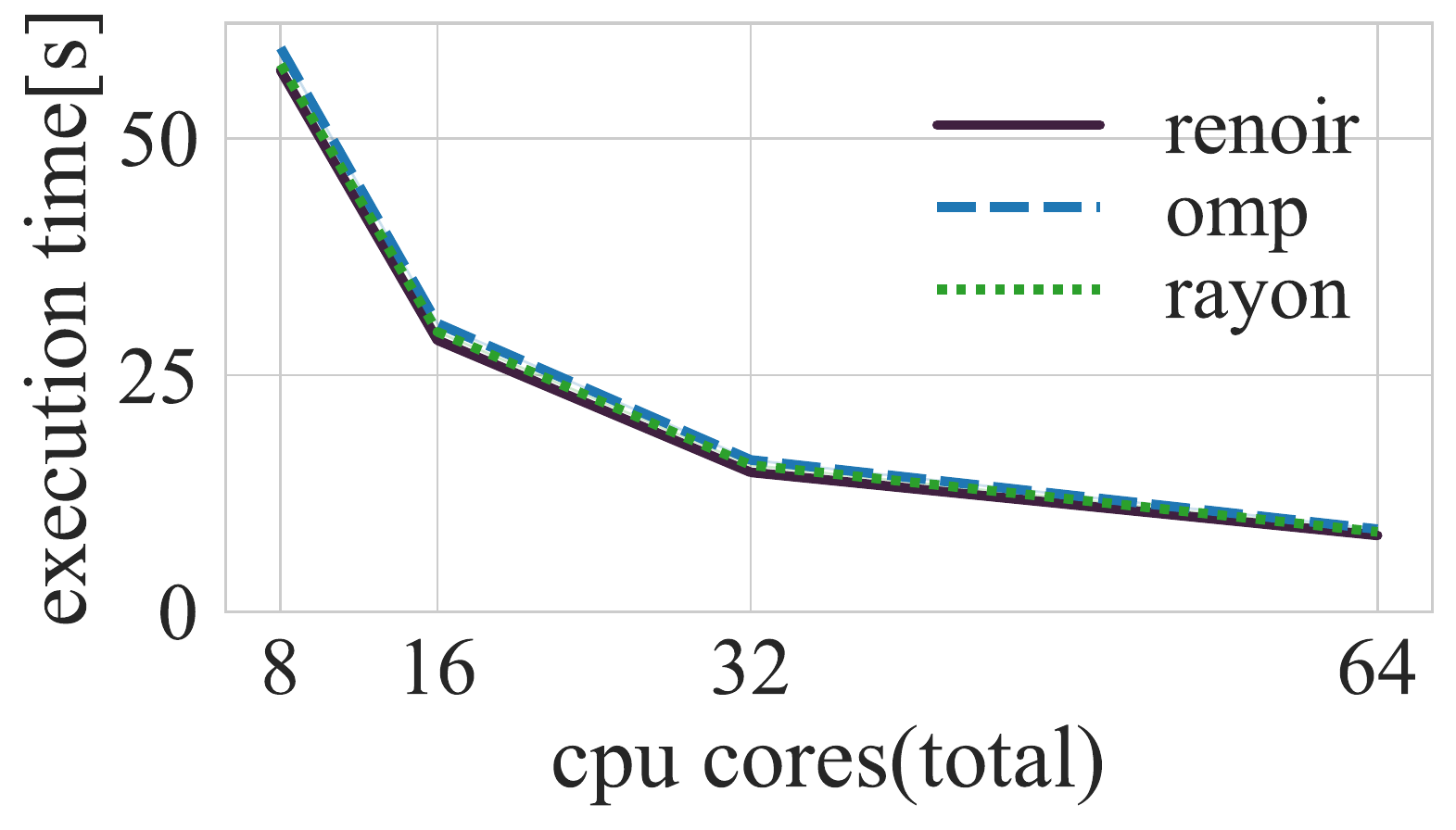}%
    \includegraphics[width=0.33\columnwidth]{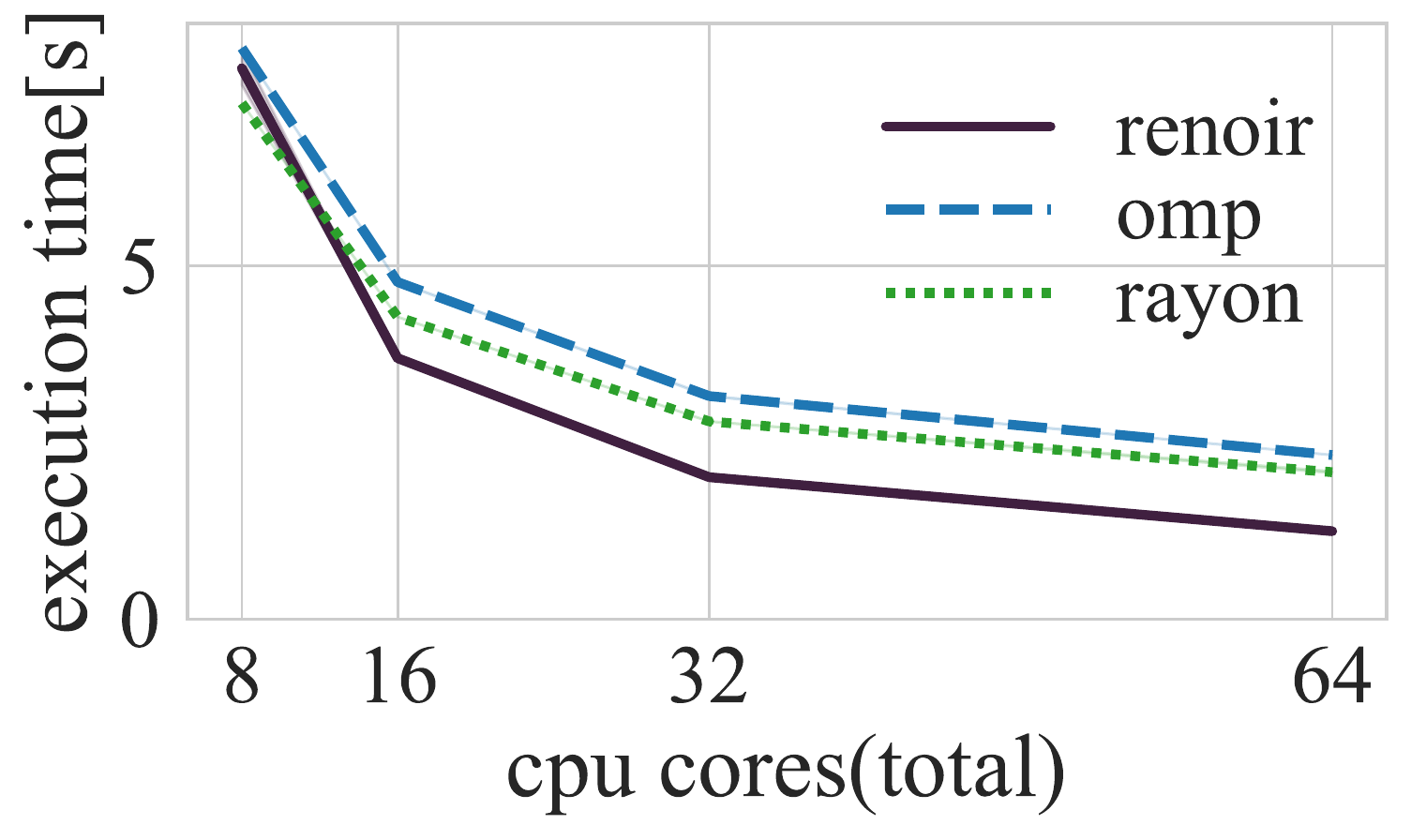}%
    \includegraphics[width=0.33\columnwidth]{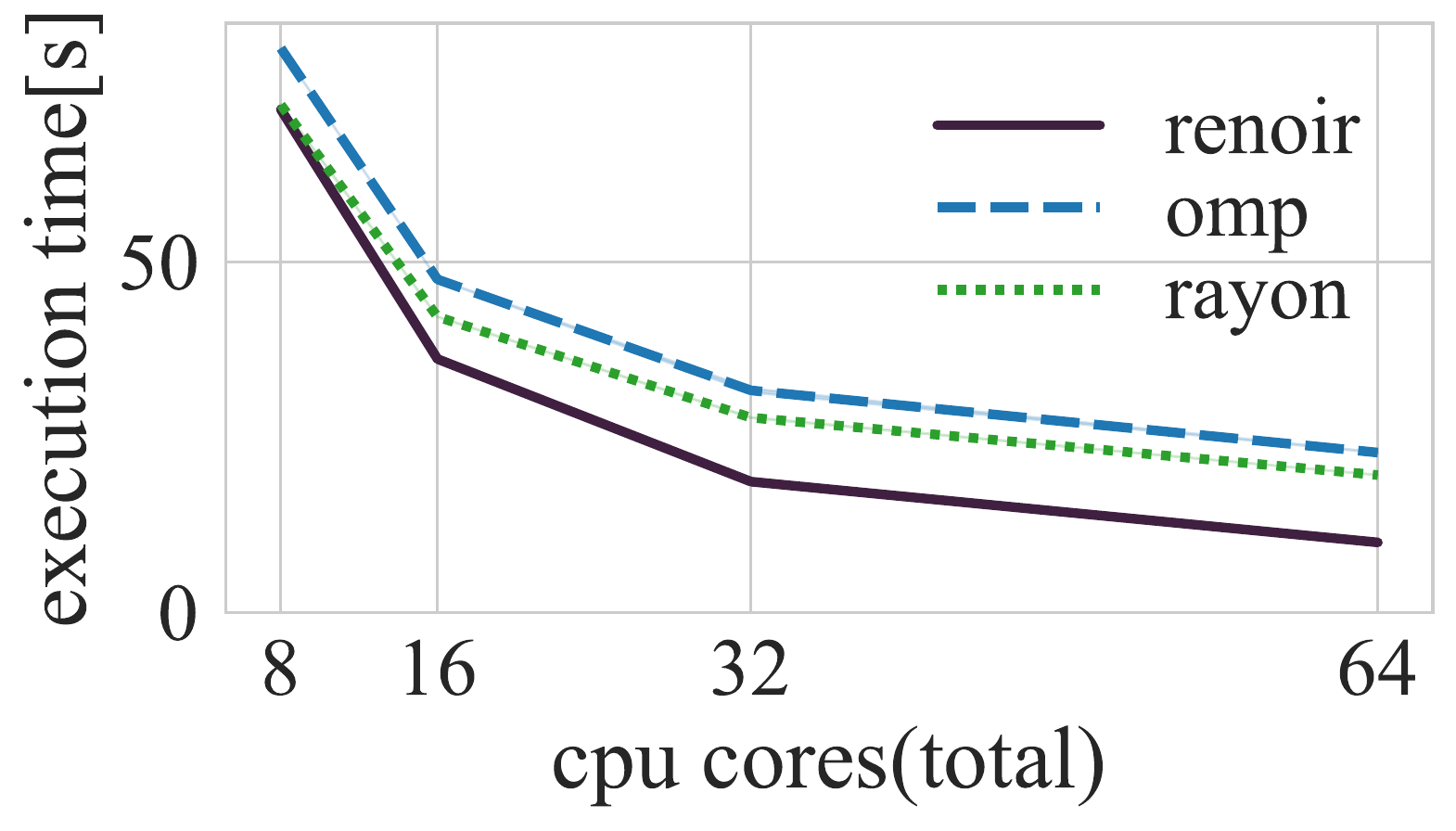}%
    \label{fig:eval:vertical:kmeans}}
  \caption{Scalability on a single host.}
  \label{fig:eval:vertical}
\end{figure}

\systemtt is designed to enable large-scale data analysis on a cluster of
machines.  However, it can also be used as an efficient library for parallel
computations within a single process.  This section focuses on this capability
and compares the performance of \systemtt with alternative state-of-the-art
solutions for parallel processing: \omp and \rayon.
With \omp, developers annotate blocks of C/C++ code that can be executed in
parallel, the compiler then generates the parallel implementation using a
fork-join model of execution.  With \rayon, developers use an iterator-like
api that is translated to a set of tasks executed by a fixed size thread pool.
All three systems are configured to use a level of parallelism equal to the
number of processors available.

For the experiments presented in this section, we use \texttt{c6a} EC2
instances, which offer 3rd generation AMD EPYC processors.  For each
experiment, we use a single instance and we configure the systems under test
for optimal exploitation of the compute resources available.  We start from a
\texttt{c6a.2xlarge} instance (8 CPUs, 16GB RAM) and run the experiments
scaling up to a \texttt{c6a.16xlarge} instance (64 CPUs, 128GB RAM).

We use the \collatz, \wc and \kmeans benchmarks to cover different scenarios.
\collatz is an iterative algorithm that process numbers in parallel. It does
not require synchronization across numbers, but presents highly heterogeneous
execution times for different numbers.
\wc includes data transformations (to split lines into words) and an
associative aggregation (to count the number of occurrences).
\kmeans is an iterative algorithm that requires synchronization at each
iteration.

\fig{eval:vertical:collatz} shows the performance we measure for \collatz.
The execution time is comparable for all the systems under test: \rayon is
about 30\% faster than \omp and about 15\% faster than \systemtt, and the
relative gap remains almost constant when changing the number of cores.
Due to the unbalanced nature of the task, these differences may be attributed
at least in part to different partitioning strategies: static in \systemtt and
dynamic in \omp and \rayon.

\fig{eval:vertical:wc} shows the performance we measure for \wc.  \system
shows consistently better performance than \omp. With this workload, \rayon
exhibits a strange behavior, with execution time increasing when using more
cores. This kind of problem when using many threads has been documented and
may be caused be the
scheduler\footnote{\url{https://github.com/rayon-rs/rayon/issues/795#issuecomment-1155652842}}.

\fig{eval:vertical:kmeans} shows the performance we measure for \kmeans.  When
considering 300 centroids (\fig{eval:vertical:kmeans}, left), all systems
under test show nearly identical performance and scalability. When considering
30 centroids and 10M points (\fig{eval:vertical:kmeans}, center) both \rayon
and \systemtt show better performance than \omp.  \systemtt also scales
better, and becomes almost twice as fast as \rayon with 64 cores. The same
pattern emerges when increasing the number of points to 100M
(\fig{eval:vertical:kmeans}, right): the execution times increase, but the
relative performance between the systems remains similar.  The performance
advantage may be explained by the native support for iterations and
partitioning: while \omp and \rayon need to join to the main thread after each
iteration, \systemtt keeps the points partitioned, processes partitions
independently from each other, and only synchronizes the tasks to collect the
new centroids.

\subsection{Discussion}
\label{sec:eval:discussion}

We evaluated \systemtt with very heterogeneous workloads, ranging from batch
to streaming and even parallel computations, including iterative, graph
processing \rev{and machine learning} algorithms. We compared it with
state-of-the-art solutions in each field.

First of all, our analysis reveals the benefits of the programming interface
of \system.  Indeed, writing programs with \systemtt (and \flink, which adopts
a similar programming abstraction) was considerably simple than using \mpi or
\timely.
Frequently, the added complexity may also become detrimental for performance:
despite experimenting alternative solutions, in several benchmarks \mpi shows
higher execution times than \system. Writing efficient \mpi solutions requires
significant effort in designing parts of the code that are outside the scope
of the task at hand, such as communication and buffering, and most solutions
may be sub-optimal with respect to what \systemtt achieves automatically with
efficient default strategies.  Moreover, when we found possible optimizations
in \mpi, we could frequently replicate them in \system.
While \timely also offers a high-level programming interface, it is more
complex than the one in \system.  Even when starting from implementations
provided by \timely developers, it was difficult for us to ensure that the
final algorithm was equivalent to that of the other platforms: for instance,
in \pagerank, we could not modify the provided implementation to use floating
point numbers as rank; similarly, in \nexmark, the results where highly
dependent on the strategies used to define timestamps.
In terms of absolute performance and scalability, \systemtt was always
comparable or faster than alternative solutions in all the scenarios we
tested, showing that it is suitable for a wide range of problems and diverse
hardware configurations.

In summary, our evaluation shows that \systemtt provides a good balance
between ease of use and performance, allowing to easily develop code that
solves the problem at hand in a way that outperforms more complex solutions,
developed with lower-level programming primitives.

\section{Related Work}
\label{sec:related}

Our work focuses on programming models and platforms for parallel, distributed
data processing.  In this context we can demark four main categories of
related systems.

\fakeparagraph{General purpose dataflow platforms}
The dataflow model we consider in this paper has attracted increasing
attention over the last several years, and most mainstream data processing
platforms rely on this model~\cite{akidau:VLDB:2013:MillWheel,
  toshniwal:SIGMOD:2014:Storm, akidau:VLDB:2015:dataflow,
  kulkarni:SIGMOD:2015:heron,rocklin_dask_2015}.
The Flink system we use for our evaluation is a mature commercial product,
representative of these platforms, and often cited for its good level of
performance~\cite{marcu:CLUSTER:16:SparkVsFlink}.
Dataflow platforms provide general API that work well with structured and
unstructured data, but they have large margins of improvements in terms of
performance, as shown in our evaluation.

\fakeparagraph{Specialized libraries for dataflow platforms}
To simplify the implementation of complex algorithms, some platforms also
offer higher-level libraries for specific domains.  Prominent examples are the
libraries to process structured data~\cite{armbrust:SIGMOD:2015:SparkSQL},
which convert declarative queries from SQL-like languages to dataflow
programs, often providing unified abstractions for batch and stream processing
of structured data~\cite{sax:BIRTE:2018:StreamsAndTables,
  begoli:SIGMOD:2019:OneSQL}.  The conversion enables automated query
optimizations, which are common in database systems.
Other examples of libraries range from machine
learning~\cite{meng:JMLR:2016:MLlib} to graph
processing~\cite{gonzalez:SOSP:2014:GraphX}, to pattern recognition in streams
of events~\cite{giatrakos:VLDB:2020:CER}.
As all these libraries generate programs that are executed on an existing
dataflow engine, their performance is limited by that of the underlying
platform.  This kind of abstractions could also be implemented on top of
\system.  Given the performance advantages we demonstrated, we believe \system
has the potential to bring significant improvements in domains like structured
data or graph data processing.
\rev{We are currently developing a library for machine learning on top of
  \system.  Preliminary experiments show promising results, in line with those
  measured for the k-means machine learning algorithm we analyzed in
  \s{eval}.}

\fakeparagraph{Specialized engines}
To overcome the performance limitations of the previously cited systems, an
alternative is to build an optimized engine using a compiled language and
expose a domain-specific high-level API.  Some systems choose a structured
query language~\cite{dageville_snowflake_2016,rlink,surrealdb}, others define
operations on dataframes~\cite{petersohn:2020:modin}.  Previous works also
showed promising results using a single multi-core
machine~\cite{polars,raasveldt_duckdb_2019}.
By restricting to a specific domain, this kind of systems trades
expressiveness in favor of performance: high-level abstractions are translated
to calls to a set of prepared, optimized components that often outperform
current general purpose dataflow systems.
\system chooses a more general approach, allowing user defined functions that
can work effectively in multiple domains and with both structured and
unstructured data.

\fakeparagraph{Single machine dataflow platforms}
Some systems optimize the use of resources on a single machine.  For instance,
StreamBox and BriskStream target multi-core
machines~\cite{miao:ATC:2017:StreamBox,zhang_briskstream_2019}, while SABER
considers heterogeneous hardware platforms consisting of multi-core CPUs and
GPUs, which are increasingly available in modern heterogeneous
servers~\cite{koliousis:SIGMOD:2016:SABER}.  By building on a compiled
language, \system simplifies the access to hardware resources with respect to
JVM-based systems.  In fact, we already experimented with OpenCL-based
implementations of operators that exploit GPUs, and we plan to further explore
this line of research in future work.

\fakeparagraph{Extensions to the dataflow programming model}
Fernandez et al.~\cite{fernandez:ATC:2014:StateExplicit} introduce an
imperative programming model with explicit mutable state and annotations for
data partitioning and replication.  TSpoon~\cite{affetti:JPDC:2020:TSpoon}
extends the dataflow abstraction with additional features and guarantees, such
as transactional semantics.
These efforts are orthogonal to our work, which mostly targets efficient
system design and implementation, rather than investigating new models. 
Naiad~\cite{murray:SOSP:2013:naiad} and its timely dataflow model enrich
dataflow computations with explicit timestamps that enable implementing
efficient coordination mechanisms.  We compared \system with the Rust
implementation of the timely dataflow model in \s{eval}.
Finally, modern data processing systems provide fault-tolerance mechanisms to
recover from software and hardware failures.
In \system we are implementing fault-tolerance mechanisms building on
consolidated approaches such as asynchronous
snapshots~\cite{carbone:2015:snapshots}, which bring negligible runtime
overhead.  However, as fault-tolerance is still in an experimental state
in \system, we disabled it in all the systems used in our
evaluation (see \s{eval}) for a fair comparison.

\section{Conclusions}
\label{sec:conclusions}

This paper introduced \system, a novel data processing framework written in
Rust.  \system provides all core features of state-of-the-art data processing
platforms -- unified batch and stream processing, iterative computations,
windowing, time-based data analytics -- within the same, high-level processing
model.  At the same time, its design and implementation choices -- compiled
language, efficient memory and communication management, task allocation that
maximizes the use of processing resources -- yields up to more than an order
of magnitude improvements in throughput with respect to existing data
processing systems, rivaling and even outperforming custom MPI solutions in
some workloads.

Our research shows that the advantages of a high-level programming model are
not restricted to the simplicity in defining data analysis tasks.  If the
model is supported by an efficient execution platform, it can unlock
performance improvement that are hard to achieve with custom, manual
optimizations.

Moving from this observation, we plan to further contribute to the research on
data processing platforms focusing both on enriching the programming model,
for instance to support domain-specific operators, and on improving the
capabilities of the processing engines, such as supporting hardware
accelerators and dynamic scaling.


\bibliographystyle{elsarticle-num}
\bibliography{biblio}

\end{document}